\documentclass[superscriptaddress,11pt,preprintnumbers,nofootinbib]{revtex4}
\pdfoutput=1
\usepackage{multirow}
\usepackage[dvipdf,dvips]{graphicx}
\usepackage{color}
\usepackage{hyperref}
\usepackage{url}
\usepackage{subfigure}
\usepackage[usenames,dvipsnames]{xcolor}
\usepackage{amsmath}
\usepackage{amsfonts}
\usepackage{float} 
\usepackage{amssymb}
\usepackage{epsfig}
\usepackage{graphics}
\usepackage{euscript}
\usepackage{epstopdf}
\usepackage[utf8]{inputenc}
\usepackage[T1]{fontenc} 
\usepackage{slashed}
\usepackage{upgreek}
\usepackage{ctable}
\usepackage{booktabs}

\allowdisplaybreaks

\expandafter\ifx\csname ProvidesPackage\endcsname\relax\toks0=\expandafter{%
\fi\ProvidesPackage{diagrams}[2014/12/31 v3.94 Paul Taylor's commutative
diagrams]
\toks0=\bgroup}
\ifx\diagram\isundefined\else\expandafter\endinput
\fi

\edef\cdrestoreat{
\noexpand\catcode`\noexpand\@=\the\catcode`\@
\noexpand\catcode`\noexpand\#=\the\catcode`\#
\noexpand\catcode`\noexpand\$=\the\catcode`\$
\noexpand\catcode`\noexpand\<=\the\catcode`\<
\noexpand\catcode`\noexpand\>=\the\catcode`\>
\noexpand\catcode`\noexpand\:=\the\catcode`\:
\noexpand\catcode`\noexpand\;=\the\catcode`\;
\noexpand\catcode`\noexpand\!=\the\catcode`\!
\noexpand\catcode`\noexpand\?=\the\catcode`\?
\noexpand\catcode`\noexpand\+=\the\catcode'53
}\catcode`\@=11 \catcode`\#=6 \catcode`\<=12 \catcode`\>=12 \catcode'53=12
\catcode`\:=12 \catcode`\;=12 \catcode`\!=12 \catcode`\?=12

\ifx\diagram@help@messages\CD@qK\let\diagram@help@messages y\fi

\def\cdps@Rokicki#1{\special{ps:#1}}\let\cdps@dvips\cdps@Rokicki\let
\cdps@RadicalEye\cdps@Rokicki\let\CD@HB\cdps@Rokicki\let\CD@IK\cdps@Rokicki
\let\CD@HB\cdps@Rokicki
\def\cdps@Bechtolsheim#1{\special{dvitps: Literal "#1"}}%
\let\cdps@dvitps\cdps@Bechtolsheim\let\cdps@IntegratedComputerSystems
\cdps@Bechtolsheim
\def\cdps@Clark#1{\special{dvitops: inline #1}}
\let\cdps@dvitops\cdps@Clark
\let\cdps@OzTeX\empty\let\cdps@oztex\empty\let\cdps@Trevorrow\empty
\def\cdps@Coombes#1{\special{ps-string #1}}

\count@=\year\multiply\count@12 \advance\count@\month
\ifnum\count@>24240 
\ifnum\count@>24243 
\errhelp{You may press RETURN and carry on for the time being.}\errmessage{! remain
compatible with your files. Please get it NOW.}\fi\fi

\def\CD@DE{\global\let}\def\CD@RH{\outer\def}

{\escapechar\m@ne\xdef\CD@o{\string\{}\xdef\CD@yC{\string\}}
\catcode`\&=4 \CD@DE\CD@Q=&\xdef\CD@S{\string\&}
\catcode`\$=3 \CD@DE\CD@k=$\CD@DE\CD@ND=$
\xdef\CD@nC{\string\$}\gdef\CD@LG{$$}
\catcode`\_=8 \CD@DE\CD@lJ=_
\obeylines\catcode`\^=7 \CD@DE\@super=^
\ifnum\newlinechar=10 \gdef\CD@uG{^^J}
\else\ifnum\newlinechar=13 \gdef\CD@uG{^^M}
\else\ifnum\newlinechar=-1 \gdef\CD@uG{^^J}
\else\CD@DE\CD@uG\space\expandafter\message{! input error: \noexpand
\newlinechar\space is ASCII \the\newlinechar, not LF=10 or CR=13.}
\fi\fi\fi}

\mathchardef\lessthan='30474 \mathchardef\greaterthan='30476

\ifx\tenln\CD@qK
\font\tenln=line10\relax
\fi\ifx\tenlnw\CD@qK\ifx\tenln\nullfont\let\tenlnw\nullfont\else
\font\tenlnw=linew10\relax
\fi\fi

\ifx\inputlineno\CD@qK\csname newcount\endcsname\inputlineno\inputlineno\m@ne
\fi

\def\cd@shouldnt#1{\CD@KB{* THIS (#1) SHOULD NEVER HAPPEN! *}}

\def\get@round@pair#1(#2,#3){#1{#2}{#3}}
\def\get@square@arg#1[#2]{#1{#2}}
\def\CD@AE#1{\CD@PK\let\CD@DH\CD@@E\CD@@E#1,],}
\def\CD@m{[}\def\CD@RD{]}\def\commdiag#1{{\let\enddiagram\relax\diagram[]#1%
\enddiagram}}

\def\CD@BF{{\ifx\CD@EH[\aftergroup\get@square@arg\aftergroup\CD@YH\else
\aftergroup\CD@JH\fi}}
\def\CD@CF#1#2{\def\CD@YH{#1}\def\CD@JH{#2}\futurelet\CD@EH\CD@BF}

\def\CD@KK{|}

\def\CD@PB{
\tokcase\CD@DD:\CD@y\break@args;\catcase\@super:\upper@label;\catcase\CD@lJ:%
\lower@label;\tokcase{~}:\middle@label;
\tokcase<:\CD@iF;
\tokcase>:\CD@iI;
\tokcase(:\CD@BC;
\tokcase[:\optional@;
\tokcase.:\CD@JJ;
\catcase\space:\eat@space;\catcase\bgroup:\positional@;\default:\CD@@A
\break@args;\endswitch}

\def\switch@arg{
\catcase\@super:\upper@label;\catcase\CD@lJ:\lower@label;\tokcase[:\optional@
;
\tokcase.:\CD@JJ;
\catcase\space:\eat@space;\catcase\bgroup:\positional@;\tokcase{~}:%
\middle@label;
\default:\CD@y\break@args;\endswitch}

\let\CD@tJ\relax\ifx\protect\CD@qK\let\protect\relax\fi\ifx\AtEndDocument
\CD@qK\def\CD@PG{\CD@gB}\def\CD@GF#1#2{}\else\def\CD@PG#1{\edef\CD@CH{#1}%
\expandafter\CD@oC\CD@CH\CD@OD}\def\CD@oC#1\CD@OD{\AtEndDocument{\typeout{%
\CD@tA: #1}}}\def\CD@GF#1#2{\gdef#1{#2}\AtEndDocument{#1}}\fi\def\CD@ZA#1#2{%
\def#1{\CD@PG{#2\CD@mD\CD@W}\CD@DE#1\relax}}\def\CD@uF#1\repeat{\def\CD@p{#1}%
\CD@OF}\def\CD@OF{\CD@p\relax\expandafter\CD@OF\fi}\def\CD@sF#1\repeat{\def
\CD@q{#1}\CD@PF}\def\CD@PF{\CD@q\relax\expandafter\CD@PF\fi}\def\CD@tF#1%
\repeat{\def\CD@r{#1}\CD@QF}\def\CD@QF{\CD@r\relax\expandafter\CD@QF\fi}\def
\CD@tG#1#2#3{\def#2{\let#1\iftrue}\def#3{\let#1\iffalse}#3}\if y%
\diagram@help@messages\def\CD@rG#1#2{\csname newtoks\endcsname#1#1=%
\expandafter{\csname#2\endcsname}}\else\csname newtoks\endcsname\no@cd@help
\no@cd@help{See the manual}\def\CD@rG#1#2{\let#1\no@cd@help}\fi\chardef\CD@lF
=1 \chardef\CD@lI=2 \chardef\CD@MH=5 \chardef\CD@tH=6 \chardef\CD@sH=7
\chardef\CD@PC=9 \dimendef\CD@hI=2 \dimendef\CD@hF=3 \dimendef\CD@mF=4
\dimendef\CD@mI=5 \dimendef\CD@wJ=6 \dimendef\CD@tI=8 \dimendef\CD@sI=9
\skipdef\CD@uB=1 \skipdef\CD@NF=2 \skipdef\CD@tB=3 \skipdef\CD@ZE=4 \skipdef
\countdef\CD@JC=9 \countdef\CD@gD=8 \countdef\CD@A=7 \def\sdef#1#2{\def#1{#2}%
}\def\CD@L#1{\expandafter\aftergroup\csname#1\endcsname}\def\CD@RC#1{%
\expandafter\def\csname#1\endcsname}\def\CD@sD#1{\expandafter\gdef\csname#1%
\endcsname}\def\CD@vC#1{\expandafter\edef\csname#1\endcsname}\def\CD@nF#1#2{%
\expandafter\let\csname#1\expandafter\endcsname\csname#2\endcsname}\def\CD@EE
#1#2{\expandafter\CD@DE\csname#1\expandafter\endcsname\csname#2\endcsname}%
\def\CD@AK#1{\csname#1\endcsname}\def\CD@XJ#1{\expandafter\show\csname#1%
\endcsname}\def\CD@ZJ#1{\expandafter\showthe\csname#1\endcsname}\def\CD@WJ#1{%
\expandafter\showbox\csname#1\endcsname}\def\CD@tA{Commutative Diagram}\edef
\CD@kH{\string\par}\edef\CD@dC{\string\diagram}\edef\CD@HD{\string\enddiagram
}\edef\CD@EC{\string\\}\def\CD@eF{LaTeX}\ifx\@ignoretrue\CD@qK\expandafter
\CD@tG\csname if@ignore\endcsname\ignore@true\ignore@false\def\@ignoretrue{%
\global\ignore@true}\def\@ignorefalse{\global\ignore@false}\fi

\def\CD@g{{\ifnum0=`}\fi}\def\CD@wC{\ifnum0=`{\fi}}\def\catcase#1:{\ifcat
\noexpand\CD@EH#1\CD@tJ\expandafter\CD@kC\else\expandafter\CD@dJ\fi}\def
\tokcase#1:{\ifx\CD@EH#1\CD@tJ\expandafter\CD@kC\else\expandafter\CD@dJ\fi}%
\def\CD@kC#1;#2\endswitch{#1}\def\CD@dJ#1;{}\let\endswitch\relax\def\default:%
#1;#2\endswitch{#1}\ifx\at@\CD@qK\def\at@{@}\fi\edef\CD@P{\CD@o pt\CD@yC}%
\CD@RC{\CD@P>}#1>#2>{\CD@z\rTo\sp{#1}\sb{#2}\CD@z}\CD@RC{\CD@P<}#1<#2<{\CD@z
\lTo\sp{#1}\sb{#2}\CD@z}\CD@RC{\CD@P)}#1)#2){\CD@z\rTo\sp{#1}\sb{#2}\CD@z}%
\CD@RC{\CD@P(}#1(#2({\CD@z\lTo\sp{#1}\sb{#2}\CD@z}
\def\CD@O{\def\endCD{\enddiagram}\CD@RC{\CD@P A}##1A##2A{\uTo<{##1}>{##2}%
\CD@z\CD@z}\CD@RC{\CD@P V}##1V##2V{\dTo<{##1}>{##2}\CD@z\CD@z}\CD@RC{\CD@P=}{%
\CD@z\hEq\CD@z}\CD@RC{\CD@P\CD@KK}{\vEq\CD@z\CD@z}\CD@RC{\CD@P\string\vert}{%
\vEq\CD@z\CD@z}\CD@RC{\CD@P.}{\CD@z\CD@z}\let\CD@z\CD@Q}\def\CD@IE{\let\tmp
\CD@JE\ifcat A\noexpand\CD@CH\else\ifcat=\noexpand\CD@CH\else\ifcat\relax
\noexpand\CD@CH\else\let\tmp\at@\fi\fi\fi\tmp}\def\CD@JE#1{\CD@nF{tmp}{\CD@P
\string#1}\ifx\tmp\relax\def\tmp{\at@#1}\fi\tmp}\def\CD@z{}\begingroup
\aftergroup\def\aftergroup\CD@T\aftergroup{\aftergroup\def\catcode`\@\active
\aftergroup @\endgroup{\futurelet\CD@CH\CD@IE}}\newcount\CD@uA\newcount\CD@vA
\newcount\CD@wA\newcount\CD@xA\newdimen\CD@OA\newdimen\CD@PA\CD@tG\CD@gE
\CD@@A\CD@y\CD@tG\CD@hE\CD@EA\CD@BA\newdimen\CD@RA\newdimen\CD@SA\newcount
\CD@yA\newcount\CD@zA\newdimen\CD@QA\newbox\CD@DA\CD@tG\CD@lE\CD@dA\CD@bA
\newcount\CD@LH\newcount\CD@TC\def\CD@V#1#2{\ifdim#1<#2\relax#1=#2\relax\fi}%
\def\CD@X#1#2{\ifdim#1>#2\relax#1=#2\relax\fi}\newdimen\CD@XH\CD@XH=1sp
\newdimen\CD@zC\CD@zC\z@\def\CD@cJ{\ifdim\CD@zC=1em\else\CD@nJ\fi}\def\CD@nJ{%
\CD@zC1em\def\CD@NC{\fontdimen8\textfont3 }\CD@@J\CD@NJ\setbox0=\vbox{\CD@t
\noindent\CD@k\null\penalty-9993\null\CD@ND\null\endgraf\setbox0=\lastbox
\unskip\unpenalty\setbox1=\lastbox\global\setbox\CD@IG=\hbox{\unhbox0\unskip
\unskip\unpenalty\setbox0=\lastbox}\global\setbox\CD@KG=\hbox{\unhbox1\unskip
\unpenalty\setbox1=\lastbox}}}\newdimen\CD@@I\CD@@I=1true in \divide\CD@@I300
\def\CD@zH#1{\multiply#1\tw@\advance#1\ifnum#1<\z@-\else+\fi\CD@@I\divide#1%
\tw@\divide#1\CD@@I\multiply#1\CD@@I}\def\MapBreadth{\afterassignment\CD@gI
\CD@LF}\newdimen\CD@LF\newdimen\CD@oI\def\CD@gI{\CD@oI\CD@LF\CD@V\CD@@I{4%
\CD@XH}\CD@X\CD@@I\p@\CD@zH\CD@oI\ifdim\CD@LF>\z@\CD@V\CD@oI\CD@@I\fi\CD@cJ}%
\def\CD@RJ#1{\CD@zD\count@\CD@@I#1\ifnum\count@>\z@\divide\CD@@I\count@\fi
\CD@gI\CD@NJ}\def\CD@NJ{\dimen@\CD@QC\count@\dimen@\divide\count@5\divide
\count@\CD@@I\edef\CD@OC{\the\count@}}\def\CD@AJ{\CD@QJ\z@}\def\CD@QJ#1{%
\CD@tI\axisheight\advance\CD@tI#1\relax\advance\CD@tI-.5\CD@oI\CD@zH\CD@tI
\CD@sI-\CD@tI\advance\CD@tI\CD@LF}\newdimen\CD@DC\CD@DC\z@\newdimen\CD@eJ
\CD@eJ\z@\def\CD@CJ#1{\CD@sI#1\relax\CD@tI\CD@sI\advance\CD@tI\CD@LF\relax}%
\def\horizhtdp{height\CD@tI depth\CD@sI}\def\axisheight{\fontdimen22\the
\textfont\tw@}\def\script@axisheight{\fontdimen22\the\scriptfont\tw@}\def
\ss@axisheight{\fontdimen22\the\scriptscriptfont\tw@}\def\CD@NC{0.4pt}\def
\CD@VK{\fontdimen3\textfont\z@}\def\CD@UK{\fontdimen3\textfont\z@}\newdimen
\PileSpacing\newdimen\CD@nA\CD@nA\z@\def\CD@RG{\ifincommdiag1.3em\else2em\fi}%
\newdimen\CD@YB\def\CellSize{\afterassignment\CD@kB\DiagramCellHeight}%
\newdimen\DiagramCellHeight\DiagramCellHeight-\maxdimen\newdimen
\DiagramCellWidth\DiagramCellWidth-\maxdimen\def\CD@kB{\DiagramCellWidth
\DiagramCellHeight}\def\CD@QC{3em}\newdimen\MapShortFall\def\MapsAbut{%
\MapShortFall\z@\objectheight\z@\objectwidth\z@}\newdimen\CD@iA\CD@iA\z@
\CD@tG\CD@vE\CD@aB\CD@ZB\expandafter\ifx\expandafter\iftrue\csname
ifUglyObsoleteDiagrams\endcsname\CD@ZB\else\CD@aB\fi\CD@nF{%
ifUglyObsoleteDiagrams}{relax}\newif\ifUglyObsoleteDiagrams\def\CD@nK{\CD@aB
\UglyObsoleteDiagramsfalse}\def\CD@oK{\CD@ZB\UglyObsoleteDiagramstrue}\CD@vE
\CD@nK\else\CD@oK\fi\CD@tG\CD@hK\CD@dK\CD@cK\CD@cK\def\CD@sK{\ifx\pdfoutput
\CD@qK\else\ifx\pdfoutput\relax\else\ifnum\pdfoutput>\z@\CD@pK\fi\fi\fi} \def
\CD@pK{\global\CD@dK\global\CD@aB\global\UglyObsoleteDiagramsfalse\global\let
\CD@n\empty\global\let\CD@oK\relax\global\let\CD@pK\relax\global\let\CD@sK
\relax}\def\CD@tK#1{}\ifx\pdfliteral\CD@qK\else\ifx
\pdfliteral\relax\else\let\CD@tK\pdfliteral\fi\fi\ifx\XeTeXrevision\CD@qK
\else\ifx\XeTeXrevision\relax\else\ifdim\XeTeXrevision pt<.996pt \expandafter
\message{! XeTeX version \XeTeXrevision\space does not support PDF literals,
so diagonals will not work!}\else\expandafter\message{RUNNING UNDER XETEX
\XeTeXrevision}\CD@pK\fi\fi\fi\CD@sK\def\newarrowhead{\CD@mG h\CD@BG\CD@GG>}%
\def\newarrowtail{\CD@mG t\CD@BG\CD@GG>}\def\newarrowmiddle{\CD@mG m\CD@BG
\hbox@maths\empty}\def\newarrowfiller{\CD@mG f\CD@bE\CD@MK-}\def\CD@mG#1#2#3#%
4#5#6#7#8#9{\CD@RC{r#1:#5}{#2{#6}}\CD@RC{l#1:#5}{#2{#7}}\CD@RC{d#1:#5}{#3{#8}%
}\CD@RC{u#1:#5}{#3{#9}}\CD@vC{-#1:#5}{\expandafter\noexpand\csname-#1:#4%
\endcsname\noexpand\CD@MC}\CD@vC{+#1:#5}{\expandafter\noexpand\csname+#1:#4%
\endcsname\noexpand\CD@MC}}\CD@ZA\CD@MC{\CD@eF\space diagonals are used unless
PostScript is set}\def\defaultarrowhead#1{\edef\CD@sJ{#1}\CD@@J}\def\CD@@J{%
\CD@IJ\CD@sJ<>ht\CD@IJ\CD@sJ<>th}\def\CD@IJ#1#2#3#4#5{\CD@HJ{r#4}{#3}{l#5}{#2%
}{r#4:#1}\CD@HJ{r#5}{#2}{l#4}{#3}{l#4:#1}\CD@HJ{d#4}{#3}{u#5}{#2}{d#4:#1}%
\CD@HJ{d#5}{#2}{u#4}{#3}{u#4:#1}}\def\CD@HJ#1#2#3#4#5{\begingroup\aftergroup
\CD@GJ\CD@L{#1+:#2}\CD@L{#1:#2}\CD@L{#3:#4}\CD@L{#5}\endgroup}\def\CD@GJ#1#2#%
3#4{\csname newbox\endcsname#1\def#2{\copy#1}\def#3{\copy#1}\setbox#1=\box
\voidb@x}\def\CD@sJ{}\CD@@J\def\CD@GJ#1#2#3#4{\setbox#1=#4}\ifx\tenln
\nullfont\def\CD@sJ{vee}\else\let\CD@sJ\CD@eF\fi\def\CD@xF#1#2#3{\begingroup
\aftergroup\CD@wF\CD@L{#1#2:#3#3}\CD@L{#1#2:#3}\aftergroup\CD@yF\CD@L{#1#2:#3%
-#3}\CD@L{#1#2:#3}\endgroup}\def\CD@wF#1#2{\def#1{\hbox{\rlap{#2}\kern.4%
\CD@zC#2}}}\def\CD@yF#1#2{\def#1{\hbox{\rlap{#2}\kern.4\CD@zC#2\kern-.4\CD@zC
}}}\CD@xF lh>\CD@xF rt>\CD@xF rh<\CD@xF rt<\def\CD@yF#1#2{\def#1{\hbox{\kern-%
.4\CD@zC\rlap{#2}\kern.4\CD@zC#2}}}\CD@xF rh>\CD@xF lh<\CD@xF lt>\CD@xF lt<%
\def\CD@wF#1#2{\def#1{\vbox{\vbox to\z@{#2\vss}\nointerlineskip\kern.4\CD@zC#%
2}}}\def\CD@yF#1#2{\def#1{\vbox{\vbox to\z@{#2\vss}\nointerlineskip\kern.4%
\CD@zC#2\kern-.4\CD@zC}}}\CD@xF uh>\CD@xF dt>\CD@xF dh<\CD@xF dt<\def\CD@yF#1%
#2{\def#1{\vbox{\kern-.4\CD@zC\vbox to\z@{#2\vss}\nointerlineskip\kern.4%
\CD@zC#2}}}\CD@xF dh>\CD@xF ut>\CD@xF uh<\CD@xF ut<\def\CD@BG#1{\hbox{%
\mathsurround\z@\offinterlineskip\CD@k\mkern-1.5mu{#1}\mkern-1.5mu\CD@ND}}%
\def\hbox@maths#1{\hbox{\CD@k#1\CD@ND}}\def\CD@GG#1{\hbox to\CD@LF{\setbox0=%
\hbox{\offinterlineskip\mathsurround\z@\CD@k{#1}\CD@ND}\dimen0.5\wd0\advance
\dimen0-.5\CD@oI\CD@zH{\dimen0}\kern-\dimen0\unhbox0\hss}}\def\CD@sB#1{\hbox
to2\CD@LF{\hss\offinterlineskip\mathsurround\z@\CD@k{#1}\CD@ND\hss}}\def
\CD@vF#1{\hbox{\mathsurround\z@\CD@k{#1}\CD@ND}}\def\CD@bE#1{\hbox{\kern-.15%
\CD@zC\CD@k{#1}\CD@ND\kern-.15\CD@zC}}\def\CD@MK#1{\vbox{\offinterlineskip
\kern-.2ex\CD@GG{#1}\kern-.2ex}}\def\@fillh{\xleaders\vrule\horizhtdp}\def
\@fillv{\xleaders\hrule width\CD@LF}\CD@nF{rf:-}{@fillh}\CD@nF{lf:-}{@fillh}%
\CD@nF{df:-}{@fillv}\CD@nF{uf:-}{@fillv}\CD@nF{rh:}{null}\CD@nF{rm:}{null}%
\CD@nF{rt:}{null}\CD@nF{lh:}{null}\CD@nF{lm:}{null}\CD@nF{lt:}{null}\CD@nF{dh%
:}{null}\CD@nF{dm:}{null}\CD@nF{dt:}{null}\CD@nF{uh:}{null}\CD@nF{um:}{null}%
\CD@nF{ut:}{null}\CD@nF{+h:}{null}\CD@nF{+m:}{null}\CD@nF{+t:}{null}\CD@nF{-h%
:}{null}\CD@nF{-m:}{null}\CD@nF{-t:}{null}\def\CD@@D{\hbox{\vrule height 1pt
depth-1pt width 1pt}}\CD@RC{rf:}{\CD@@D}\CD@nF{lf:}{rf:}\CD@nF{+f:}{rf:}%
\CD@RC{df:}{\CD@@D}\CD@nF{uf:}{df:}\CD@nF{-f:}{df:}\def\CD@BD{\CD@U\null
\CD@@D\null\CD@@D\null}\edef\CD@lG{\string\newarrow}\def\newarrow#1#2#3#4#5#6%
{\begingroup\edef\@name{#1}\edef\CD@oJ{#2}\edef\CD@iD{#3}\edef\CD@QG{#4}\edef
\CD@jD{#5}\edef\CD@LE{#6}\let\CD@HE\CD@sG\let\CD@FK\CD@BH\let\@x\CD@AH\ifx
\CD@oJ\CD@iD\let\CD@oJ\empty\fi\ifx\CD@LE\CD@jD\let\CD@LE\empty\fi\def\CD@LI{%
r}\def\CD@SF{l}\def\CD@IC{d}\def\CD@yJ{u}\def\CD@gH{+}\def\@m{-}\ifx\CD@iD
\CD@jD\ifx\CD@QG\CD@iD\let\CD@QG\empty\fi\ifx\CD@LE\empty\ifx\CD@iD\CD@aE\let
\@x\CD@yG\else\let\@x\CD@zG\fi\fi\else\edef\CD@a{\CD@iD\CD@oJ}\ifx\CD@a\empty
\ifx\CD@QG\CD@jD\let\CD@QG\empty\fi\fi\fi\ifmmode\aftergroup\CD@kG\else\CD@@A
\CD@oB rh{head\space\space}\CD@LE\CD@oB rf{filler}\CD@iD\CD@oB rm{middle}%
\CD@QG\ifx\CD@jD\CD@iD\else\CD@oB rf{filler}\CD@jD\fi\CD@oB rt{tail\space
\space}\CD@oJ\CD@gE\CD@HE\CD@FK\@x\CD@nG l-2+2{lu}{nw}\NorthWest\CD@nG r+2+2{%
ru}{ne}\NorthEast\CD@nG l-2-2{ld}{sw}\SouthWest\CD@nG r+2-2{rd}{se}\SouthEast
\else\aftergroup\CD@b\CD@L{r\@name}\fi\fi\endgroup}\def\CD@sG{\CD@vG\CD@LI
\CD@SF rl\Horizontal@Map}\def\CD@BH{\CD@vG\CD@IC\CD@yJ du\Vertical@Map}\def
\CD@AH{\CD@vG\CD@gH\@m+-\Vector@Map}\def\CD@yG{\CD@vG\CD@gH\@m+-\Slant@Map}%
\def\CD@zG{\CD@vG\CD@gH\@m+-\Slope@Map}\catcode`\/=\active\def\CD@vG#1#2#3#4#%
5{\CD@jG#1#3#5t:\CD@oJ/f:\CD@iD/m:\CD@QG/f:\CD@jD/h:\CD@LE//\CD@jG#2#4#5h:%
\CD@LE/f:\CD@jD/m:\CD@QG/f:\CD@iD/t:\CD@oJ//}\def\CD@jG#1#2#3#4//{\edef\CD@fG
{#2}\aftergroup\sdef\CD@L{#1\@name}\aftergroup{\aftergroup#3\CD@M#4//%
\aftergroup}}\def\CD@M#1/{\edef\CD@EH{#1}\ifx\CD@EH\empty\else\CD@L{\CD@fG#1}%
\expandafter\CD@M\fi}\catcode`\/=12 \def\CD@nG#1#2#3#4#5#6#7#8{\aftergroup
\sdef\CD@L{#6\@name}\aftergroup{\CD@L{#2\@name}\if#2#4\aftergroup\CD@CI\else
\aftergroup\CD@BI\fi\CD@L{#1\@name}%
\aftergroup(\aftergroup#3\aftergroup,\aftergroup#5\aftergroup)\aftergroup}}%
\def\CD@oB#1#2#3#4{\expandafter\ifx\csname#1#2:#4\endcsname\relax\CD@y\CD@gB{%
arrow#3 "#4" undefined}\fi}\CD@rG\CD@VE{All five components must be defined
before an arrow.}\CD@rG\CD@SE{\CD@lG, unlike \string\HorizontalMap, is a
declaration.}\def\CD@b#1{\CD@YA{Arrows \string#1 etc could not be defined}%
\CD@VE}\def\CD@kG{\CD@YA{misplaced \CD@lG}\CD@SE}\def\newdiagramgrid#1#2#3{%
\CD@RC{cdgh@#1}{#2,],}
\CD@RC{cdgv@#1}{#3,],}}
\CD@tG\ifincommdiag\incommdiagtrue\incommdiagfalse\CD@tG\CD@@F\CD@IF\CD@HF
\newcount\CD@VA\CD@VA=0 \def\CD@yH{\CD@VA6 }\def\CD@OB{\CD@VA1 \global\CD@yA1
\CD@DE\CD@YF\empty}\def\CD@YF{}\def\CD@nB#1{\relax\CD@MD\edef\CD@vJ{#1}%
\begingroup\CD@rE\else\ifcase\CD@VA\ifmmode\else\CD@YG\CD@E0\fi\or\CD@cE5\or
\CD@YG\CD@F5\or\CD@YG\CD@B5\or\CD@YG\CD@B5\or\CD@YG\CD@C5\or\CD@cE7\or\CD@YG
\CD@D7\fi\fi\endgroup\xdef\CD@YF{#1}}\def\CD@pB#1#2#3#4#5{\relax\CD@MD\xdef
\CD@vJ{#4}\begingroup\ifnum\CD@VA<#1 \expandafter\CD@cE\ifcase\CD@VA0\or#2\or
#3\else#2\fi\else\ifnum\CD@VA<6 \CD@tJ\CD@YG\CD@B#2\else\CD@YG\CD@G#2\fi\fi
\endgroup\CD@DE\CD@YF\CD@vJ\ifincommdiag\let\CD@ZD#5\else\let\CD@ZD\CD@LK\fi}%
\def\CD@yI{\global\CD@yA=\ifnum\CD@VA<5 1\else2\fi\relax}\def\CD@OI{\CD@VA
\CD@yA}\def\CD@cE#1{\aftergroup\CD@VA\aftergroup#1\aftergroup\relax}\def
\CD@HH{\def\CD@nB##1{\relax}\let\CD@pB\CD@FH\let\CD@yH\relax\let\CD@OB\relax
\let\CD@yI\relax\let\CD@OI\relax}\def\CD@FH#1#2#3#4#5{\ifincommdiag\let\CD@ZD
#5\else\xdef\CD@vJ{#4}\let\CD@ZD\CD@LK\fi}\def\CD@YG#1{\aftergroup#1%
\aftergroup\relax\CD@cE}\def\CD@B{\CD@YE\CD@S\CD@ME\CD@Q}\def\CD@G{\CD@YE{%
\CD@yC\CD@S}\CD@XE\CD@QD\CD@Q}\def\CD@F{\CD@YE{*\CD@S}\CD@RE\clubsuit\CD@Q}%
\def\CD@C{\CD@YE{\CD@S*\CD@S}\CD@RE\CD@Q\clubsuit\CD@Q}\def\CD@D{\CD@YE\CD@EC
\CD@TE\\}\def\CD@E{\CD@YE\CD@nC\CD@QE\CD@k}\def\CD@LK{\CD@YA{\CD@vJ\space
ignored \CD@dH}\CD@WE}\def\CD@FE{}\def\CD@d{\CD@YA{maps must never be enclosed
in braces}\CD@OE}\def\CD@dH{outside diagram}\def\CD@FC{\string\HonV, \string
\VonH\space and \string\HmeetV}\CD@rG\CD@ME{The way that horizontal and
vertical arrows are terminated implicitly means\CD@uG that they cannot be
mixed with each other or with \CD@FC.}\CD@rG\CD@XE{\string\pile\space is for
parallel horizontal arrows; verticals can just be put together in\CD@uG a cell%
. \CD@FC\space are not meaningful in a \string\pile.}\CD@rG\CD@RE{The
horizontal maps must point to an object, not each other (I've put in\CD@uG one
which you're unlikely to want). Use \string\pile\space if you want them
parallel.}\CD@rG\CD@TE{Parallel horizontal arrows must be in separate layers
of a \string\pile.}\CD@rG\CD@QE{Horizontal arrows may be used \CD@dH s, but
must still be in maths.}\CD@rG\CD@WE{Vertical arrows, \CD@FC\space\CD@dH s don%
't know where\CD@uG where to terminate.}\CD@rG\CD@OE{This prevents them from
stretching correctly.}\def\CD@YE#1{\CD@YA{"#1" inserted \ifx\CD@YF\empty
before \CD@vJ\else between \CD@YF\ifx\CD@YF\CD@vJ s\else\space and \CD@vJ\fi
\fi}}\count@=\year\multiply\count@12 \advance\count@\month\ifnum\count@>24247
\expandafter\endinput\fi\def
\Horizontal@Map{\CD@nB{horizontal map}\CD@LC\CD@TJ\CD@qD}\def\CD@TJ{\CD@GB-%
9999 \let\CD@ZD\CD@XD\ifincommdiag\else\CD@cJ\ifinpile\else\skip2\z@ plus 1.5%
\CD@VK minus .5\CD@UK\skip4\skip2 \fi\fi\let\CD@kD\@fillh\CD@nF{fill@dot}{rf:%
.}}\def\Vector@Map{\CD@HK4}\def\Slant@Map{\CD@HK{\CD@EF255\else6\fi}}\def
\Slope@Map{\CD@HK\CD@OC}\def\CD@HK#1#2#3#4#5#6{\CD@LC\def\CD@WK{2}\def\CD@aK{%
2}\def\CD@ZK{1}\def\CD@bK{1}\let\Horizontal@Map\CD@nI\def\CD@OG{#1}\def\CD@NI
{\CD@U#2#3#4#5#6}}\def\CD@nI{\CD@TJ\CD@JB\let\CD@ZD\CD@TD\CD@qD}\CD@tG\CD@pE
\CD@rA\CD@qA\CD@rA\def\cds@missives{\CD@rA}\def\CD@TD{\CD@vE\let\CD@OG\CD@OC
\CD@x\CD@zE\CD@WF\fi\setbox0\hbox{\incommdiagfalse\CD@HI}\CD@pE\CD@aD\else
\global\CD@YC\CD@bD\fi\ifvoid6 \ifvoid7 \CD@eE\fi\fi\CD@zE\else\CD@BD\global
\CD@YC\let\CD@CG\CD@IH\CD@YD\fi\else\CD@NI\CD@MI\global\CD@YC\CD@YD\fi}\def
\CD@LC{\begingroup\dimen1=\MapShortFall\dimen2=\CD@RG\dimen5=\MapShortFall
\setbox3=\box\voidb@x\setbox6=\box\voidb@x\setbox7=\box\voidb@x\CD@pD
\mathsurround\z@\skip2\z@ plus1fill minus 1000pt\skip4\skip2 \CD@TB}\CD@tG
\CD@tE\CD@UB\CD@TB\def\CD@U#1#2#3#4#5{\let\CD@oJ#1\let\CD@iD#2\let\CD@QG#3%
\let\CD@jD#4\let\CD@LE#5\CD@TB\ifx\CD@iD\CD@jD\CD@UB\fi}\def\CD@qD#1#2#3#4#5{%
\CD@U#1#2#3#4#5\CD@tD}\def\Vertical@Map{\CD@pB433{vertical map}\CD@cD\CD@LC
\CD@GB-9995 \let\CD@kD\@fillv\CD@nF{fill@dot}{df:.}\CD@qD}\def\break@args{%
\def\CD@tD{\CD@ZD}\CD@ZD\endgroup\aftergroup\CD@FE}\def\CD@MJ{\setbox1=\CD@oJ
\setbox5=\CD@LE\ifvoid3 \ifx\CD@QG\null\else\setbox3=\CD@QG\fi\fi\CD@@G2%
\CD@iD\CD@@G4\CD@jD}\def\CD@pF#1{\ifvoid1\else\CD@oF1#1\fi\ifvoid2\else\CD@oF
2#1\fi\ifvoid3\else\CD@oF3#1\fi\ifvoid4\else\CD@oF4#1\fi\ifvoid5\else\CD@oF5#%
1\fi} \def\CD@oF#1#2{\setbox#1\vbox{\offinterlineskip\box#1\dimen@\prevdepth
\advance\dimen@-#2\relax\setbox0\null\dp0\dimen@\ht0-\dimen@\box0}}\def\CD@@G
#1#2{\ifx#2\CD@kD\setbox#1=\box\voidb@x\else\setbox#1=#2\def#2{\xleaders\box#%
1}\fi}\CD@ZA\CD@BK{\string\HorizontalMap, \string\VerticalMap\space and
\string\DiagonalMap\CD@uG are obsolete - use \CD@lG\space to pre-define maps}%
\def\HorizontalMap#1#2#3#4#5{\CD@BK\CD@nB{old horizontal map}\CD@LC\CD@TJ\def
\CD@oJ{\CD@UH{#1}}\CD@SH\CD@iD{#2}\def\CD@QG{\CD@UH{#3}}\CD@SH\CD@jD{#4}\def
\CD@LE{\CD@UH{#5}}\CD@tD}\def\VerticalMap#1#2#3#4#5{\CD@BK\CD@pB433{vertical
map}\CD@cD\CD@LC\CD@GB-9995 \let\CD@kD\@fillv\def\CD@oJ{\CD@GG{#1}}\CD@VH
\CD@iD{#2}\def\CD@QG{\CD@GG{#3}}\CD@VH\CD@jD{#4}\def\CD@LE{\CD@GG{#5}}\CD@tD}%
\def\DiagonalMap#1#2#3#4#5{\CD@BK\CD@LC\def\CD@OG{4}\let\CD@kD\CD@qK\let
\CD@ZD\CD@YD\def\CD@WK{2}\def\CD@aK{2}\def\CD@ZK{1}\def\CD@bK{1}\def\CD@QG{%
\CD@vF{#3}}\ifPositiveGradient\let\mv\raise\def\CD@oJ{\CD@vF{#5}}\def\CD@iD{%
\CD@vF{#4}}\def\CD@jD{\CD@vF{#2}}\def\CD@LE{\CD@vF{#1}}\else\let\mv\lower\def
\CD@oJ{\CD@vF{#1}}\def\CD@iD{\CD@vF{#2}}\def\CD@jD{\CD@vF{#4}}\def\CD@LE{%
\CD@vF{#5}}\fi\CD@tD}\def\CD@aE{-}\def\CD@AD{\empty}\def\CD@SH{\CD@EG\CD@bE
\CD@aE\@fillh}\def\CD@VH{\CD@EG\CD@MK\CD@KK\@fillv}\def\CD@EG#1#2#3#4#5{\def
\CD@CH{#5}\ifx\CD@CH#2\let#4#3\else\let#4\null\ifx\CD@CH\empty\else\ifx\CD@CH
\CD@AD\else\let#4\CD@CH\fi\fi\fi}\def\CD@UH#1{\hbox{\mathsurround\z@
\offinterlineskip\def\CD@CH{#1}\ifx\CD@CH\empty\else\ifx\CD@CH\CD@AD\else
\CD@k\mkern-1.5mu{\CD@CH}\mkern-1.5mu\CD@ND\fi\fi}}\def\CD@yD#1#2{\setbox#1=%
\hbox\bgroup\setbox0=\hbox{\CD@k\labelstyle()\CD@ND}
\setbox1=\null\ht1\ht0\dp1\dp0\box1 \kern.1\CD@zC\CD@k\bgroup\labelstyle
\aftergroup\CD@LD\CD@xD}\def\CD@LD{\CD@ND\kern.1\CD@zC\egroup\CD@tD}\def
\CD@xD{\futurelet\CD@EH\CD@mJ}\def\CD@mJ{
\catcase\bgroup:\CD@v;\catcase\egroup:\missing@label;\catcase\space:\CD@TF;%
\tokcase[:\CD@XF;
\default:\CD@zJ;\endswitch}\def\CD@v{\let\CD@MD\CD@c\let\CD@CH}\def\CD@zJ#1{%
\let\CD@UF\egroup{\let\actually@braces@missing@around@macro@in@label\CD@ZH
\let\CD@MD\CD@xC\let\CD@UF\CD@VF#1%
\actually@braces@missing@around@macro@in@label}\CD@UF}\def
\actually@braces@missing@around@macro@in@label{\let\CD@CH=}\def\missing@label
{\egroup\CD@YA{missing label}\CD@PE}\def\CD@xC{\egroup\missing@label}\outer
\def\CD@ZH{}\def\CD@UF{}\def\CD@VF{\CD@wC\CD@UF}\def\CD@MD{}\def\CD@XF{\let
\CD@N\CD@xD\get@square@arg\CD@AE}\CD@rG\CD@PE{The text which has just been
read is not allowed within map labels.}\def\CD@c{\egroup\CD@YA{missing \CD@yC
\space inserted after label}\CD@PE}\def\upper@label{\CD@oD\CD@yD6}\def
\lower@label{\def\positional@{\CD@@A\break@args}\CD@yD7}\def\middle@label{%
\CD@yD3}\CD@tG\CD@yE\CD@pD\CD@oD\def\CD@iF{\ifPositiveGradient\CD@tJ
\expandafter\upper@label\else\expandafter\lower@label\fi}\def\CD@iI{%
\ifPositiveGradient\CD@tJ\expandafter\lower@label\else\expandafter
\upper@label\fi}\def\positional@{\CD@gB{labels as positional arguments are
obsolete}\CD@yE\CD@tJ\expandafter\upper@label\else\expandafter\lower@label\fi
-}\def\CD@tD{\futurelet\CD@EH\switch@arg}\def\eat@space{\afterassignment
\CD@tD\let\CD@EH= }\def\CD@TF{\afterassignment\CD@xD\let\CD@EH= }\def\CD@BC{%
\get@round@pair\CD@uD}\def\CD@uD#1#2{\def\CD@WK{#1}\def\CD@aK{#2}\CD@tD}\def
\optional@{\let\CD@N\CD@tD\get@square@arg\CD@AE}\def\CD@JJ.{\CD@sC\CD@tD}\def
\CD@sC{\let\CD@iD\fill@dot\let\CD@jD\fill@dot\def\CD@MI{\let\CD@iD\dfdot\let
\CD@jD\dfdot}}\def\CD@MI{}\def\CD@@E#1,{\CD@nH#1,\begingroup\ifx\@name\CD@RD
\CD@FF\aftergroup\CD@e\fi\aftergroup\CD@jC\else\expandafter\def\expandafter
\CD@RF\expandafter{\csname\@name\endcsname}\expandafter\CD@vD\CD@RF\CD@KD\ifx
\CD@RF\empty\aftergroup\CD@pC\expandafter\aftergroup\csname\CD@FB\@name
\endcsname\expandafter\aftergroup\csname\CD@FB @\@name\endcsname\else\gdef
\CD@GE{#1}\CD@gB{\string\relax\space inserted before `[\CD@GE'}\message{(I was
trying to read this as a \CD@tA\ option.)}\aftergroup\CD@H\fi\fi\endgroup}%
\def\CD@vD#1#2\CD@KD{\def\CD@RF{#2}}\def\CD@jC{\let\CD@CH\CD@N\let\CD@N\relax
\CD@CH}\def\CD@H#1],{
\CD@jC\relax\def\CD@RF{#1}\ifx\CD@RF\empty\def\CD@RF{[\CD@GE]}%
\else\def\CD@RF{[\CD@GE,#1]}
\fi\CD@RF}\def\CD@pC#1#2{\ifx#2\CD@qK\ifx#1\CD@qK\CD@gB{option `\@name'
undefined}\else#1\fi\else\CD@FF\expandafter#2\CD@GK\CD@PK\else\CD@QK\fi\fi
\CD@DH}\CD@tG\CD@FF\CD@QK\CD@PK\def\CD@nH#1,{\CD@FF\ifx\CD@GK\CD@qK\CD@e\else
\expandafter\CD@oH\CD@GK,#1,(,),(,)[]%
\fi\fi\CD@FF\else\CD@mH#1==,\fi}\def\CD@e{\CD@gB{option `\@name' needs (x,y)
value}\CD@PK\let\@name\empty}\def\CD@mH#1=#2=#3,{\def\@name{#1}\def\CD@GK{#2}%
\def\CD@RF{#3}\ifx\CD@RF\empty\let\CD@GK\CD@qK\fi}%
\def\CD@oH#1(#2,#3)#4,(#5,#6)#7[]{\def\CD@GK{{#2}{#3}}\def\CD@RF{#1#4#5#6}%
\ifx\CD@RF\empty\def\CD@RF{#7}\ifx\CD@RF\empty\CD@e\fi\else\CD@e\fi}\def
\CD@FB{cds@}\let\CD@N\relax\def\CD@zD#1{\ifx\CD@GK\CD@qK\CD@gB{option `\@name
' needs a value}\else#1\CD@GK\relax\fi}\def\CD@BE#1#2{\ifx\CD@GK\CD@qK#1#2%
\relax\else#1\CD@GK\relax\fi}\def\cds@@showpair#1#2{\message{x=#1,y=#2}}\def
\cds@@diagonalbase#1#2{\edef\CD@ZK{#1}\edef\CD@bK{#2}}\def\CD@DI#1{\def\CD@CH
{#1}\CD@nF{@x}{cdps@#1}\ifx\CD@CH\empty\CD@f\CD@CH{cannot be used}\else\ifx
\CD@CH\relax\CD@f\CD@CH{unknown}\else\let\CD@IK\@x\fi\fi}\def\CD@f#1#2{\CD@gB
{PostScript translator `#1' #2}}\def\CD@PH{}\def\CD@PJ{\CD@fA\edef\CD@PH{%
\noexpand\CD@KB{\@name\space ignored within maths}}}\def\diagramstyle{\CD@cJ
\let\CD@N\relax\CD@CF\CD@AE\CD@AE}\CD@tG\CD@sE
\CD@SB\CD@RB\CD@tG\CD@qE\CD@EB\CD@DB\CD@tG\CD@oE\CD@pA\CD@oA\CD@tG\CD@iE
\CD@HA\CD@GA\CD@HA\CD@tG\CD@jE\CD@JA\CD@IA\CD@tG\CD@kE\CD@LA\CD@KA\CD@tG
\CD@EF\CD@DK\CD@CK\CD@tG\CD@rE\CD@JB\CD@IB\CD@tG\CD@mE\CD@gA\CD@fA\CD@tG
\CD@nE\CD@kA\CD@jA\CD@tG\CD@AF\CD@iG\CD@hG\CD@RC{cds@ }{}\CD@RC{cds@}{}\CD@RC
{cds@1em}{\CellSize1\CD@zC}\CD@RC{cds@1.5em}{\CellSize1.5\CD@zC}\CD@RC{cds@2%
em}{\CellSize2\CD@zC}\CD@RC{cds@2.5em}{\CellSize2.5\CD@zC}\CD@RC{cds@3em}{%
\CellSize3\CD@zC}\CD@RC{cds@3.5em}{\CellSize3.5\CD@zC}\CD@RC{cds@4em}{%
\CellSize4\CD@zC}\CD@RC{cds@4.5em}{\CellSize4.5\CD@zC}\CD@RC{cds@5em}{%
\CellSize5\CD@zC}\CD@RC{cds@6em}{\CellSize6\CD@zC}\CD@RC{cds@7em}{\CellSize7%
\CD@zC}\CD@RC{cds@8em}{\CellSize8\CD@zC}\def\cds@abut{\MapsAbut\dimen1\z@
\dimen5\z@}\def\cds@alignlabels{\CD@IA\CD@KA}\def\cds@amstex{\ifincommdiag
\CD@O\else\def\CD{\diagram[amstex]}
\fi\CD@T\catcode`\@\active}\def\cds@b{\let\CD@dB\CD@bB}\def\cds@balance{\let
\CD@hA\CD@AA}\let\cds@bottom\cds@b\def\cds@center{\cds@vcentre\cds@nobalance}%
\let\cds@centre\cds@center\def\cds@centerdisplay{\CD@HA\CD@PJ\cds@balance}%
\let\cds@centredisplay\cds@centerdisplay\def\cds@crab{\CD@BE\CD@DC{.5%
\PileSpacing}}\CD@RC{cds@crab-}{\CD@DC-.5\PileSpacing}\CD@RC{cds@crab+}{%
\CD@DC.5\PileSpacing}\CD@RC{cds@crab++}{\CD@DC1.5\PileSpacing}\CD@RC{cds@crab%
--}{\CD@DC-1.5\PileSpacing}\def\cds@defaultsize{\CD@BE{\let\CD@QC}{3em}\CD@NJ
}\def\cds@displayoneliner{\CD@DB}\let\cds@dotted\CD@sC\def\cds@dpi{\CD@RJ{1%
truein}}\def\cds@dpm{\CD@RJ{100truecm}}\let\CD@XA\CD@qK\def\cds@eqno{\let
\CD@XA\CD@GK\let\CD@EJ\empty}\def\cds@fixed{\CD@qA}\CD@tG\CD@fE\CD@J\CD@I\def
\cds@flushleft{\CD@I\CD@GA\CD@PJ\cds@nobalance\CD@BE\CD@nA\CD@nA}\def\cds@gap
{\CD@AJ\setbox3=\null\ht3=\CD@tI\dp3=\CD@sI\CD@BE{\wd3=}\MapShortFall} \def
\cds@grid{\ifx\CD@GK\CD@qK\let\h@grid\relax\let\v@grid\relax\else\CD@nF{%
h@grid}{cdgh@\CD@GK}\CD@nF{v@grid}{cdgv@\CD@GK}\ifx\h@grid\relax\CD@gB{%
unknown grid `\CD@GK'}\else\CD@WB\fi\fi}\let\h@grid\relax\let\v@grid\relax
\def\cds@gridx{\ifx\CD@GK\CD@qK\else\cds@grid\fi\let\CD@CH\h@grid\let\h@grid
\v@grid\let\v@grid\CD@CH}\def\cds@h{\CD@zD\DiagramCellHeight}\def\cds@hcenter
{\let\CD@hA\CD@aA}\let\cds@hcentre\cds@hcenter\def\cds@heads{\CD@BE{\let
\CD@sJ}\CD@sJ\CD@@J\CD@vE\else\ifx\CD@sJ\CD@eF\else\CD@MC\fi\fi}\let
\cds@height\cds@h\let\cds@hmiddle\cds@balance\def\cds@htriangleheight{\CD@BE
\DiagramCellHeight\DiagramCellHeight\DiagramCellWidth1.73205%
\DiagramCellHeight}\def\cds@htrianglewidth{\CD@BE\DiagramCellWidth
\DiagramCellWidth\DiagramCellHeight.57735\DiagramCellWidth}\CD@tG\CD@zE\CD@eE
\CD@dE\CD@eE\def\cds@hug{\CD@eE} \def\cds@inline{\CD@gA\let\CD@PH\empty}\def
\cds@inlineoneliner{\CD@EB}\CD@RC{cds@l>}{\CD@zD{\let\CD@RG}\dimen2=\CD@RG}%
\def\cds@labelstyle{\CD@zD{\let\labelstyle}}\def\cds@landscape{\CD@kA}\def
\cds@large{\CellSize5\CD@zC}\let\CD@EJ\empty\def\CD@FJ{\refstepcounter{%
equation}\def\CD@XA{\hbox{\@eqnnum}}}\def\cds@LaTeXeqno{\let\CD@EJ\CD@FJ}\def
\cds@lefteqno{\CD@pA}\def\cds@leftflush{\cds@flushleft\CD@J}\def
\cds@leftshortfall{\CD@zD{\dimen1 }}\def\cds@lowershortfall{%
\ifPositiveGradient\cds@leftshortfall\else\cds@rightshortfall\fi}\def
\cds@loose{\CD@VB}\def\cds@midhshaft{\CD@JA}\def\cds@midshaft{\CD@JA}\def
\cds@midvshaft{\CD@LA}\def\cds@moreoptions{\CD@@A}\let\cds@nobalance
\cds@hcenter\def\cds@nohcheck{\CD@HH}\def\cds@nohug{\CD@dE} \def
\cds@nooptions{\def\CD@aC{\CD@WD}}\let\cds@noorigin\cds@nobalance\def
\cds@nopixel{\CD@@I4\CD@XH\CD@cJ}\def\cds@UO{\CD@oK\global\let\CD@n\empty}%
\def\cds@UglyObsolete{\cds@UO\let\cds@PS\empty}\def\CD@rK#1{\CD@gB{option `#1%
' renamed as `UglyObsolete'}}\def\cds@noPostScript{\CD@rK{noPostScript}}\def
\cds@noPS{\CD@rK{noPostScript}}\def\cds@notextflow{\CD@RB}\def\cds@noTPIC{%
\CD@CK}\def\cds@objectstyle{\CD@zD{\let\objectstyle}}\def\cds@origin{\let
\CD@hA\CD@iB}\def\cds@p{\CD@zD\PileSpacing}\let\cds@pilespacing\cds@p\def
\cds@pixelsize{\CD@zD\CD@@I\CD@gI}\def\cds@portrait{\CD@jA}\def
\cds@PostScript{\CD@nK\global\let\CD@n\empty\CD@BE\CD@DI\empty}\def\cds@PS{%
\CD@nK\global\let\CD@n\empty}\CD@GF\CD@n{\typeout{\CD@tA: try the PostScript
option for better results}}\def\cds@repositionpullbacks{\let\make@pbk\CD@fH
\let\CD@qH\CD@pH}\def\cds@righteqno{\CD@oA}\def\cds@rightshortfall{\CD@zD{%
\dimen5 }}\def\cds@ruleaxis{\CD@zD{\let\axisheight}}\def\cds@cmex{\let\CD@GG
\CD@sB\let\CD@QJ\CD@CJ}\def\cds@s{\cds@height\DiagramCellWidth
\DiagramCellHeight}\def\cds@scriptlabels{\let\labelstyle\scriptstyle}\def
\cds@shortfall{\CD@zD\MapShortFall\dimen1\MapShortFall\dimen5\MapShortFall}%
\def\cds@showfirstpass{\CD@BE{\let\CD@nD}\z@}\def\cds@silent{\def\CD@KB##1{}%
\def\CD@gB##1{}}\let\cds@size\cds@s\def\cds@small{\CellSize2\CD@zC}\def
\cds@snake{\CD@BE\CD@eJ\z@}\def\cds@t{\let\CD@dB\CD@fB}\def\cds@textflow{%
\CD@SB\CD@PJ}\def\cds@thick{\let\CD@rF\tenlnw\CD@LF\CD@NC\CD@BE\MapBreadth{2%
\CD@LF}\CD@@J}\def\cds@thin{\let\CD@rF\tenln\CD@BE\MapBreadth{\CD@NC}\CD@@J}%
\def\cds@tight{\CD@WB}\let\cds@top\cds@t\def\cds@TPIC{\CD@DK}\def
\cds@uppershortfall{\ifPositiveGradient\cds@rightshortfall\else
\cds@leftshortfall\fi}\def\cds@vcenter{\let\CD@dB\CD@cB}\let\cds@vcentre
\cds@vcenter\def\cds@vtriangleheight{\CD@BE\DiagramCellHeight
\DiagramCellHeight\DiagramCellWidth.577035\DiagramCellHeight}\def
\cds@vtrianglewidth{\CD@BE\DiagramCellWidth\DiagramCellWidth
\DiagramCellHeight1.73205\DiagramCellWidth}\def\cds@vmiddle{\let\CD@dB\CD@eB}%
\def\cds@w{\CD@zD\DiagramCellWidth}\let\cds@width\cds@w\def\diagram{\relax
\protect\CD@bC}\def\enddiagram{\protect\CD@SG}\def\CD@bC{\CD@g\CD@uI
\incommdiagtrue\edef\CD@wI{\the\CD@NB}\global\CD@NB\z@\boxmaxdepth\maxdimen
\everycr{}\CD@sK\everymath{}\everyhbox{}\ifx\pdfsyncstop\CD@qK\else
\pdfsyncstop\fi\CD@aC}\def\CD@aC{\CD@y\let\CD@N\CD@ZC\CD@CF\CD@AE\CD@WD}\def
\CD@ZC{\CD@gE\expandafter\CD@aC\else\expandafter\CD@WD\fi}\def\CD@WD{\let
\CD@EH\relax\CD@nE\CD@vE\else\CD@hK\else\CD@KB{landscape ignored without
PostScript}\CD@jA\fi\fi\fi\CD@EJ\setbox2=\vbox\bgroup\CD@JF\CD@VD}\def\CD@cH{%
\CD@nE\CD@fB\else\CD@dB\fi\CD@hA\nointerlineskip\setbox0=\null\ht0-\CD@pI\dp0%
\CD@pI\wd0\CD@kI\box0 \global\CD@QA\CD@kF\global\CD@yA\CD@XB\ifx\CD@NK\CD@qK
\global\CD@RA\CD@kF\else\global\CD@RA\CD@NK\fi\egroup\CD@zF\CD@nE\setbox2=%
\hbox to\dp2{\vrule height\wd2 depth\CD@QA width\z@\global\CD@QA\ht2\ht2\z@
\dp2\z@\wd2\z@\CD@hK\CD@tK{q 0 1 -1 0 0 0 cm}\else\global\CD@iG\CD@IK{0 1
bturn}\fi\box2\CD@gK\hss}\CD@DB\fi\ifnum\CD@yA=1 \else\CD@DB\fi\global
\@ignorefalse\CD@mE\leavevmode\fi\ifvmode\CD@TA\else\ifmmode\CD@PH\CD@GI\else
\CD@qE\CD@gA\fi\ifinner\CD@gA\fi\CD@mE\CD@GI\else\CD@sE\CD@QB\else\CD@TA\fi
\fi\fi\fi\CD@dD}\def\CD@dD{\global\CD@NB\CD@wI\relax\CD@xE\global\CD@ID\else
\aftergroup\CD@mC\fi\if@ignore\aftergroup\ignorespaces\fi\CD@wC\ignorespaces}%
\def\CD@fB{\advance\CD@pI\dimen1\relax}\def\CD@eB{\advance\CD@pI.5\dimen1%
\relax}\def\CD@bB{}\def\CD@cB{\CD@fB\advance\CD@pI\CD@YB\divide\CD@pI2
\advance\CD@pI-\axisheight\relax}\def\CD@aA{}\def\CD@iB{\CD@kF\z@}\def\CD@AA{%
\ifdim\dimen2>\CD@kF\CD@kF\dimen2 \else\dimen2\CD@kF\CD@kI\dimen0 \advance
\CD@kI\dimen2 \fi}\def\CD@QB{\skip0\z@\relax\loop\skip1\lastskip\ifdim\skip1>%
\z@\unskip\advance\skip0\skip1 \repeat\vadjust{\prevdepth\dp\strutbox\penalty
\predisplaypenalty\vskip\abovedisplayskip\CD@UA\penalty\postdisplaypenalty
\vskip\belowdisplayskip}\ifdim\skip0=\z@\else\hskip\skip0 \global\@ignoretrue
\fi}\def\CD@TA{\CD@LG\kern-\displayindent\CD@UA\CD@LG\global\@ignoretrue}\def
\CD@UA{\hbox to\hsize{\CD@fE\ifdim\CD@RA=\z@\else\advance\CD@QA-\CD@RA\setbox
2=\hbox{\kern\CD@RA\box2}\fi\fi\setbox1=\hbox{\ifx\CD@XA\CD@qK\else\CD@k
\CD@XA\CD@ND\fi}\CD@oE\CD@iE\else\advance\CD@QA\wd1 \fi\wd1\z@\box1 \fi\dimen
0\wd2 \advance\dimen0\wd1 \advance\dimen0-\hsize\ifdim\dimen0>-\CD@nA\CD@HA
\fi\advance\dimen0\CD@QA\ifdim\dimen0>\z@\CD@KB{wider than the page by \the
\dimen0 }\CD@HA\fi\CD@iE\hss\else\CD@V\CD@QA\CD@nA\fi\CD@GI\hss\kern-\wd1\box
1 }}\def\CD@GI{\CD@AF\CD@@F\else\CD@SC\global\CD@hG\fi\fi\kern\CD@QA\box2 }%
\CD@tG\CD@wE\CD@YC\CD@XC\def\CD@JF{\CD@cJ\ifdim\DiagramCellHeight=-\maxdimen
\DiagramCellHeight\CD@QC\fi\ifdim\DiagramCellWidth=-\maxdimen
\DiagramCellWidth\CD@QC\fi\global\CD@XC\CD@IF\let\CD@FE\empty\let\CD@z\CD@Q
\let\overprint\CD@eH\let\CD@s\CD@rJ\let\enddiagram\CD@ED\let\\\CD@cC\let\par
\CD@jH\let\CD@MD\empty\let\switch@arg\CD@PB\let\shift\CD@iA\baselineskip
\DiagramCellHeight\lineskip\z@\lineskiplimit\z@\mathsurround\z@\tabskip\z@
\CD@OB}\def\CD@VD{\penalty-123 \begingroup\CD@jA\aftergroup\CD@K\halign
\bgroup\global\advance\CD@NB1 \vadjust{\penalty1}\global\CD@FA\z@\CD@OB\CD@j#%
#\CD@DD\CD@Q\CD@Q\CD@OI\CD@j##\CD@DD\cr}\def\CD@ED{\CD@MD\CD@GD\crcr\egroup
\global\CD@JD\endgroup}\def\CD@j{\global\advance\CD@FA1 \futurelet\CD@EH\CD@i
}\def\CD@i{\ifx\CD@EH\CD@DD\CD@tJ\hskip1sp plus 1fil \relax\let\CD@DD\relax
\CD@vI\else\hfil\CD@k\objectstyle\let\CD@FE\CD@d\fi}\def\CD@DD{\CD@MD\relax
\CD@yI\CD@vI\global\CD@QA\CD@iA\penalty-9993 \CD@ND\hfil\null\kern-2\CD@QA
\null}\def\CD@cC{\cr}\def\across#1{\span\omit\mscount=#1 \global\advance
\CD@FA\mscount\global\advance\CD@FA\m@ne\CD@sF\ifnum\mscount>2 \CD@fJ\repeat
\ignorespaces}\def\CD@fJ{\relax\span\omit\advance\mscount\m@ne}\def\CD@qJ{%
\ifincommdiag\ifx\CD@iD\@fillh\ifx\CD@jD\@fillh\ifdim\dimen3>\z@\else\ifdim
\dimen2>93\CD@@I\ifdim\dimen2>18\p@\ifdim\CD@LF>\z@\count@\CD@bJ\advance
\count@\m@ne\ifnum\count@<\z@\count@20\let\CD@aJ\CD@uJ\fi\xdef\CD@bJ{\the
\count@}\fi\fi\fi\fi\fi\fi\fi}\def\CD@cG#1{\vrule\horizhtdp width#1\dimen@
\kern2\dimen@}\def\CD@uJ{\rlap{\dimen@\CD@@I\CD@V\dimen@{.182\p@}\CD@zH
\dimen@\advance\CD@tI\dimen@\CD@cG0\CD@cG0\CD@cG2\CD@cG6\CD@cG6\CD@cG2\CD@cG0%
\CD@cG0\CD@cG2\CD@cG6\CD@cG0\CD@cG0\CD@cG2\CD@cG2\CD@cG6\CD@cG0\CD@cG0\CD@cG2%
\CD@cG6\CD@cG2\CD@cG2\CD@cG0\CD@cG0}}\def\CD@bJ{10}\def\CD@aJ{}\def\CD@XD{%
\CD@gE\CD@TB\fi\CD@x\CD@WF\CD@HI}\def\CD@x{\CD@QJ\CD@DC\CD@MJ\ifdim\CD@DC=\z@
\else\CD@pF\CD@DC\fi\ifvoid3 \setbox3=\null\ht3\CD@tI\dp3\CD@sI\else\CD@V{\ht
3}\CD@tI\CD@V{\dp3}\CD@sI\fi\dimen3=.5\wd3 \ifdim\dimen3=\z@\CD@tE\else\dimen
3-\CD@XH\fi\else\CD@TB\fi\CD@V{\dimen2}{\wd7}\CD@V{\dimen2}{\wd6}\CD@qJ
\advance\dimen2-2\dimen3 \dimen4.5\dimen2 \dimen2\dimen4 \advance\dimen2%
\CD@eJ\advance\dimen4-\CD@eJ\advance\dimen2-\wd1 \advance\dimen4-\wd5 \ifvoid
2 \else\CD@V{\ht3}{\ht2}\CD@V{\dp3}{\dp2}\CD@V{\dimen2}{\wd2}\fi\ifvoid4 \else
\CD@V{\ht3}{\ht4}\CD@V{\dp3}{\dp4}\CD@V{\dimen4}{\wd4}\fi\advance\skip2\dimen
2 \advance\skip4\dimen4 \CD@tE\advance\skip2\skip4 \dimen0\dimen5 \advance
\dimen0\wd5 \skip3-\skip4 \advance\skip3-\dimen0 \let\CD@jD\empty\else\skip3%
\z@\relax\dimen0\z@\fi}\def\CD@WF{\offinterlineskip\lineskip.2\CD@zC\ifvoid6
\else\setbox3=\vbox{\hbox to2\dimen3{\hss\box6\hss}\box3}\fi\ifvoid7 \else
\setbox3=\vtop{\box3 \hbox to2\dimen3{\hss\box7\hss}}\fi}\def\CD@HI{\kern
\dimen1 \box1 \CD@aJ\CD@iD\hskip\skip2 \kern\dimen0 \ifincommdiag\CD@jE
\penalty1\fi\kern\dimen3 \penalty\CD@GB\hskip\skip3 \null\kern-\dimen3 \else
\hskip\skip3 \fi\box3 \CD@jD\hskip\skip4 \box5 \kern\dimen5}\def\CD@MF{\ifnum
\CD@LH>\CD@TC\CD@V{\dimen1}\objectheight\CD@V{\dimen5}\objectheight\else\CD@V
{\dimen1}\objectwidth\CD@V{\dimen5}\objectwidth\fi}\def\CD@Y{\begingroup
\ifdim\dimen7=\z@\kern\dimen8 \else\ifdim\dimen6=\z@\kern\dimen9 \else\dimen5%
\dimen6 \dimen6\dimen9 \CD@KJ\dimen4\dimen2 \CD@dG{\dimen4}\dimen6\dimen5
\dimen7\dimen8 \CD@KJ\CD@iC{\dimen2}\ifdim\dimen2<\dimen4 \kern\dimen2 \else
\kern\dimen4 \fi\fi\fi\endgroup}\def\CD@jJ{\CD@JI\setbox\z@\hbox{\lower
\axisheight\hbox to\dimen2{\CD@DF\ifPositiveGradient\dimen8\ht\CD@MH\dimen9%
\CD@mI\else\dimen8\dp3 \dimen9\dimen1 \fi\else\dimen8 \ifPositiveGradient
\objectheight\else\z@\fi\dimen9\objectwidth\fi\advance\dimen8
\ifPositiveGradient-\fi\axisheight\CD@Y\unhbox\z@\CD@DF\ifPositiveGradient
\dimen8\dp3 \dimen9\dimen0 \else\dimen8\ht\CD@MH\dimen9\CD@mF\fi\else\dimen8
\ifPositiveGradient\z@\else\objectheight\fi\dimen9\objectwidth\fi\advance
\dimen8 \ifPositiveGradient\else-\fi\axisheight\CD@Y}}}\def\CD@bD{\dimen6
\CD@aK\DiagramCellHeight\dimen7 \CD@WK\DiagramCellWidth\CD@jJ
\ifPositiveGradient\advance\dimen7-\CD@ZK\DiagramCellWidth\else\dimen7 \CD@ZK
\DiagramCellWidth\dimen6\z@\fi\advance\dimen6-\CD@bK\DiagramCellHeight\CD@mK
\setbox0=\rlap{\kern-\dimen7 \lower\dimen6\box\z@}\ht0\z@\dp0\z@\raise
\axisheight\box0 }\def\CD@mK{\setbox0\hbox{\ht\z@\z@\dp\z@\z@\wd\z@\z@\CD@hK
\expandafter\CD@tK{q \CD@eK\space\CD@lK\space\CD@kK\space\CD@eK\space0 0 cm}%
\else\global\CD@iG\CD@eD{\the\CD@TC\space\ifPositiveGradient\else-\fi\the
\CD@LH\space bturn}\fi\box\z@\CD@gK}}\def\CD@vB{\advance\CD@hF-\CD@mI\CD@wJ
\CD@hF\advance\CD@wJ\CD@hI\ifvoid\CD@sH\ifdim\CD@wJ<.1em\ifnum\CD@gD=\@m\else
\CD@aG h\CD@wJ<.1em:objects overprint:\CD@FA\CD@gD\fi\fi\else\ifhbox\CD@sH
\CD@SK\else\CD@TK\fi\advance\CD@wJ\CD@mI\CD@bH{-\CD@mI}{\box\CD@sH}{\CD@wJ}%
\z@\fi\CD@hF-\CD@mF\CD@gD\CD@FA\CD@hI\z@}\def\CD@SK{\setbox\CD@sH=\hbox{%
\unhbox\CD@sH\unskip\unpenalty}\setbox\CD@tH=\hbox{\unhbox\CD@tH\unskip
\unpenalty}\setbox\CD@sH=\hbox to\CD@wJ{\CD@OA\wd\CD@sH\unhbox\CD@sH\CD@PA
\lastkern\unkern\ifdim\CD@PA=\z@\CD@UB\advance\CD@OA-\wd\CD@tH\else\CD@TB\fi
\ifnum\lastpenalty=\z@\else\CD@JA\unpenalty\fi\kern\CD@PA\ifdim\CD@hF<\CD@OA
\CD@JA\fi\ifdim\CD@hI<\wd\CD@tH\CD@JA\fi\CD@jE\CD@hI\CD@wJ\advance\CD@hI-%
\CD@OA\advance\CD@hI\wd\CD@tH\ifdim\CD@hI<2\wd\CD@tH\CD@aG h\CD@hI<2\wd\CD@tH
:arrow too short:\CD@FA\CD@gD\fi\divide\CD@hI\tw@\CD@hF\CD@wJ\advance\CD@hF-%
\CD@hI\fi\CD@tE\kern-\CD@hI\fi\hbox to\CD@hI{\unhbox\CD@tH}\CD@HG}}\CD@tG
\ifinpile\inpiletrue\inpilefalse\inpilefalse\def\pile{\protect\CD@UJ\protect
\CD@uH}\def\CD@uH#1{\CD@l#1\CD@QD}\def\CD@UJ{\CD@nB{pile}\setbox0=\vtop
\bgroup\aftergroup\CD@lD\inpiletrue\let\CD@FE\empty\let\pile\CD@KF\let\CD@QD
\CD@PD\let\CD@GD\CD@FD\CD@yH\baselineskip.5\PileSpacing\lineskip.1\CD@zC
\relax\lineskiplimit\lineskip\mathsurround\z@\tabskip\z@\let\\\CD@wH}\def
\CD@l{\CD@DE\CD@YF\empty\halign\bgroup\hfil\CD@k\let\CD@FE\CD@d\let\\\CD@vH##%
\CD@MD\CD@ND\hfil\CD@Q\CD@R##\cr}\CD@rG\CD@NE{pile only allows one column.}%
\CD@rG\CD@UE{you left it out!}\def\CD@R{\CD@QD\CD@Q\relax\CD@YA{missing \CD@yC
\space inserted after \string\pile}\CD@NE}\def\CD@PD{\CD@MD\crcr\egroup
\egroup}\def\CD@GD{\CD@MD}\def\CD@FD{\CD@MD\relax\CD@QD\CD@YA{missing \CD@yC
\space inserted between \string\pile\space and \CD@HD}\CD@UE}\def\CD@QD{%
\CD@MD}\def\CD@lD{\vbox{\dimen1\dp0 \unvbox0 \setbox0=\lastbox\advance\dimen1%
\dp0 \nointerlineskip\box0 \nointerlineskip\setbox0=\null\dp0.5\dimen1\ht0-%
\dp0 \box0}\ifincommdiag\CD@tJ\penalty-9998 \fi\xdef\CD@YF{pile}}\def\CD@vH{%
\cr}\def\CD@wH{\noalign{\skip@\prevdepth\advance\skip@-\baselineskip
\prevdepth\skip@}}\def\CD@KF#1{#1}\def\CD@TK{\setbox\CD@sH=\vbox{\unvbox
\CD@sH\setbox1=\lastbox\setbox0=\box\voidb@x\CD@tF\setbox\CD@sH=\lastbox
\ifhbox\CD@sH\CD@rC\repeat\unvbox0 \global\CD@QA\CD@ZE}\CD@ZE\CD@QA}\def
\CD@rC{\CD@jE\setbox\CD@sH=\hbox{\unhbox\CD@sH\unskip\setbox\CD@sH=\lastbox
\unskip\unhbox\CD@sH}\ifdim\CD@wJ<\wd\CD@sH\CD@aG h\CD@wJ<\wd\CD@sH:arrow in
pile too short:\CD@FA\CD@gD\else\setbox\CD@sH=\hbox to\CD@wJ{\unhbox\CD@sH}%
\fi\else\CD@gJ\fi\setbox0=\vbox{\box\CD@sH\nointerlineskip\ifvoid0 \CD@tJ\box
1 \else\vskip\skip0 \unvbox0 \fi}\skip0=\lastskip\unskip}\def\CD@gJ{\penalty7
\noindent\unhbox\CD@sH\unskip\setbox\CD@sH=\lastbox\unskip\unhbox\CD@sH
\endgraf\setbox\CD@tH=\lastbox\unskip\setbox\CD@tH=\hbox{\CD@JG\unhbox\CD@tH
\unskip\unskip\unpenalty}\ifcase\prevgraf\cd@shouldnt P\or\ifdim\CD@wJ<\wd
\CD@tH\CD@aG h\CD@wJ<\wd\CD@sH:object in pile too wide:\CD@FA\CD@gD\setbox
\CD@sH=\hbox to\CD@wJ{\hss\unhbox\CD@tH\hss}\else\setbox\CD@sH=\hbox to\CD@wJ
{\hss\kern\CD@hF\unhbox\CD@tH\kern\CD@hI\hss}\fi\or\setbox\CD@sH=\lastbox
\unskip\CD@SK\else\cd@shouldnt Q\fi\unskip\unpenalty}\def\CD@cD{\CD@MJ\ifvoid
3 \setbox3=\null\ht3\axisheight\dp3-\ht3 \dimen3.5\CD@LF\else\dimen4\dp3
\dimen3.5\wd3 \setbox3=\CD@GG{\box3}\dp3\dimen4 \ifdim\ht3=-\dp3 \else\CD@TB
\fi\fi\dimen0\dimen3 \advance\dimen0-.5\CD@LF\setbox0\null\ht0\ht3\dp0\dp3\wd
0\wd3 \ifvoid6\else\setbox6\hbox{\unhbox6\kern\dimen0\kern2pt}\dimen0\wd6 \fi
\ifvoid7\else\setbox7\hbox{\kern2pt\kern\dimen3\unhbox7}\dimen3\wd7 \fi
\setbox3\hbox{\ifvoid6\else\kern-\dimen0\unhbox6\fi\unhbox3 \ifvoid7\else
\unhbox7\kern-\dimen3\fi}\ht3\ht0\dp3\dp0\wd3\wd0 \CD@tE\dimen4=\ht\CD@MH
\advance\dimen4\dp5 \advance\dimen4\dimen1 \let\CD@jD\empty\else\dimen4\ht3
\fi\setbox0\null\ht0\dimen4 \offinterlineskip\setbox8=\vbox spread2ex{\kern
\dimen5 \box1 \CD@iD\vfill\CD@tE\else\kern\CD@eJ\fi\box0}\ht8=\z@\setbox9=%
\vtop spread2ex{\kern-\ht3 \kern-\CD@eJ\box3 \CD@jD\vfill\box5 \kern\dimen1}%
\dp9=\z@\hskip\dimen0plus.0001fil \box9 \kern-\CD@LF\box8 \CD@kE\penalty2 \fi
\CD@tE\penalty1 \fi\kern\PileSpacing\kern-\PileSpacing\kern-.5\CD@LF\penalty
\CD@GB\null\kern\dimen3}\def\CD@cI{\ifhbox\CD@VA\CD@KB{clashing verticals}\ht
\CD@MH.5\dp\CD@VA\dp\CD@MH-\ht5 \CD@yB\ht\CD@MH\z@\dp\CD@MH\z@\fi\dimen1\dp
\CD@VA\CD@xA\prevgraf\unvbox\CD@VA\CD@wA\lastpenalty\unpenalty\setbox\CD@VA=%
\null\setbox\CD@lI=\hbox{\CD@JG\unhbox\CD@lI\unskip\unpenalty\dimen0\lastkern
\unkern\unkern\unkern\kern\dimen0 \CD@HG}\setbox\CD@lF=\hbox{\unhbox\CD@lF
\dimen0\lastkern\unkern\unkern\global\CD@QA\lastkern\unkern\kern\dimen0 }%
\CD@tF\ifnum\CD@xA>4 \CD@zI\repeat\unskip\unskip\advance\CD@mF.5\wd\CD@VA
\advance\CD@mF\wd\CD@lF\advance\CD@mI.5\wd\CD@VA\advance\CD@mI\wd\CD@lI\ifnum
\CD@FA=\CD@lA\CD@OA.5\wd\CD@VA\edef\CD@NK{\the\CD@OA}\fi\setbox\CD@VA=\hbox{%
\kern-\CD@mF\box\CD@lF\unhbox\CD@VA\box\CD@lI\kern-\CD@mI\penalty\CD@wA
\penalty\CD@NB}\ht\CD@VA\dimen1 \dp\CD@VA\z@\wd\CD@VA\CD@tB\CD@vB}\def\CD@zI{%
\ifdim\wd\CD@lF<\CD@QA\setbox\CD@lF=\hbox to\CD@QA{\CD@JG\unhbox\CD@lF}\fi
\advance\CD@xA\m@ne\setbox\CD@VA=\hbox{\box\CD@lF\unhbox\CD@VA}\unskip\setbox
\CD@lF=\lastbox\setbox\CD@lF=\hbox{\unhbox\CD@lF\unskip\unpenalty\dimen0%
\lastkern\unkern\unkern\global\CD@QA\lastkern\unkern\kern\dimen0 }}\def\CD@yB
{\dimen1\dp\CD@VA\ifhbox\CD@VA\CD@xB\else\CD@zB\fi\setbox\CD@VA=\vbox{%
\penalty\CD@NB}\dp\CD@VA-\dp\CD@MH\wd\CD@VA\CD@tB}\def\CD@zB{\unvbox\CD@VA
\CD@wA\lastpenalty\unpenalty\ifdim\dimen1<\ht\CD@MH\CD@aG v\dimen1<\ht\CD@MH:%
rows overprint:\CD@NB\CD@wA\fi}\def\CD@xB{\dimen0=\ht\CD@VA\setbox\CD@VA=%
\hbox\bgroup\advance\dimen1-\ht\CD@MH\unhbox\CD@VA\CD@xA\lastpenalty
\unpenalty\CD@wA\lastpenalty\unpenalty\global\CD@RA-\lastkern\unkern\setbox0=%
\lastbox\CD@tF\setbox\CD@VA=\hbox{\box0\unhbox\CD@VA}\setbox0=\lastbox\ifhbox
0 \CD@kJ\repeat\global\CD@SA-\lastkern\unkern\global\CD@QA\CD@JK\unhbox\CD@VA
\egroup\CD@JK\CD@QA\CD@bH{\CD@SA}{\box\CD@VA}{\CD@RA}{\dimen1}}\def\CD@kJ{%
\setbox0=\hbox to\wd0\bgroup\unhbox0 \unskip\unpenalty\dimen7\lastkern\unkern
\ifnum\lastpenalty=1 \unpenalty\CD@UB\else\CD@TB\fi\ifnum\lastpenalty=2
\unpenalty\dimen2.5\dimen0\advance\dimen2-.5\dimen1\advance\dimen2-%
\axisheight\else\dimen2\z@\fi\setbox0=\lastbox\dimen6\lastkern\unkern\setbox1%
=\lastbox\setbox0=\vbox{\unvbox0 \CD@tE\kern-\dimen1 \else\ifdim\dimen2=\z@
\else\kern\dimen2 \fi\fi}\ifdim\dimen0<\ht0 \CD@aG v\dimen0<\ht0:upper part of
vertical too short:{\CD@tE\CD@NB\else\CD@wA\fi}\CD@xA\else\setbox0=\vbox to%
\dimen0{\unvbox0}\fi\setbox1=\vtop{\unvbox1}\ifdim\dimen1<\dp1 \CD@aG v\dimen
1<\dp1:lower part of vertical too short:\CD@NB\CD@wA\else\setbox1=\vtop to%
\dimen1{\ifdim\dimen2=\z@\else\kern-\dimen2 \fi\unvbox1 }\fi\box1 \kern\dimen
6 \box0 \kern\dimen7 \CD@HG\global\CD@QA\CD@JK\egroup\CD@JK\CD@QA\relax}%
\countdef\CD@u=14 \newcount\CD@CA\newcount\CD@XB\newcount\CD@NB\let\CD@LB
\insc@unt\newcount\CD@FA\newcount\CD@lA\let\CD@mA\CD@XB\newcount\CD@MB\CD@tG
\CD@DF\CD@bI\CD@aI\CD@aI\def\CD@nD{-1}\def\CD@K{\ifnum\CD@nD<\z@\else
\begingroup\scrollmode\showboxdepth\CD@nD\showboxbreadth\maxdimen\showlists
\endgroup\fi\CD@bI\CD@zF\CD@CA=\CD@u\advance\CD@CA1 \CD@XB=\CD@CA\ifnum\CD@NB
=1 \CD@JA\fi\advance\CD@XB\CD@NB\dimen1\z@\skip0\z@\count@=\insc@unt\advance
\count@\CD@u\divide\count@2 \ifnum\CD@XB>\count@\CD@KB{The diagram has too
many rows! It can't be reformatted.}\else\CD@NG\CD@WI\fi\CD@cH}\def\CD@NG{%
\CD@NB\CD@CA\CD@uF\ifnum\CD@NB<\CD@XB\setbox\CD@NB\box\voidb@x\advance\CD@NB1%
\relax\repeat\CD@NB\CD@CA\skip\z@\z@\CD@uF\CD@GB\lastpenalty\unpenalty\ifnum
\CD@GB>\z@\CD@KE\repeat\ifnum\CD@GB=-123 \CD@tJ\unpenalty\else\cd@shouldnt D%
\fi\ifx\v@grid\relax\else\CD@NB\CD@XB\advance\CD@NB\m@ne\expandafter\CD@VJ
\v@grid\fi\CD@MB\CD@mA\CD@tB\z@\CD@XG\ifx\h@grid\relax\else\expandafter\CD@LJ
\h@grid\fi\count@\CD@XB\advance\count@\m@ne\CD@YB\ht\count@}\def\CD@KE{%
\ifcase\CD@GB\or\CD@MG\else\CD@uA-\lastpenalty\unpenalty\CD@vA\lastpenalty
\unpenalty\setbox0=\lastbox\CD@WG\fi\CD@wD}\def\CD@wD{\skip1\lastskip\unskip
\advance\skip0\skip1 \ifdim\skip1=\z@\else\expandafter\CD@wD\fi}\def\CD@MG{%
\setbox0=\lastbox\CD@pI\dp0 \advance\CD@pI\skip\z@\skip\z@\z@\advance\CD@NF
\CD@pI\CD@uE\ifnum\CD@NB>\CD@CA\CD@NF\DiagramCellHeight\CD@pI\CD@NF\advance
\CD@pI-\CD@qI\fi\fi\CD@qI\ht0 \CD@NF\CD@qI\setbox\CD@NB\hbox{\unhbox\CD@NB
\unhbox0}\dp\CD@NB\CD@pI\ht\CD@NB\CD@qI\advance\CD@NB1 }\def\CD@WG{\ifnum
\CD@uA<\z@\advance\CD@uA\CD@XB\ifnum\CD@uA<\CD@CA\CD@UG\else\CD@OA\dp\CD@uA
\CD@PA\ht\CD@uA\setbox\CD@uA\hbox{\box\z@\penalty\CD@vA\penalty\CD@GB\unhbox
\CD@uA}\dp\CD@uA\CD@OA\ht\CD@uA\CD@PA\fi\else\CD@UG\fi}\def\CD@UG{\CD@KB{%
diagonal goes outside diagram (lost)}}\def\CD@fI{\advance\CD@uA\CD@XB\ifnum
\CD@uA<\CD@CA\CD@UG\else\ifnum\CD@uA=\CD@NB\CD@VG\else\ifnum\CD@uA>\CD@NB
\cd@shouldnt M\else\CD@OA\dp\CD@uA\CD@PA\ht\CD@uA\setbox\CD@uA\hbox{\box\z@
\penalty\CD@vA\penalty\CD@GB\unhbox\CD@uA}\dp\CD@uA\CD@OA\ht\CD@uA\CD@PA\fi
\fi\fi}\def\CD@WI{\CD@t\CD@AJ\setbox\CD@PC=\hbox{\CD@k A\@super f\CD@lJ f%
\CD@ND}\CD@ZE\z@\CD@JK\z@\CD@kI\z@\CD@kF\z@\CD@NB=\CD@XB\CD@NF\z@\CD@uB\z@
\CD@uF\ifnum\CD@NB>\CD@CA\advance\CD@NB\m@ne\CD@qI\ht\CD@NB\CD@pI\dp\CD@NB
\advance\CD@NF\CD@qI\CD@rI\advance\CD@uB\CD@NF\CD@KC\CD@ZI\CD@w\ht\CD@NB
\CD@qI\dp\CD@NB\CD@pI\nointerlineskip\box\CD@NB\CD@NF\CD@pI\setbox\CD@NB\null
\ht\CD@NB\CD@uB\repeat\CD@wB\nointerlineskip\box\CD@NB\CD@gG\CD@ZE
\DiagramCellWidth{width}\CD@gG\CD@JK\DiagramCellHeight{height}\CD@VA\CD@LB
\advance\CD@VA-\CD@lA\advance\CD@VA\m@ne\advance\CD@VA\CD@mA\dimen0\wd\CD@VA
\CD@tI\axisheight\dimen1\CD@uB\advance\dimen1-\CD@YB\dimen2\CD@kI\advance
\dimen2-\dimen0 \advance\CD@XB-\CD@CA\advance\CD@LB-\CD@lA}\count@\year
\multiply\count@12 \advance\count@\month\ifnum\count@>24254 \loop\iftrue
\repeat\fi\def\CD@wB{\CD@qI-\CD@NF\CD@pI\CD@NF
\setbox\CD@MH=\null\dp\CD@MH\CD@NF\ht\CD@MH-\CD@NF\CD@mF\z@\CD@mI\z@\CD@lA
\CD@LB\advance\CD@lA-\CD@MB\advance\CD@lA\CD@mA\CD@FA\CD@LB\CD@VA\CD@MB\CD@sF
\ifnum\CD@FA>\CD@lA\advance\CD@FA\m@ne\advance\CD@VA\m@ne\CD@tB\wd\CD@VA
\setbox\CD@FA=\box\voidb@x\CD@yB\repeat\CD@w\ht\CD@NB\CD@qI\dp\CD@NB\CD@pI}%
\def\CD@gG#1#2#3{\ifdim#1>.01\CD@zC\CD@PA#2\relax\advance\CD@PA#1\relax
\advance\CD@PA.99\CD@zC\count@\CD@PA\divide\count@\CD@zC\CD@KB{increase cell #%
3 to \the\count@ em}\fi}\def\CD@rI{\CD@FA=\CD@LB\penalty4 \noindent\unhbox
\CD@NB\CD@sF\unskip\setbox0=\lastbox\ifhbox0 \advance\CD@FA\m@ne\setbox\CD@FA
\hbox to\wd0{\null\penalty-9990\null\unhbox0}\repeat\CD@lA\CD@FA\advance
\CD@FA\CD@MB\advance\CD@FA-\CD@mA\ifnum\CD@FA<\CD@LB\count@\CD@FA\advance
\count@\m@ne\dimen0=\wd\count@\count@\CD@MB\advance\count@\m@ne\CD@tB\wd
\count@\CD@sF\ifnum\CD@FA<\CD@LB\CD@DJ\CD@XG\dimen0\wd\CD@FA\advance\CD@FA1
\repeat\fi\CD@sF\CD@GB\lastpenalty\unpenalty\ifnum\CD@GB>\z@\CD@vA
\lastpenalty\unpenalty\CD@VG\repeat\endgraf\unskip\ifnum\lastpenalty=4
\unpenalty\else\cd@shouldnt S\fi}\def\CD@VG{\advance\CD@vA\CD@lA\advance
\CD@vA\m@ne\setbox0=\lastbox\ifnum\CD@vA<\CD@LB\setbox\CD@vA\hbox{\box0%
\penalty\CD@GB\unhbox\CD@vA}\else\CD@UG\fi}\def\CD@bG{}\CD@tG\CD@uE\CD@WB
\CD@VB\def\CD@DJ{\advance\dimen0\wd\CD@FA\divide\dimen0\tw@\CD@uE\dimen0%
\DiagramCellWidth\else\CD@V{\dimen0}\DiagramCellWidth\CD@pJ\fi\advance\CD@tB
\dimen0 }\def\CD@XG{\setbox\CD@MB=\vbox{}\dp\CD@MB=\CD@uB\wd\CD@MB\CD@tB
\advance\CD@MB1 }\def\CD@LJ#1,{\def\CD@GK{#1}\ifx\CD@GK\CD@RD\else\advance
\CD@tB\CD@GK\DiagramCellWidth\CD@XG\expandafter\CD@LJ\fi}\def\CD@VJ#1,{\def
\CD@GK{#1}\ifx\CD@GK\CD@RD\else\ifnum\CD@NB>\CD@CA\CD@NF\CD@GK
\DiagramCellHeight\advance\CD@NF-\dp\CD@NB\advance\CD@NB\m@ne\ht\CD@NB\CD@NF
\fi\expandafter\CD@VJ\fi}\def\CD@pJ{\CD@wE\CD@OA\dimen0 \advance\CD@OA-%
\DiagramCellWidth\ifdim\CD@OA>2\MapShortFall\CD@KB{badly drawn diagonals (see
manual)}\let\CD@pJ\empty\fi\else\let\CD@pJ\empty\fi}\def\CD@KC{\CD@VA\CD@mA
\CD@sF\ifnum\CD@VA<\CD@MB\dimen0\dp\CD@VA\advance\dimen0\CD@NF\dp\CD@VA\dimen
0 \advance\CD@VA1 \repeat}\def\CD@bH#1#2#3#4{\ifnum\CD@FA<\CD@LB\CD@OA=#1%
\relax\setbox\CD@FA=\hbox{\setbox0=#2\dimen7=#4\relax\dimen8=#3\relax\ifhbox
\CD@FA\unhbox\CD@FA\advance\CD@OA-\lastkern\unkern\fi\ifdim\CD@OA=\z@\else
\kern-\CD@OA\fi\raise\dimen7\box0 \kern-\dimen8 }\ifnum\CD@FA=\CD@lA\CD@V
\CD@kF\CD@OA\fi\else\cd@shouldnt O\fi}\def\CD@w{\setbox\CD@NB=\hbox{\CD@FA
\CD@lA\CD@VA\CD@mA\CD@PA\z@\relax\CD@sF\ifnum\CD@FA<\CD@LB\CD@tB\wd\CD@VA
\relax\CD@eI\advance\CD@FA1 \advance\CD@VA1 \repeat}\CD@V\CD@kI{\wd\CD@NB}\wd
\CD@NB\z@}\def\CD@eI{\ifhbox\CD@FA\CD@OA\CD@tB\relax\advance\CD@OA-\CD@PA
\relax\ifdim\CD@OA=\z@\else\kern\CD@OA\fi\CD@PA\CD@tB\advance\CD@PA\wd\CD@FA
\relax\unhbox\CD@FA\advance\CD@PA-\lastkern\unkern\fi}\def\CD@ZI{\setbox
\CD@sH=\box\voidb@x\CD@VA=\CD@MB\CD@FA\CD@LB\CD@VA\CD@mA\advance\CD@VA\CD@FA
\advance\CD@VA-\CD@lA\advance\CD@VA\m@ne\CD@tB\wd\CD@VA\count@\CD@LB\advance
\count@\m@ne\CD@hF.5\wd\count@\advance\CD@hF\CD@tB\CD@A\m@ne\CD@gD\@m\CD@sF
\ifnum\CD@FA>\CD@lA\advance\CD@FA\m@ne\advance\CD@hF-\CD@tB\CD@PI\wd\CD@VA
\CD@tB\advance\CD@hF\CD@tB\advance\CD@VA\m@ne\CD@tB\wd\CD@VA\repeat\CD@mF
\CD@kF\CD@mI-\CD@mF\CD@vB}\newcount\CD@GB\def\CD@s{}\def\CD@t{\mathsurround
\z@\hsize\z@\rightskip\z@ plus1fil minus\maxdimen\parfillskip\z@\linepenalty
9000 \looseness0 \hfuzz\maxdimen\hbadness10000 \clubpenalty0 \widowpenalty0
\displaywidowpenalty0 \interlinepenalty0 \predisplaypenalty0
\postdisplaypenalty0 \interdisplaylinepenalty0 \interfootnotelinepenalty0
\floatingpenalty0 \brokenpenalty0 \everypar{}\leftskip\z@\parskip\z@
\parindent\z@\pretolerance10000 \tolerance10000 \hyphenpenalty10000
\exhyphenpenalty10000 \binoppenalty10000 \relpenalty10000 \adjdemerits0
\doublehyphendemerits0 \finalhyphendemerits0 \baselineskip\z@\CD@IA\prevdepth
\z@}\newbox\CD@KG\newbox\CD@IG\def\CD@JG{\unhcopy\CD@KG}\def\CD@HG{\unhcopy
\CD@IG}\def\CD@iJ{\hbox{}\penalty1\nointerlineskip}\def\CD@PI{\penalty5
\noindent\setbox\CD@MH=\null\CD@mF\z@\CD@mI\z@\ifnum\CD@FA<\CD@LB\ht\CD@MH\ht
\CD@FA\dp\CD@MH\dp\CD@FA\unhbox\CD@FA\skip0=\lastskip\unskip\else\CD@OK\skip0%
=\z@\fi\endgraf\ifcase\prevgraf\cd@shouldnt Y \or\cd@shouldnt Z \or\CD@RI\or
\CD@XI\else\CD@QI\fi\unskip\setbox0=\lastbox\unskip\unskip\unpenalty\noindent
\unhbox0\setbox0\lastbox\unpenalty\unskip\unskip\unpenalty\setbox0\lastbox
\CD@tF\CD@GB\lastpenalty\unpenalty\ifnum\CD@GB>\z@\setbox\z@\lastbox\CD@lB
\repeat\endgraf\unskip\unskip\unpenalty}\def\CD@YJ{\CD@uA\CD@XB\advance\CD@uA
-\CD@NB\CD@vA\CD@FA\advance\CD@vA-\CD@lA\advance\CD@vA1 \expandafter\message{%
prevgraf=\the\prevgraf at (\the\CD@uA,\the\CD@vA)}}\def\CD@XI{\CD@CE\setbox
\CD@lI=\lastbox\setbox\CD@lI=\hbox{\unhbox\CD@lI\unskip\unpenalty}\unskip
\ifdim\ht\CD@lI>\ht\CD@PC\setbox\CD@MH=\copy\CD@lI\else\ifdim\dp\CD@lI>\dp
\CD@PC\setbox\CD@MH=\copy\CD@lI\else\CD@FG\CD@lI\fi\fi\advance\CD@mF.5\wd
\CD@lI\advance\CD@mI.5\wd\CD@lI\setbox\CD@lI=\hbox{\unhbox\CD@lI\CD@HG}\CD@bH
\CD@mF{\box\CD@lI}\CD@mI\z@\CD@yB\CD@vB}\def\CD@CE{\ifnum\CD@A>0 \advance
\dimen0-\CD@tB\CD@iA-.5\dimen0 \CD@A-\CD@A\else\CD@A0 \CD@iA\z@\fi\setbox
\CD@MH=\lastbox\setbox\CD@MH=\hbox{\unhbox\CD@MH\unskip\unskip\unpenalty
\setbox0=\lastbox\global\CD@QA\lastkern\unkern}\advance\CD@iA-.5\CD@QA\unskip
\setbox\CD@MH=\null\CD@mI\CD@iA\CD@mF-\CD@iA}\def\CD@Z{\ht\CD@MH\CD@tI\dp
\CD@MH\CD@sI}\def\CD@FG#1{\setbox\CD@MH=\hbox{\CD@V{\ht\CD@MH}{\ht#1}\CD@V{%
\dp\CD@MH}{\dp#1}\CD@V{\wd\CD@MH}{\wd#1}\vrule height\ht\CD@MH depth\dp\CD@MH
width\wd\CD@MH}}\def\CD@QI{\CD@CE\CD@Z\setbox\CD@lI=\lastbox\unskip\setbox
\CD@lF=\lastbox\unskip\setbox\CD@lF=\hbox{\unhbox\CD@lF\unskip\global\CD@yA
\lastpenalty\unpenalty}\advance\CD@yA9999 \ifcase\CD@yA\CD@VI\or\CD@YI\or
\CD@TI\or\CD@dI\or\CD@cI\or\CD@SI\else\cd@shouldnt9\fi}\def\CD@VI{\CD@FG
\CD@lI\CD@UI\setbox\CD@sH=\box\CD@lF\setbox\CD@tH=\box\CD@lI}\def\CD@YI{%
\CD@FG\CD@lF\setbox\CD@lI\hbox{\penalty8 \unhbox\CD@lI\unskip\unpenalty\ifnum
\lastpenalty=8 \else\CD@xH\fi}\CD@UI\setbox\CD@lF=\hbox{\unhbox\CD@lF\unskip
\unpenalty\global\setbox\CD@DA=\lastbox}\ifdim\wd\CD@lF=\z@\else\CD@xH\fi
\setbox\CD@sH=\box\CD@DA}\def\CD@xH{\CD@KB{extra material in \string\pile
\space cell (lost)}}\def\CD@UI{\CD@yB\ifvoid\CD@sH\else\CD@KB{Clashing
horizontal arrows}\CD@mI.5\CD@hF\CD@mF-\CD@mI\CD@vB\CD@mI\z@\CD@mF\z@\fi
\CD@hI\CD@hF\advance\CD@hI-\CD@mI\CD@hF-\CD@mF\CD@JC\CD@FA}\def\CD@RI{\setbox
0\lastbox\unskip\CD@iA\z@\CD@Z\ifdim\skip0>\z@\CD@tJ\CD@A0 \else\ifnum\CD@A<1
\CD@A0 \dimen0\CD@tB\fi\advance\CD@A1 \fi}\def\VonH{\CD@MA46\VonH{.5\CD@LF}}%
\def\HonV{\CD@MA57\HonV{.5\CD@LF}}\def\HmeetV{\CD@MA44\HmeetV{-\MapShortFall}%
}\def\CD@MA#1#2#3#4{\CD@pB34#1{\string#3}\CD@SD\CD@GB-999#2 \dimen0=#4\CD@tI
\dimen0\advance\CD@tI\axisheight\CD@sI\dimen0\advance\CD@sI-\axisheight\CD@CF
\CD@HC\CD@ZD}\def\CD@HC#1{\setbox0=\hbox{\CD@k#1\CD@ND}\dimen0.5\wd0 \CD@tI
\ht0 \CD@sI\dp0 \CD@ZD}\def\CD@SD{\setbox0=\null\ht0=\CD@tI\dp0=\CD@sI\wd0=%
\dimen0 \copy0\penalty\CD@GB\box0 }\def\CD@TI{\CD@GC\CD@yB}\def\CD@dI{\CD@GC
\CD@vB}\def\CD@SI{\CD@GC\CD@yB\CD@vB}\def\CD@GC{\setbox\CD@lI=\hbox{\unhbox
\CD@lI}\setbox\CD@lF=\hbox{\unhbox\CD@lF\global\setbox\CD@DA=\lastbox}\ht
\CD@MH\ht\CD@DA\dp\CD@MH\dp\CD@DA\advance\CD@mF\wd\CD@DA\advance\CD@mI\wd
\CD@lI}\CD@tG\ifPositiveGradient\CD@CI\CD@BI\CD@CI\CD@tG\ifClimbing\CD@rB
\CD@qB\CD@rB\newcount\DiagonalChoice\DiagonalChoice\m@ne\ifx\tenln\nullfont
\CD@tJ\def\CD@qF{\CD@KH\ifPositiveGradient/\else\CD@k\backslash\CD@ND\fi}%
\else\def\CD@qF{\CD@rF\char\count@}\fi\let\CD@rF\tenln\def\Use@line@char#1{%
\hbox{#1\CD@rF\ifPositiveGradient\else\advance\count@64 \fi\char\count@}}\def
\CD@cF{\Use@line@char{\count@\CD@TC\multiply\count@8\advance\count@-9\advance
\count@\CD@LH}}\def\CD@ZF{\Use@line@char{\ifcase\DiagonalChoice\CD@gF\or
\CD@fF\or\CD@fF\else\CD@gF\fi}}\def\CD@gF{\ifnum\CD@TC=\z@\count@'33 \else
\count@\CD@TC\multiply\count@\sixt@@n\advance\count@-9\advance\count@\CD@LH
\advance\count@\CD@LH\fi}\def\CD@fF{\count@'\ifcase\CD@LH55\or\ifcase\CD@TC66%
\or22\or52\or61\or72\fi\or\ifcase\CD@TC66\or25\or22\or63\or52\fi\or\ifcase
\CD@TC66\or16\or36\or22\or76\fi\or\ifcase\CD@TC66\or27\or25\or67\or22\fi\fi
\relax}\def\CD@uC#1{\hbox{#1\setbox0=\Use@line@char{#1}\ifPositiveGradient
\else\raise.3\ht0\fi\copy0 \kern-.7\wd0 \ifPositiveGradient\raise.3\ht0\fi
\box0}}\def\CD@jF#1{\hbox{\setbox0=#1\kern-.75\wd0 \vbox to.25\ht0{%
\ifPositiveGradient\else\vss\fi\box0 \ifPositiveGradient\vss\fi}}}\def\CD@jI#%
1{\hbox{\setbox0=#1\dimen0=\wd0 \vbox to.25\ht0{\ifPositiveGradient\vss\fi
\box0 \ifPositiveGradient\else\vss\fi}\kern-.75\dimen0 }}\CD@RC{+h:>}{%
\Use@line@char\CD@fF}\CD@RC{-h:>}{\Use@line@char\CD@gF}\CD@nF{+t:<}{-h:>}%
\CD@nF{-t:<}{+h:>}\CD@RC{+t:>}{\CD@jF{\Use@line@char\CD@fF}}\CD@RC{-t:>}{%
\CD@jI{\Use@line@char\CD@gF}}\CD@nF{+h:<}{-t:>}\CD@nF{-h:<}{+t:>}\CD@RC{+h:>>%
}{\CD@uC\CD@fF}\CD@RC{-h:>>}{\CD@uC\CD@gF}\CD@nF{+t:<<}{-h:>>}\CD@nF{-t:<<}{+%
h:>>}\CD@nF{+h:>->}{+h:>>}\CD@nF{-h:>->}{-h:>>}\CD@nF{+t:<-<}{-h:>>}\CD@nF{-t%
:<-<}{+h:>>}\CD@RC{+t:>>}{\CD@jF{\CD@uC\CD@fF}}\CD@RC{-t:>>}{\CD@jI{\CD@uC
\CD@gF}}\CD@nF{+h:<<}{-t:>>}\CD@nF{-h:<<}{+t:>>}\CD@nF{+t:>->}{+t:>>}\CD@nF{-%
t:>->}{-t:>>}\CD@nF{+h:<-<}{-t:>>}\CD@nF{-h:<-<}{+t:>>}\CD@RC{+f:-}{\CD@EF
\null\else\CD@cF\fi}\CD@nF{-f:-}{+f:-}\def\CD@tC#1#2{\vbox to#1{\vss\hbox to#%
2{\hss.\hss}\vss}}\def\hfdot{\CD@tC{2\axisheight}{.5em}}%
\def\vfdot{\CD@tC{1ex}\z@}
\def\CD@bF{\hbox{\dimen0=.3\CD@zC\dimen1\dimen0 \ifnum\CD@LH>\CD@TC\CD@iC{%
\dimen1}\else\CD@dG{\dimen0}\fi\CD@tC{\dimen0}{\dimen1}}}\newarrowfiller{.}%
\hfdot\hfdot\vfdot\vfdot\def\dfdot{\CD@bF\CD@CK}\CD@RC{+f:.}{\dfdot}\CD@RC{-f%
:.}{\dfdot}\def\CD@@K#1{\hbox\bgroup\def\CD@CH{#1\egroup}\afterassignment
\CD@CH
\count@='}\def\lnchar{\CD@@K\CD@qF}\def\CD@dF#1{\setbox#1=\hbox{\dimen5\dimen
#1 \setbox8=\box#1 \dimen1\wd8 \count@\dimen5 \divide\count@\dimen1 \ifnum
\count@=0 \box8 \ifdim\dimen5<.95\dimen1 \CD@gB{diagonal line too short}\fi
\else\dimen3=\dimen5 \advance\dimen3-\dimen1 \divide\dimen3\count@\dimen4%
\dimen3 \CD@dG{\dimen4}\ifPositiveGradient\multiply\dimen4\m@ne\fi\dimen6%
\dimen1 \advance\dimen6-\dimen3 \loop\raise\count@\dimen4\copy8 \ifnum\count@
>0 \kern-\dimen6 \advance\count@\m@ne\repeat\fi}}\def\CD@CG#1{\CD@EF\CD@xJ{#1%
}\else\CD@dF{#1}\fi}\def\CD@IH#1{}\newdimen\objectheight\objectheight1.8ex
\newdimen\objectwidth\objectwidth1em \def\CD@YD{\dimen6=\CD@aK
\DiagramCellHeight\dimen7=\CD@WK\DiagramCellWidth\CD@KJ\ifnum\CD@LH>0 \ifnum
\CD@TC>0 \CD@aF\else\aftergroup\CD@VC\fi\else\aftergroup\CD@UC\fi}\def\CD@VC{%
\CD@YA{diagonal map is nearly vertical}\CD@NA}\def\CD@UC{\CD@YA{diagonal map
is nearly horizontal}\CD@NA}\CD@rG\CD@NA{Use an orthogonal map instead}\def
\CD@aF{\CD@MJ\dimen3\dimen7\dimen7\dimen6\CD@iC{\dimen7}\advance\dimen3-%
\dimen7 \CD@MF\ifnum\CD@LH>\CD@TC\advance\dimen6-\dimen1\advance\dimen6-%
\dimen5 \CD@iC{\dimen1}\CD@iC{\dimen5}\else\dimen0\dimen1\advance\dimen0%
\dimen5\CD@dG{\dimen0}\advance\dimen6-\dimen0 \fi\dimen2.5\dimen7\advance
\dimen2-\dimen1 \dimen4.5\dimen7\advance\dimen4-\dimen5 \ifPositiveGradient
\dimen0\dimen5 \advance\dimen1-\CD@WK\DiagramCellWidth\advance\dimen1 \CD@ZK
\DiagramCellWidth\setbox6=\llap{\unhbox6\kern.1\ht2}\setbox7=\rlap{\kern.1\ht
2\unhbox7}\else\dimen0\dimen1 \advance\dimen1-\CD@ZK\DiagramCellWidth\setbox7%
=\llap{\unhbox7\kern.1\ht2}\setbox6=\rlap{\kern.1\ht2\unhbox6}\fi\setbox6=%
\vbox{\box6\kern.1\wd2}\setbox7=\vtop{\kern.1\wd2\box7}\CD@dG{\dimen0}%
\advance\dimen0-\axisheight\advance\dimen0-\CD@bK\DiagramCellHeight\dimen5-%
\dimen0 \advance\dimen0\dimen6 \advance\dimen1.5\dimen3 \ifdim\wd3>\z@\ifdim
\ht3>-\dp3\CD@TB\fi\fi\dimen3\dimen2 \dimen7\dimen2\advance\dimen7\dimen4
\ifvoid3 \else\CD@tE\else\dimen8\ht3\advance\dimen8-\axisheight\CD@iC{\dimen8%
}\CD@X{\dimen8}{.5\wd3}\dimen9\dp3\advance\dimen9\axisheight\CD@iC{\dimen9}%
\CD@X{\dimen9}{.5\wd3}\ifPositiveGradient\advance\dimen2-\dimen9\advance
\dimen4-\dimen8 \else\advance\dimen4-\dimen9\advance\dimen2-\dimen8 \fi\fi
\advance\dimen3-.5\wd3 \fi\dimen9=\CD@aK\DiagramCellHeight\advance\dimen9-2%
\DiagramCellHeight\CD@tE\advance\dimen2\dimen4 \CD@CG{2}\dimen2-\dimen0%
\advance\dimen2\dp2 \else\CD@CG{2}\CD@CG{4}\ifPositiveGradient\dimen2-\dimen0%
\advance\dimen2\dp2 \dimen4\dimen5\advance\dimen4-\ht4 \else\dimen4-\dimen0%
\advance\dimen4\dp4 \dimen2\dimen5\advance\dimen2-\ht2 \fi\fi\setbox0=\hbox to%
\z@{\kern\dimen1 \ifvoid1 \else\ifPositiveGradient\advance\dimen0-\dp1 \lower
\dimen0 \else\advance\dimen5-\ht1 \raise\dimen5 \fi\rlap{\unhbox1}\fi\raise
\dimen2\rlap{\unhbox2}\ifvoid3 \else\lower.5\dimen9\rlap{\kern\dimen3\unhbox3%
}\fi\kern.5\dimen7 \lower.5\dimen9\box6 \lower.5\dimen9\box7 \kern.5\dimen7
\CD@tE\else\raise\dimen4\llap{\unhbox4}\fi\ifvoid5 \else\ifPositiveGradient
\advance\dimen5-\ht5 \raise\dimen5 \else\advance\dimen0-\dp5 \lower\dimen0 \fi
\llap{\unhbox5}\fi\hss}\ht0=\axisheight\dp0=-\ht0\box0 }\def\NorthWest{\CD@BI
\CD@rB\DiagonalChoice0 }\def\NorthEast{\CD@CI\CD@rB\DiagonalChoice1 }\def
\SouthWest{\CD@CI\CD@qB\DiagonalChoice3 }\def\SouthEast{\CD@BI\CD@qB
\DiagonalChoice2 }\def\CD@aD{\vadjust{\CD@uA\CD@FA\advance\CD@uA
\ifPositiveGradient\else-\fi\CD@ZK\relax\CD@vA\CD@NB\advance\CD@vA-\CD@bK
\relax\hbox{\advance\CD@uA\ifPositiveGradient-\fi\CD@WK\advance\CD@vA\CD@aK
\hbox{\box6 \kern\CD@DC\kern\CD@eJ\penalty1 \box7 \box\z@}\penalty\CD@uA
\penalty\CD@vA}\penalty\CD@uA\penalty\CD@vA\penalty104}}\def\CD@eH#1{\relax
\vadjust{\hbox@maths{#1}\penalty\CD@FA\penalty\CD@NB\penalty\tw@}}\def\CD@lB{%
\ifcase\CD@GB\or\or\CD@bH{.5\wd0}{\box0}{.5\wd0}\z@\or\unhbox\z@\setbox\z@
\lastbox\CD@bH{.5\wd0}{\box0}{.5\wd0}\z@\unpenalty\unpenalty\setbox\z@
\lastbox\or\CD@TG\else\advance\CD@GB-100 \ifnum\CD@GB<\z@\cd@shouldnt B\fi
\setbox\z@\hbox{\kern\CD@mF\copy\CD@MH\kern\CD@mI\CD@uA\CD@XB\advance\CD@uA-%
\CD@NB\penalty\CD@uA\CD@uA\CD@FA\advance\CD@uA-\CD@lA\penalty\CD@uA\unhbox\z@
\global\CD@yA\lastpenalty\unpenalty\global\CD@zA\lastpenalty\unpenalty}\CD@uA
-\CD@yA\CD@vA\CD@zA\CD@fI\fi}\def\CD@TG{\unhbox\z@\setbox\z@\lastbox\CD@uA
\lastpenalty\unpenalty\advance\CD@uA\CD@mA\CD@vA\CD@XB\advance\CD@vA-%
\lastpenalty\unpenalty\dimen1\lastkern\unkern\setbox3\lastbox\dimen0\lastkern
\unkern\setbox0=\hbox to\z@{\unhbox0\setbox0\lastbox\setbox7\lastbox
\unpenalty\CD@eJ\lastkern\unkern\CD@DC\lastkern\unkern\setbox6\lastbox\dimen7%
\CD@tB\advance\dimen7-\wd\CD@uA\ifdim\dimen7<\z@\CD@CI\multiply\dimen7\m@ne
\let\mv\empty\else\CD@BI\def\mv{\raise\ht1}\kern-\dimen7 \fi\ifnum\CD@vA>%
\CD@NB\dimen6\CD@uB\advance\dimen6-\ht\CD@vA\else\dimen6\z@\fi\CD@jJ\CD@mK
\setbox1\null\ht1\dimen6\wd1\dimen7 \dimen7\dimen2 \dimen6\wd1 \CD@KJ\CD@uA
\CD@LH\CD@vA\CD@TC\dimen6\ht1 \CD@KJ\setbox2\null\divide\dimen2\tw@\advance
\dimen2\CD@eJ\CD@eG{\dimen2}\wd2\dimen2 \dimen0.5\dimen7 \advance\dimen0%
\ifPositiveGradient\else-\fi\CD@eJ\CD@dG{\dimen0}\advance\dimen0-\axisheight
\ht2\dimen0 \dimen0\CD@DC\CD@eG{\dimen0}\advance\dimen0\ht2\ht2\dimen0 \dimen
0\ifPositiveGradient-\fi\CD@DC\CD@dG{\dimen0}\advance\dimen0\wd2\wd2\dimen0
\setbox4\null\dimen0 .6\CD@zC\CD@eG{\dimen0}\ht4\dimen0 \dimen0 .2\CD@zC
\CD@dG{\dimen0}\wd4\dimen0 \dimen0\wd2 \ifvoid6\else\dimen1\ht4 \advance
\dimen1\ht2 \CD@CC6+-\raise\dimen1\rlap{\ifPositiveGradient\advance\dimen0-%
\wd6\advance\dimen0-\wd4 \else\advance\dimen0\wd4 \fi\kern\dimen0\box6}\fi
\dimen0\wd2 \ifvoid7\else\dimen1\ht4 \advance\dimen1-\ht2 \CD@CC7-+\lower
\dimen1\rlap{\ifPositiveGradient\advance\dimen0\wd4 \else\advance\dimen0-\wd7%
\advance\dimen0-\wd4 \fi\kern\dimen0\box7}\fi\mv\box0\hss}\ht0\z@\dp0\z@
\CD@bH{\z@}{\box\z@}{\z@}{\axisheight}}\def\CD@CC#1#2#3{\dimen4.5\wd#1 \ifdim
\dimen4>.25\dimen7\dimen4=.25\dimen7\fi\ifdim\dimen4>\CD@zC\dimen4.4\dimen4
\advance\dimen4.6\CD@zC\fi\CD@eG{\dimen4}\dimen5\axisheight\CD@dG{\dimen5}%
\advance\dimen4-\dimen5 \dimen5\dimen4\CD@eG{\dimen5}\advance\dimen0%
\ifPositiveGradient#2\else#3\fi\dimen5 \CD@dG{\dimen4}\advance\dimen1\dimen4 }
\def\CD@eD#1{\expandafter\CD@IK{#1}}\CD@ZA\CD@EK{output is PostScript
dependent}\def\CD@SC{\CD@IK{/bturn {gsave currentpoint currentpoint translate
4 2 roll neg exch atan rotate neg exch neg exch translate } def /eturn {%
currentpoint grestore moveto} def}}\def\CD@gK{\relax\CD@hK\CD@tK{Q}\else
\CD@IK{eturn}\fi} \def\CD@OJ#1{\count@#1\relax\multiply\count@7\advance
\count@16577\divide\count@33154 }\def\CD@fD#1{\expandafter\special{#1}} \def
\CD@xJ#1{\setbox#1=\hbox{\dimen0\dimen#1\CD@dG{\dimen0}\CD@OJ{\dimen0}\setbox
0=\null\ifPositiveGradient\count@-\count@\ht0\dimen0 \else\dp0\dimen0 \fi\box
0 \CD@uA\count@\CD@OJ\CD@LF\CD@fD{pn \the\count@}\CD@fD{pa 0 0}\CD@OJ{\dimen#%
1}\CD@fD{pa \the\count@\space\the\CD@uA}\CD@fD{fp}\kern\dimen#1}}\def\CD@JI{%
\CD@KJ\begingroup\ifdim\dimen7<\dimen6 \dimen2=\dimen6 \dimen6=\dimen7 \dimen
7=\dimen2 \count@\CD@LH\CD@LH\CD@TC\CD@TC\count@\else\dimen2=\dimen7 \fi
\ifdim\dimen6>.01\p@\CD@KI\global\CD@QA\dimen0 \else\global\CD@QA\dimen7 \fi
\endgroup\dimen2\CD@QA\CD@iK\CD@lK{\ifPositiveGradient\else-\fi\dimen6}\CD@iK
\CD@kK{\ifPositiveGradient-\fi\dimen6}\CD@iK\CD@eK{\dimen7}}\def\CD@KI{\CD@hJ
\ifdim\dimen7>1.73\dimen6 \divide\dimen2 4 \multiply\CD@TC2 \else\dimen2=0.%
353553\dimen2 \advance\CD@LH-\CD@TC\multiply\CD@TC4 \fi\dimen0=4\dimen2 \CD@ZG
4\CD@ZG{-2}\CD@ZG2\CD@ZG{-2.5}}\def\CD@AI{\begingroup\count@\dimen0 \dimen2 45%
pt \divide\count@\dimen2 \ifdim\dimen0<\z@\advance\count@\m@ne\fi\ifodd
\count@\advance\count@1\CD@@A\else\CD@y\fi\advance\dimen0-\count@\dimen2
\CD@gE\multiply\dimen0\m@ne\fi\ifnum\count@<0 \multiply\count@-7 \fi\dimen3%
\dimen1 \dimen6\dimen0 \dimen7 3754936sp \ifdim\dimen0<6\p@\def\CD@OG{4000}%
\fi\CD@KJ\dimen2\dimen3\CD@dG{\dimen2}\CD@hJ\multiply\CD@TC-6 \dimen0\dimen2
\CD@ZG1\CD@ZG{0.3}\dimen1\dimen0 \dimen2\dimen3 \dimen0\dimen3 \CD@ZG3\CD@ZG{%
1.5}\CD@ZG{0.3}\divide\count@2 \CD@gE\multiply\dimen1\m@ne\fi\ifodd\count@
\dimen2\dimen1\dimen1\dimen0\dimen0-\dimen2 \fi\divide\count@2 \ifodd\count@
\multiply\dimen0\m@ne\multiply\dimen1\m@ne\fi\global\CD@QA\dimen0\global
\CD@RA\dimen1\endgroup\dimen6\CD@QA\dimen7\CD@RA}\def\CD@OC{255}\let\CD@OG
\CD@OC\def\CD@KJ{\begingroup\ifdim\dimen7<\dimen6 \dimen9\dimen7\dimen7\dimen
6\dimen6\dimen9\CD@@A\else\CD@y\fi\dimen2\z@\dimen3\CD@XH\dimen4\CD@XH\dimen0%
\z@\dimen8=\CD@OG\CD@XH\CD@lC\global\CD@yA\dimen\CD@gE0\else3\fi\global\CD@zA
\dimen\CD@gE3\else0\fi\endgroup\CD@LH\CD@yA\CD@TC\CD@zA}\def\CD@lC{\count@
\dimen6 \divide\count@\dimen7 \advance\dimen6-\count@\dimen7 \dimen9\dimen4
\advance\dimen9\count@\dimen0 \ifdim\dimen9>\dimen8 \CD@@C\else\CD@AC\ifdim
\dimen6>\z@\dimen9\dimen6 \dimen6\dimen7 \dimen7\dimen9 \expandafter
\expandafter\expandafter\CD@lC\fi\fi}\def\CD@@C{\ifdim\dimen0=\z@\ifdim\dimen
9<2\dimen8 \dimen0\dimen8 \fi\else\advance\dimen8-\dimen4 \divide\dimen8%
\dimen0 \ifdim\count@\CD@XH<2\dimen8 \count@\dimen8 \dimen9\dimen4 \advance
\dimen9\count@\dimen0 \CD@AC\fi\fi}\def\CD@AC{\dimen4\dimen0 \dimen0\dimen9
\advance\dimen2\count@\dimen3 \dimen9\dimen2 \dimen2\dimen3 \dimen3\dimen9 }%
\def\CD@ZG#1{\CD@dG{\dimen2}\advance\dimen0 #1\dimen2 }\def\CD@dG#1{\divide#1%
\CD@TC\multiply#1\CD@LH}\def\CD@eG#1{\divide#1\CD@vA\multiply#1\CD@uA}\def
\CD@iC#1{\divide#1\CD@LH\multiply#1\CD@TC}\def\CD@hJ{\dimen6\CD@LH\CD@XH
\multiply\dimen6\CD@LH\dimen7\CD@TC\CD@XH\multiply\dimen7\CD@TC\CD@KJ}\def
\CD@iK#1#2{\begingroup\dimen@#2\relax\loop\ifdim\dimen2<.4\maxdimen\multiply
\dimen2\tw@\multiply\dimen@\tw@\repeat\divide\dimen2\@cclvi\divide\dimen@
\dimen2\relax\multiply\dimen@\@cclvi\expandafter\CD@jK\the\dimen@\endgroup
\let#1\CD@fK}{\catcode`p=12 \catcode`0=12 \catcode`.=12 \catcode`t=12 \gdef
\CD@jK#1pt{\gdef\CD@fK{#1}}}\ifx\errorcontextlines\CD@qK\CD@tJ\let\CD@GH
\relax\else\def\CD@GH{\errorcontextlines\m@ne}\fi\ifnum\inputlineno<0 \let
\CD@CD\empty\let\CD@W\empty\let\CD@mD\relax\let\CD@uI\relax\let\CD@vI\relax
\let\CD@zF\relax\else\def\CD@W{ at line \number\inputlineno}\def\CD@mD{ - first occurred%
}\def\CD@uI{\edef\CD@h{\the\inputlineno}\global\let\CD@jB\CD@h}\def\CD@h{9999%
}\def\CD@vI{\xdef\CD@jB{\the\inputlineno}}\def\CD@jB{\CD@h}\def\CD@zF{\ifnum
\CD@h<\inputlineno\edef\CD@CD{\space at lines \CD@h--\the\inputlineno}\else
\edef\CD@CD{\CD@W}\fi}\fi\let\CD@CD\empty\def\CD@YA#1#2{\CD@GH\errhelp=#2%
\expandafter\errmessage{\CD@tA: #1}}\def\CD@KB#1{\begingroup\expandafter
\message{! \CD@tA: #1\CD@CD}\ifnum\CD@XB>\CD@NB\ifnum\CD@CA>\CD@NB\else\ifnum
\CD@lA>\CD@FA\else\ifnum\CD@LB>\CD@FA\advance\CD@XB-\CD@NB\advance\CD@FA-%
\CD@lA\advance\CD@FA1\relax\expandafter\message{! (error detected at row \the
\CD@XB, column \the\CD@FA, but probably caused elsewhere)}\fi\fi\fi\fi
\endgroup}\def\CD@gB#1{{\expandafter\message{\CD@tA\space Warning: #1\CD@W}}}%
\def\CD@CB#1#2{\CD@gB{#1 \string#2 is obsolete\CD@mD}}\def\CD@AB#1{\CD@CB{%
Dimension}{#1}\CD@DE#1\CD@BB\CD@BB}\def\CD@BB{\CD@OA=}\def\CD@@B#1{\CD@CB{%
Count}{#1}\CD@DE#1\CD@OH\CD@OH}\def\CD@OH{\count@=}\def\HorizontalMapLength{%
\CD@AB\HorizontalMapLength}\def\VerticalMapHeight{\CD@AB\VerticalMapHeight}%
\def\VerticalMapDepth{\CD@AB\VerticalMapDepth}\def\VerticalMapExtraHeight{%
\CD@AB\VerticalMapExtraHeight}\def\VerticalMapExtraDepth{\CD@AB
\VerticalMapExtraDepth}\def\DiagonalLineSegments{\CD@@B\DiagonalLineSegments}%
\ifx\tenln\nullfont\CD@ZA\CD@KH{\CD@eF\space diagonal line and arrow font not
available}\else\let\CD@KH\relax\fi\def\CD@aG#1#2<#3:#4:#5#6{\begingroup\CD@PA
#3\relax\advance\CD@PA-#2\relax\ifdim.1em<\CD@PA\CD@uA#5\relax\CD@vA#6\relax
\ifnum\CD@uA<\CD@vA\count@\CD@vA\advance\count@-\CD@uA\CD@KB{#4 by \the\CD@PA
}\if#1v\let\CD@CH\CD@JK\edef\tmp{\the\CD@uA--\the\CD@vA,\the\CD@FA}\else
\advance\count@\count@\if#1l\advance\count@-\CD@A\else\if#1r\advance\count@
\CD@A\fi\fi\advance\CD@PA\CD@PA\let\CD@CH\CD@ZE\edef\tmp{\the\CD@NB,\the
\CD@uA--\the\CD@vA}\fi\divide\CD@PA\count@\ifdim\CD@CH<\CD@PA\global\CD@CH
\CD@PA\fi\fi\fi\endgroup}\CD@tG\CD@xE\CD@JD\CD@ID\CD@rG\CD@xI{See the message
above.}\CD@rG\CD@lH{Perhaps you've forgotten to end the diagram before
resuming the text, in\CD@uG which case some garbage may be added to the
diagram, but we should be ok now.\CD@uG Alternatively you've left a blank line
in the middle - TeX will now complain\CD@uG that the remaining \CD@S s are
misplaced - so please use comments for layout.}\CD@rG\CD@hD{You have already
closed too many brace pairs or environments; an \CD@HD\CD@uG command was (%
over)due.}\CD@rG\CD@hH{\CD@dC\space and \CD@HD\space commands must match.}%
\def\CD@jH{\ifnum\inputlineno=0 \else\expandafter\CD@iH\fi}\def\CD@iH{\CD@MD
\CD@GD\crcr\CD@YA{missing \CD@HD\space inserted before \CD@kH- type "h"}%
\CD@lH\enddiagram\CD@AG\CD@kH\par}\def\CD@AG#1{\edef\enddiagram{\noexpand
\CD@rD{#1\CD@W}}}\def\CD@rD#1{\CD@YA{\CD@HD\space(anticipated by #1) ignored}%
\CD@xI\let\enddiagram\CD@SG}\def\CD@SG{\CD@YA{misplaced \CD@HD\space ignored}%
\CD@hH}\def\CD@mC{\CD@YA{missing \CD@HD\space inserted.}\CD@hD\CD@AG{closing
group}}\ifx\ifx
\fi\fi

\catcode`\$=3 
\def\vboxtoz{\vbox to\z@}

\def\scriptaxis#1{\@scriptaxis{$\scriptstyle#1$}}
\def\ssaxis#1{\@ssaxis{$\scriptscriptstyle#1$}}
\def\@scriptaxis#1{\dimen0\axisheight\advance\dimen0-\ss@axisheight\raise
\dimen0\hbox{#1}}\def\@ssaxis#1{\dimen0\axisheight\advance\dimen0-%
\ss@axisheight\raise\dimen0\hbox{#1}}

\ifx\boldmath\CD@qK
\let\boldscriptaxis\scriptaxis
\def\boldscript#1{\hbox{$\scriptstyle#1$}}
\else\def\boldscriptaxis#1{\@scriptaxis{\boldmath$\scriptstyle#1$}}
\def\boldscript#1{\hbox{\boldmath$\scriptstyle#1$}}
\fi

\def\raisehook#1#2#3{\hbox{\setbox3=\hbox{#1$\scriptscriptstyle#3$}%
\dimen0\ss@axisheight
\dimen1\axisheight\advance\dimen1-\dimen0
\dimen2\ht3\advance\dimen2-\dimen0%
\advance\dimen2-0.021em\advance\dimen1 #2\dimen2%
\raise\dimen1\box3}}
\def\shifthook#1#2#3{\setbox1=\hbox{#1$\scriptscriptstyle#3$}\dimen0\wd1%
\divide\dimen0 12\CD@zH{\dimen0}
\dimen1\wd1\advance\dimen1-2\dimen0 \advance\dimen1-2\CD@oI\CD@zH{\dimen1}%
\kern#2\dimen1\box1}

\def\@cmex{\mathchar"03}

\def\make@pbk#1{\setbox\tw@\hbox to\z@{#1}\ht\tw@\z@\dp\tw@\z@\box\tw@}\def
\CD@fH#1{\overprint{\hbox to\z@{#1}}}\def\CD@qH{\kern0.11em}\def\CD@pH{\kern0%
.35em}

\def\dblvert{\def\CD@rH{\kern.5\PileSpacing}}\def\CD@rH{}

\def\SEpbk{\make@pbk{\CD@qH\CD@rH\vrule depth 2.87ex height -2.75ex width 0.%
95em \vrule height -0.66ex depth 2.87ex width 0.05em \hss}}

\def\SWpbk{\make@pbk{\hss\vrule height -0.66ex depth 2.87ex width 0.05em
\vrule depth 2.87ex height -2.75ex width 0.95em \CD@qH\CD@rH}}

\def\NEpbk{\make@pbk{\CD@qH\CD@rH\vrule depth -3.81ex height 4.00ex width 0.%
95em \vrule height 4.00ex depth -1.72ex width 0.05em \hss}}

\def\NWpbk{\make@pbk{\hss\vrule height 4.00ex depth -1.72ex width 0.05em
\vrule depth -3.81ex height 4.00ex width 0.95em \CD@qH\CD@rH}}

\def\puncture{{\setbox0\hbox{A}\vrule height.53\ht0 depth-.47\ht0 width.35\ht
0 \kern.12\ht0 \vrule height\ht0 depth-.65\ht0 width.06\ht0 \kern-.06\ht0
\vrule height.35\ht0 depth0pt width.06\ht0 \kern.12\ht0 \vrule height.53\ht0
depth-.47\ht0 width.35\ht0 }}

\def\NEclck{\overprint{\raise2.5ex\rlap{ \CD@rH$\scriptstyle\searrow$}}}
\def\NEanti{\overprint{\raise2.5ex\rlap{ \CD@rH$\scriptstyle\nwarrow$}}}
\def\NWclck{\overprint{\raise2.5ex\llap{$\scriptstyle\nearrow$ \CD@rH}}}
\def\NWanti{\overprint{\raise2.5ex\llap{$\scriptstyle\swarrow$ \CD@rH}}}
\def\SEclck{\overprint{\lower1ex\rlap{ \CD@rH$\scriptstyle\swarrow$}}}
\def\SEanti{\overprint{\lower1ex\rlap{ \CD@rH$\scriptstyle\nearrow$}}}
\def\SWclck{\overprint{\lower1ex\llap{$\scriptstyle\nwarrow$ \CD@rH}}}
\def\SWanti{\overprint{\lower1ex\llap{$\scriptstyle\searrow$ \CD@rH}}}

\def\rhvee{\mkern-10mu\greaterthan}
\def\lhvee{\lessthan\mkern-10mu}
\def\dhvee{\vboxtoz{\vss\hbox{$\vee$}\kern0pt}}
\def\uhvee{\vboxtoz{\hbox{$\wedge$}\vss}}
\newarrowhead{vee}\rhvee\lhvee\dhvee\uhvee

\def\dhlvee{\vboxtoz{\vss\hbox{$\scriptstyle\vee$}\kern0pt}}
\def\uhlvee{\vboxtoz{\hbox{$\scriptstyle\wedge$}\vss}}
\newarrowhead{littlevee}{\mkern1mu\scriptaxis\rhvee}{\scriptaxis\lhvee}%
\dhlvee\uhlvee\ifx\boldmath\CD@qK
\newarrowhead{boldlittlevee}{\mkern1mu\scriptaxis\rhvee}{\scriptaxis\lhvee}%
\dhlvee\uhlvee\else
\def\dhblvee{\vboxtoz{\vss\boldscript\vee\kern0pt}}
\def\uhblvee{\vboxtoz{\boldscript\wedge\vss}}
\newarrowhead{boldlittlevee}{\mkern1mu\boldscriptaxis\rhvee}{\boldscriptaxis
\lhvee}\dhblvee\uhblvee
\fi

\def\rhcvee{\mkern-10mu\succ}
\def\lhcvee{\prec\mkern-10mu}
\def\dhcvee{\vboxtoz{\vss\hbox{$\curlyvee$}\kern0pt}}
\def\uhcvee{\vboxtoz{\hbox{$\curlywedge$}\vss}}
\newarrowhead{curlyvee}\rhcvee\lhcvee\dhcvee\uhcvee

\def\rhvvee{\mkern-13mu\gg}
\def\lhvvee{\ll\mkern-13mu}
\def\dhvvee{\vboxtoz{\vss\hbox{$\vee$}\kern-.6ex\hbox{$\vee$}\kern0pt}}
\def\uhvvee{\vboxtoz{\hbox{$\wedge$}\kern-.6ex \hbox{$\wedge$}\vss}}
\newarrowhead{doublevee}\rhvvee\lhvvee\dhvvee\uhvvee

\def\rhtriangle{\triangleright\mkern1.2mu}
\def\lhtriangle{\triangleleft\mkern.8mu}
\def\uhtriangle{\vbox{\kern-.2ex \hbox{$\scriptscriptstyle\bigtriangleup$}%
\kern-.25ex}}
\def\dhtriangle{\vbox{\kern-.28ex \hbox{$\scriptscriptstyle\bigtriangledown$}%
\kern-.1ex}}
\def\dhblack{\vbox{\kern-.25ex\nointerlineskip\hbox{$\blacktriangledown$}}}%
\def\uhblack{\vbox{\kern-.25ex\nointerlineskip\hbox{$\blacktriangle$}}}%
\def\dhlblack{\vbox{\kern-.25ex\nointerlineskip\hbox{$\scriptstyle
\blacktriangledown$}}}
\def\uhlblack{\vbox{\kern-.25ex\nointerlineskip\hbox{$\scriptstyle
\blacktriangle$}}}
\newarrowhead{triangle}\rhtriangle\lhtriangle\dhtriangle\uhtriangle
\newarrowhead{blacktriangle}{\mkern-1mu\blacktriangleright\mkern.4mu}{%
\blacktriangleleft}\dhblack\uhblack\newarrowhead{littleblack}{\mkern-1mu%
\scriptaxis\blacktriangleright}{\scriptaxis\blacktriangleleft\mkern-2mu}%
\dhlblack\uhlblack

\def\rhla{\hbox{\setbox0=\lnchar55\dimen0=\wd0\kern-.6\dimen0\ht0\z@\raise
\axisheight\box0\kern.1\dimen0}}
\def\lhla{\hbox{\setbox0=\lnchar33\dimen0=\wd0\kern.05\dimen0\ht0\z@\raise
\axisheight\box0\kern-.5\dimen0}}
\def\dhla{\vboxtoz{\vss\rlap{\lnchar77}}}
\def\uhla{\vboxtoz{\setbox0=\lnchar66 \wd0\z@\kern-.15\ht0\box0\vss}}
\newarrowhead{LaTeX}\rhla\lhla\dhla\uhla

\def\lhlala{\lhla\kern.3em\lhla}
\def\rhlala{\rhla\kern.3em\rhla}
\def\uhlala{\hbox{\uhla\raise-.6ex\uhla}}
\def\dhlala{\hbox{\dhla\lower-.6ex\dhla}}
\newarrowhead{doubleLaTeX}\rhlala\lhlala\dhlala\uhlala

\def\hhO{\scriptaxis\bigcirc\mkern.4mu} \def\hho{{\circ}\mkern1.2mu}%
\newarrowhead{o}\hho\hho\circ\circ
\newarrowhead{O}\hhO\hhO{\scriptstyle\bigcirc}{\scriptstyle\bigcirc}

\def\rhtimes{\mkern-5mu{\times}\mkern-.8mu}\def\lhtimes{\mkern-.8mu{\times}%
\mkern-5mu}\def\uhtimes{\setbox0=\hbox{$\times$}\ht0\axisheight\dp0-\ht0%
\lower\ht0\box0 }\def\dhtimes{\setbox0=\hbox{$\times$}\ht0\axisheight\box0 }%
\newarrowhead{X}\rhtimes\lhtimes\dhtimes\uhtimes\newarrowhead+++++

\newarrowhead{Y}{\mkern-3mu\Yright}{\Yleft\mkern-3mu}\Ydown\Yup

\newarrowhead{->}\rightarrow\leftarrow\downarrow\uparrow

\newarrowhead{=>}\Rightarrow\Leftarrow{\@cmex7F}{\@cmex7E}

\newarrowhead{harpoon}\rightharpoonup\leftharpoonup\downharpoonleft
\upharpoonleft

\def\twoheaddownarrow{\rlap{$\downarrow$}\raise-.5ex\hbox{$\downarrow$}}
\def\twoheaduparrow{\rlap{$\uparrow$}\raise.5ex\hbox{$\uparrow$}}
\newarrowhead{->>}\twoheadrightarrow\twoheadleftarrow\twoheaddownarrow
\twoheaduparrow

\def\ltvee{\mkern-1mu{\lessthan}\mkern.4mu}
\newarrowtail{vee}\greaterthan\ltvee\vee\wedge

\newarrowtail{littlevee}{\scriptaxis\greaterthan}{\mkern-1mu\scriptaxis
\lessthan}{\scriptstyle\vee}{\scriptstyle\wedge}\ifx\boldmath\CD@qK
\newarrowtail{boldlittlevee}{\scriptaxis\greaterthan}{\mkern-1mu\scriptaxis
\lessthan}{\scriptstyle\vee}{\scriptstyle\wedge}\else\newarrowtail{%
boldlittlevee}{\boldscriptaxis\greaterthan}{\mkern-1mu\boldscriptaxis
\lessthan}{\boldscript\vee}{\boldscript\wedge}\fi

\newarrowtail{curlyvee}\succ{\mkern-1mu{\prec}\mkern.4mu}\curlyvee\curlywedge

\def\rttriangle{\mkern1.2mu\triangleright}
\newarrowtail{triangle}\rttriangle\lhtriangle\dhtriangle\uhtriangle
\newarrowtail{blacktriangle}\blacktriangleright{\mkern-1mu\blacktriangleleft
\mkern.4mu}\dhblack\uhblack\newarrowtail{littleblack}{\scriptaxis
\blacktriangleright\mkern-2mu}{\mkern-1mu\scriptaxis\blacktriangleleft}%
\dhlblack\uhlblack

\def\rtla{\hbox{\setbox0=\lnchar55\dimen0=\wd0\kern-.5\dimen0\ht0\z@\raise
\axisheight\box0\kern-.2\dimen0}}
\def\ltla{\hbox{\setbox0=\lnchar33\dimen0=\wd0\kern-.15\dimen0\ht0\z@\raise
\axisheight\box0\kern-.5\dimen0}}
\def\dtla{\vbox{\setbox0=\rlap{\lnchar77}\dimen0=\ht0\kern-.7\dimen0\box0%
\kern-.1\dimen0}}
\def\utla{\vbox{\setbox0=\rlap{\lnchar66}\dimen0=\ht0\kern-.1\dimen0\box0%
\kern-.6\dimen0}}
\newarrowtail{LaTeX}\rtla\ltla\dtla\utla

\def\rtvvee{\gg\mkern-3mu}
\def\ltvvee{\mkern-3mu\ll}
\def\dtvvee{\vbox{\hbox{$\vee$}\kern-.6ex \hbox{$\vee$}\vss}}
\def\utvvee{\vbox{\vss\hbox{$\wedge$}\kern-.6ex \hbox{$\wedge$}\kern\z@}}
\newarrowtail{doublevee}\rtvvee\ltvvee\dtvvee\utvvee

\def\ltlala{\ltla\kern.3em\ltla}
\def\rtlala{\rtla\kern.3em\rtla}
\def\utlala{\hbox{\utla\raise-.6ex\utla}}
\def\dtlala{\hbox{\dtla\lower-.6ex\dtla}}
\newarrowtail{doubleLaTeX}\rtlala\ltlala\dtlala\utlala

\def\utbar{\vrule height 0.093ex depth0pt width 0.4em}
\let\dtbar\utbar
\def\rtbar{\mkern1.5mu\vrule height 1.1ex depth.06ex width .04em\mkern1.5mu}%
\let\ltbar\rtbar
\newarrowtail{mapsto}\rtbar\ltbar\dtbar\utbar
\newarrowtail{|}\rtbar\ltbar\dtbar\utbar

\def\rthooka{\raisehook{}+\subset\mkern-1mu}
\def\lthooka{\mkern-1mu\raisehook{}+\supset}
\def\rthookb{\raisehook{}-\subset\mkern-2mu}
\def\lthookb{\mkern-1mu\raisehook{}-\supset}

\def\dthooka{\shifthook{}+\cap}
\def\dthookb{\shifthook{}-\cap}
\def\uthooka{\shifthook{}+\cup}
\def\uthookb{\shifthook{}-\cup}

\newarrowtail{hooka}\rthooka\lthooka\dthooka\uthooka\newarrowtail{hookb}%
\rthookb\lthookb\dthookb\uthookb

\ifx\boldmath\CD@qK\newarrowtail{boldhooka}\rthooka\lthooka\dthooka\uthooka
\newarrowtail{boldhookb}\rthookb\lthookb\dthookb\uthookb\newarrowtail{%
boldhook}\rthooka\lthooka\dthookb\uthooka\else\def\rtbhooka{\raisehook
\boldmath+\subset\mkern-1mu}
\def\ltbhooka{\mkern-1mu\raisehook\boldmath+\supset}
\def\rtbhookb{\raisehook\boldmath-\subset\mkern-2mu}
\def\ltbhookb{\mkern-1mu\raisehook\boldmath-\supset}
\def\dtbhooka{\shifthook\boldmath+\cap}
\def\dtbhookb{\shifthook\boldmath-\cap}
\def\utbhooka{\shifthook\boldmath+\cup}
\def\utbhookb{\shifthook\boldmath-\cup}
\newarrowtail{boldhooka}\rtbhooka\ltbhooka\dtbhooka\utbhooka\newarrowtail{%
boldhookb}\rtbhookb\ltbhookb\dtbhookb\utbhookb\newarrowtail{boldhook}%
\rtbhooka\ltbhooka\dtbhooka\utbhooka\fi

\def\dtsqhooka{\shifthook{}+\sqcap}
\def\ltsqhooka{\mkern-1mu\raisehook{}+\sqsupset}
\def\rtsqhooka{\raisehook{}+\sqsubset\mkern-1mu}
\def\utsqhooka{\shifthook{}+\sqcup}
\newarrowtail{sqhook}\rtsqhooka\ltsqhooka\dtsqhooka\utsqhooka

\newarrowtail{hook}\rthooka\lthookb\dthooka\uthooka\newarrowtail{C}\rthooka
\lthookb\dthooka\uthooka

\newarrowtail{o}\hho\hho\circ\circ
\newarrowtail{O}\hhO\hhO{\scriptstyle\bigcirc}{\scriptstyle\bigcirc}

\newarrowtail{X}\lhtimes\rhtimes\uhtimes\dhtimes\newarrowtail+++++

\newarrowtail{Y}\Yright\Yleft\Ydown\Yup

\newarrowtail{harpoon}\leftharpoondown\rightharpoondown\upharpoonright
\downharpoonright

\newarrowtail{<=}\Leftarrow\Rightarrow{\@cmex7E}{\@cmex7F}

\newarrowfiller{=}=={\@cmex77}{\@cmex77}
\def\vfthree{\mid\!\!\!\mid\!\!\!\mid}
\newarrowfiller{3}\equiv\equiv\vfthree\vfthree

\def\vfdashstrut{\vrule width0pt height1.3ex depth0.7ex}
\def\vfthedash{\vrule width\CD@LF height0.6ex depth 0pt}
\def\hfthedash{\CD@AJ\vrule\horizhtdp width 0.26em}
\def\hfdash{\mkern5.5mu\hfthedash\mkern5.5mu}
\def\vfdash{\vfdashstrut\vfthedash}
\newarrowfiller{dash}\hfdash\hfdash\vfdash\vfdash

\newarrowmiddle+++++

\iffalse
\newarrow{To}----{vee}
\newarrow{Arr}----{LaTeX}
\newarrow{Dotsto}....{vee}
\newarrow{Dotsarr}....{LaTeX}
\newarrow{Dashto}{}{dash}{}{dash}{vee}
\newarrow{Dasharr}{}{dash}{}{dash}{LaTeX}
\newarrow{Mapsto}{mapsto}---{vee}
\newarrow{Mapsarr}{mapsto}---{LaTeX}
\newarrow{IntoA}{hooka}---{vee}
\newarrow{IntoB}{hookb}---{vee}
\newarrow{Embed}{vee}---{vee}
\newarrow{Emarr}{LaTeX}---{LaTeX}
\newarrow{Onto}----{doublevee}
\newarrow{Dotsonarr}....{doubleLaTeX}
\newarrow{Dotsonto}....{doublevee}
\newarrow{Dotsonarr}....{doubleLaTeX}
\else
\newarrow{To}---->
\newarrow{Arr}---->
\newarrow{Dotsto}....>
\newarrow{Dotsarr}....>
\newarrow{Dashto}{}{dash}{}{dash}>
\newarrow{Dasharr}{}{dash}{}{dash}>
\newarrow{Mapsto}{mapsto}--->
\newarrow{Mapsarr}{mapsto}--->
\newarrow{IntoA}{hooka}--->
\newarrow{IntoB}{hookb}--->
\newarrow{Embed}>--->
\newarrow{Emarr}>--->
\newarrow{Onto}----{>>}
\newarrow{Dotsonarr}....{>>}
\newarrow{Dotsonto}....{>>}
\newarrow{Dotsonarr}....{>>}
\fi

\newarrow{Implies}===={=>}
\newarrow{Project}----{triangle}
\newarrow{Pto}----{harpoon}
\newarrow{Relto}{harpoon}---{harpoon}

\newarrow{Eq}=====
\newarrow{Line}-----
\newarrow{Dots}.....
\newarrow{Dashes}{}{dash}{}{dash}{}

\newarrow{SquareInto}{sqhook}--->

\newarrowhead{cmexbra}{\@cmex7B}{\@cmex7C}{\@cmex3B}{\@cmex38}
\newarrowtail{cmexbra}{\@cmex7A}{\@cmex7D}{\@cmex39}{\@cmex3A}
\newarrowmiddle{cmexbra}{\braceru\bracelu}{\bracerd\braceld}{\vcenter{%
\hbox@maths{\@cmex3D\mkern-2mu}}}
{\vcenter{\hbox@maths{\mkern2mu\@cmex3C}}}
\newarrow{@brace}{cmexbra}-{cmexbra}-{cmexbra}
\newarrow{@parenth}{cmexbra}---{cmexbra}
\def\rightBrace{\d@brace[thick,cmex]}
\def\leftBrace{\u@brace[thick,cmex]}
\def\upperBrace{\r@brace[thick,cmex]}
\def\lowerBrace{\l@brace[thick,cmex]}
\def\rightParenth{\d@parenth[thick,cmex]}
\def\leftParenth{\u@parenth[thick,cmex]}
\def\upperParenth{\r@parenth[thick,cmex]}
\def\lowerParenth{\l@parenth[thick,cmex]}

\let\hEq\rEq
\let\vEq\uEq

\def\labelstyle{
\ifincommdiag
\textstyle
\else
\scriptstyle
\fi}
\let\objectstyle\displaystyle

\newdiagramgrid{pentagon}{0.618034,0.618034,1,1,1,1,0.618034,0.618034}{1.%
17557,1.17557,1.902113,1.902113}

\newdiagramgrid{perspective}{0.75,0.75,1.1,1.1,0.9,0.9,0.95,0.95,0.75,0.75}{0%
.75,0.75,1.1,1.1,0.9,0.9}

\diagramstyle[
dpi=300,
vmiddle,nobalance,
loose,
thin,
pilespacing=10pt,%
shortfall=4pt,
]

\ifx\CD@qK\else\CD@PK\relax\CD@FF\CD@e\fi\fi

\CD@vE\CD@hK\else\fi\else\fi\cdrestoreat
\newcommand{\noi}{\noindent}
\newcommand{\Cur}{\Omega}
\newcommand{\cl}{\mathfrak{C}\ell}

\newcommand{\mr}{\mathring}

\newcommand{\CC}{\mathbb{C}}

\hypersetup{hidelinks,backref=true,pagebackref=true,hyperindex=true,colorlinks=true,breaklinks=true,urlcolor= blue}
\hypersetup{%
  colorlinks = true,
  linkcolor  = blue
}
\definecolor{green1}{RGB}{0,128,0} 
\hypersetup{hidelinks,backref=true,pagebackref=true,hyperindex=true,colorlinks=true,breaklinks=true,urlcolor= blue}
\hypersetup{%
  colorlinks = true,
  linkcolor  = blue,
  citecolor = green1,
}
\usepackage{bookmark}
\usepackage{hyperref,color,xcolor}
\hypersetup{hidelinks,backref=true,pagebackref=true,hyperindex=true,colorlinks=true,breaklinks=true,urlcolor= blue}
\hypersetup{%
  colorlinks = true,
  linkcolor  = blue
}

\newcommand{\beqs}{\begin{equation*}}
\newcommand{\beq}{\begin{equation}}
\newcommand{\eeqs}{\end{equation*}}
\newcommand{\eeq}{\end{equation}}

\newcommand{\beqas}{\begin{eqnarray*}}
\newcommand{\beqa}{\begin{eqnarray}}

\newcommand{\eeqas}{\end{eqnarray*}}
\newcommand{\eeqa}{\end{eqnarray}}

\newcommand{\blist}{\begin{itemize}}

\newcommand{\elist}{\end{itemize}}

\providecommand{\href}[2]{#2}

\newcommand{\Dir}{\slash{\!\!\!\!D}}

\newcommand{\iM}{\int_M {\textrm d}^nx\,\sqrt{g}}

\begin{document}

\title{Hearing the shape of inequivalent spin structures and exotic Dirac operators}

\author{R. da Rocha}
\email{roldao.rocha@ufabc.edu.br}
\affiliation{Center of Mathematics, Federal University of ABC, Santo Andr\'e-S\~ao Paulo, Brazil, 09580-210}
\author{A. A. Tomaz}
\email{anderson.tomaz@ufabc.edu.br}
\affiliation{Center of Mathematics, Federal University of ABC, Santo Andr\'e-S\~ao Paulo, Brazil}
\affiliation{Institute of Physics, Fluminense Federal University, Av. Litoran\^ea, Niter\'oi, Rio de Janeiro, Brazil}

\begin{abstract}
    Exotic spinor fields arise from inequivalent spin structures on non-trivial  topological manifolds, $M$. This induces an additional term in the Dirac operator, defined by the cohomology group $H^1(M,\mathbb{Z}_2)$ that rules a $\check{\textrm{C}}$ech cohomology class. This formalism is extended for manifolds of any finite dimension, endowed with a metric of arbitrary signature. The exotic corrections to heat kernel coefficients, relating spectral properties of exotic Dirac operators to the geometric invariants of $M$, are derived and scrutinized.
\end{abstract}


\maketitle

\section{Introduction}
\label{sec:intro}
Based on the Lounesto's spinor field classification \cite{Lounesto:2001zz,Vaz:2016qyw}, new spinor fields, beyond the well known Dirac, Weyl and Majorana ones, have been proposed, discussed and scrutinized in the last 15 years.  There is a  comprehensive list of prominent papers in the field, among which one can highlight the most influential ones. In the context of supergravity compactifications, Refs. \cite{Bonora:2015ppa,Bonora:2014dfa} shed light on new classes of spinor fields on the $S^7$ sphere. Besides, new spinor classes in superstring theory were derived and analyzed in Ref. \cite{Lopes:2018cvu}. The compactification process in AdS/CFT then induced new spinor classes in the bulk in Ref. \cite{deBrito:2016qzl}. 
In addition,
new spinor classifications have been proposed \cite{Cavalcanti:2014wia,Fabbri:2016msm}, that are reciprocal to the Lounesto's one, also encompassing new gauge theoretical aspects.  Refs. \cite{Fabbri:2011mi,Vignolo:2011qt,Fabbri:2010pk,daRocha:2007sd}
 studied new spinors in a gravity background. Singular spinor fields and their relationship with their appropriate analogue of Lounesto's classification, were also studied in Refs. \cite{daSilvaa:2019kkt,daRocha:2005ti,Rogerio:2019tpu,Rogerio:2019dje,Rogerio:2019xcu,Beghetto:2019dsf,Vaz:2017fac,Lee:2015sqj,Ahluwalia:2009rh,Villalobos:2018emr,Rogerio:2017gvr,HoffdaSilva:2016ffx} (and references therein). 

On non-trivial  topological $M$ manifolds, inequivalent spin structures are categorized by the cohomology group $H^1(M,\mathbb{Z}_2)$  \cite{daRocha:2011yr}.
If $M$ does satisfy all the Geroch's theorem hypotheses \cite{Geroch}, it has at least one spin structure, whose  unicity is ruled by the topology of $M$. In fact, manifolds that are simply connected hold an unique spin structure, since in this case the fundamental group is indeed trivial. This does not happen to manifolds that are multiply connected, which can have many  spin structures, labeled by the cohomology group 
$H^1(M,\mathbb{Z}_2)$.
Inequivalent spin structures yielding distinct spin connections  were more precisely discussed in Refs. \cite{isham1,petry}. 
Ref. \cite{petry} used inequivalent spin structures to describe  Cooper pairs in superconductivity, through exotic spinor fields. 

On multiply connected manifolds, the existence of spin structures, that are not equivalent to the trivial spin structure,  yield Dirac operators with an additional, exotic,  term. Up to the results in Ref. \cite{Beghetto:2018fel}, the exotic term in the Dirac operator was thought to be experimentally unrealizable. In fact, for Dirac and other regular fermionic fields, the exotic term is driven by elements of the cohomology group $H^1(M, \mathbb{Z}_2)$. Hence, modifications of the Dirac operator, containing exotic terms,  were thought to be usually encrypted as a shift of the gauge potential in the corresponding Dirac equation.
 However, Ref. \cite{Beghetto:2018fel} showed that when 4D QFT takes in, the dispersion relation
obtained in this case is different from that one obtained by taking into account the gauge field interaction. Therefore, the exotic term can be also physically realized for mass dimension three-halves fermionic fields in 4D multiply connected spacetimes. Refs. \cite{daRocha:2011yr,Bernardini:2012sc} already shown that mass dimension one spinor fields, in 4D spacetimes, can probe the exotic terms in the Dirac operator ruling the equations of motion of such spinor fields.  Several more applications of inequivalent  spin structures in physics were studied, including  superconductivity and condensed matter. Ref. \cite{daSilva:2016htz} proposed black hole physics as an origin of spacetime exoticness, naturally generating a non-trivial topology. In this context, the Hawking radiation of exotic fermions was computed, showing that exotic terms in the Dirac operator do alter  black hole evaporation rates.
Exotic spinor fields were scrutinized in Refs. \cite{daRocha:2011yr,Bernardini:2012sc,Dantas:2015mfi,petry,avis0,Geroch,isham1,ford,Beghetto:2018fel}, where various physical applications were investigated.

Inequivalent  spin structures and exotic spinor fields, arising from manifolds having non-trivial topologies, can engender deep modifications on  the heat kernel coefficients associated to the manifold. Due to the important status of this research line currently, we want to investigate spectral properties of 
the so called exotic fermionic fields, relating them to the geometric invariants of the manifold that describes the spacetime, where these fermionic fields live in. For it, the heat kernel coefficients and their exotic counterparts will make possible to define a quantity, the exotic deviation coefficient, that can probe the spectral properties of  spacetimes with non-trivial topology.

Heat kernels will be used in this work, where exotic heat kernel coefficients, encompassing the exotic corrections to the Dirac operator in multiply connected manifolds in any dimension, will be derived and discussed. Related to zeta functions, heat kernel coefficients are tools to study path integrals in QFT  as well as processes of diffusion and  partition
functions in statistical mechanics \cite{8,15}.  The very core scheme underlying heat kernel coefficients  is to devise spectral properties of 
bosonic and/or fermionic fields, that propagate on a given spacetime, to the inherent geometric invariants of the manifold that represents the spacetime. The heat kernel can be thought of as being  a particular case of spectral functions, having an intimate relationship to the zeta function \cite{4,8}. 
\textcolor{black}{Ka${\check{\textrm c}}$ argued in Ref. \cite{4} whether two given domains are congruent, once they do have the same sequence of eigenvalues. His celebrated quotation, whether one can one hear the shape of a drum \cite{4}, consists of retrieving the geometric and topological properties of a manifold  from the spectrum of a given differential operator. Calculating the heat kernel coefficients of a differential operator on a manifold 
provides the answer to this relevant question. In this context, the heat kernel coefficients associated to exotic Dirac operators on  non-trivial topological manifolds can be also derived and studied. On the other hand, the heat kernel is also an adequate tool to study the Atiyah--Singer index theorem.  In Ref. \cite{triangle} a simple example, using triangles, elucidated Ka${\check{\textrm c}}$'s posed question. This problem can be extended to compact Riemannian manifolds without or with boundaries \cite{8}. With the exotic Dirac operator, computing the deviations around the standard heat kernel coefficients of trivial manifolds can inform us about the features non-trivial topological manifolds. Hence, it can clarify about the structure of inequivalent generalized spin structures.}

This paper is organized as follows: Sect.~\ref{w2} is devoted to  present the Clifford bundle on manifolds of any finite dimension, equipped with a metric of arbitrary signature, approaching inequivalent spin structures and the exotic Dirac operator. Sect. \ref{hkc0} is dedicated to introduce the heat kernel coefficients and to review the way they can be locally computable with respect to the geometric invariants of M. \textcolor{black}{The exotic heat kernel coefficients and the respective deviations from the standard heat kernel coefficients have their  calculations detailed in Sect.~\ref{hkc}, being also discussed and scrutinized. Our final remarks are displayed in Sect.~\ref{concl}.}

\section{Preliminaries: exotic spin structures}
\label{w2}
 
Inequivalent spin structures on Clifford spinor bundles 
are here reviewed and extended, in order to define the 
complete exotic Dirac operator. 
The 5-tuple $(M, g,\nabla,\uptau_g,\uptau_t)$ denotes the
spacetime structure \cite{moro}, for  $M$ denoting a $n$-dimensional,  compact, paracompact, pseudo-Riemannian
spin manifold; $g$ stands for the spacetime metric, of signature $p-q$, where $p+q=n$; $\nabla$ is the connection associated to $g$; $\uptau_g$ defines a
spacetime orientation, whereas $\uptau_t$ denotes a future-pointing temporal orientation.  $T^{\ast}M$ [$TM$] consists of the
cotangent [tangent] bundle on $M$. The tangent bundle admits the splitting $TM = (TM)^p\oplus (TM)^q$, into timelike and spacelike subbundles.  $\Upomega (M)$ is the exterior bundle, constructed on the tangent bundle on $M$. Let $\upphi_1,\upphi_2, \upphi_3 \in \Upomega (M)$ be differential form fields on $M$. The left contraction can be defined in terms of the metric and the exterior product, denoted by ``$\wedge$'' by ${g}(\upphi_1 \lrcorner \upphi_2 ,\upphi_3)={g}(\upphi_2 ,\tilde{
\upphi_1}\wedge \upphi_3)$, where the tilde denotes the reversion defined on $\Upomega (M)$. Let $F(M)$ be the bundle of frames on $M$ and  $\mathcal{P}_{\mathrm{SO}_{p,q}{}}(M)$ the orthonormal coframe bundle, whereas $\mathcal{P}_{\mathrm{Spin}%
_{p,q}{}}(M\mathbf{)}$ is the spin coframe bundle. The Clifford product a vector field $v\in TM\simeq \Upomega^1 (TM)$ and a form field $\phi\in\Upomega (TM)$ reads ${{v}}\phi ={{v}}\wedge \phi
+{{v}}\lrcorner \phi $. The Grassmann bundle $(\Upomega (M),g)$,  equipped with the Clifford product, is the Clifford bundle  $\mathfrak{C}\ell (M,g)$, with Clifford algebras $\mathfrak{C}\ell _{p,q}$ as sections. 
 The tangent  bundle $TM$ has the orthogonal group O$_{p,q}$ as its structure group.
$F(M)$ can be endowed with a set of (transition) functions $u_{ij} : U_i\cap U_j \to {\textrm O}_{p,q}$, for $\{U_k\}$ representing  open sets in $F(M)$. Besides, functions $f(U_i,U_j) =~ \det u_{ij} = \pm 1$ in a $\check{\textrm C}$ech chain define a so called cocycle, as $u_{ij}\circ u_{jk}\circ u_{ki} = {\textrm id}_{{\textrm O}_{p,q}}$. This set of functions is an element of the $\check{\textrm C}$ech cohomology class as well, representing the first Stiefel--Whitney class $w_1$.
The tangent bundle $TM$  admits a spin bundle structure if and only if $w_1(M)=\{0\}=w_2(M)$, for $w_2(M)$ being the second Stiefel--Whitney class. 

More precisely, a spin structure on
$M$ consists of a principal bundle 
$\mathbf{\uppi}_{s}:\mathcal{P}_{\mathrm{Spin}_{p,q}{}}(M)\rightarrow M$,
with group $\mathrm{Spin}_{p,q}{}$, and the 2-fold mapping \cite{Bonora:2009ta} 
\beq
s: \mathcal{P}_{\mathrm{Spin}_{p,q}{}}(M)\rightarrow \mathcal{P}_{\mathrm{SO}_{p,q}{}%
}(M),\nonumber
\eeq
satisfying $s(x \upphi)=s(x)\mathrm{ad}_{\upphi}$, for all $x\in
\mathcal{P}_{\mathrm{Spin}_{p,q}{}}(M)$. The adjoint mapping reads $\mathrm{ad}:\mathrm{Spin}_{p,q}{}\rightarrow\mathrm{Aut}(\cl_{p,q}),$
$\mathrm{ad}_{\upphi}:\Upxi\mapsto
\upphi\Upxi\upphi^{-1}\in\cl_{p,q}$ \cite{moro}, where $\mathrm{Aut}(\cl_{p,q})$ denotes the automorphism group of $\cl_{p,q}$.
This is equivalent to the commutation of the diagram
\begin{diagram}
&&M&&\\
&\ruTo_{\uppi_s} &&\luTo_{\uppi}& \\
\mathcal{P}_{\mathrm{Spin}_{p,q}{}}(M)&&\rTo^s&&\mathcal{P}_{\mathrm{SO}_{p,q}{}}(M)
\end{diagram}
There is also an onto mapping $\upsigma:\mathrm{Spin}_{p,q}{}\rightarrow\mathrm{SO}_{p,q}{}$, whose kernel consists of the group $\mathbb{Z}_{2}$. Therefore ad$({-1})= {\textrm id}_{\mathrm{Spin}_{p,q}}$, yielding the adjoint mapping  to descend to a representation of $%
\mathrm{SO}_{p,q}{}$. One regards $\mathrm{ad}^{\prime}$ this representation, consisting of  $\mathrm{ad}^{\prime}:\mathrm{SO}_{p,q}{}\rightarrow%
\mathrm{Aut}(\mathfrak{C}\ell_{p,q})$. Hence, $\mathrm{ad}%
_{\sigma(\upphi)}^{\prime}\Upxi=\mathrm{ad}_{\upphi}\Upxi=\upphi\Upxi\upphi^{-1}$.
In this context, the Clifford bundle  $\mathcal{C\ell (}M,g)$, can be made a vector bundle, whose sections are differential form fields. In addition, $\mathcal{C\ell(}M,g)=\mathcal{P}_{\mathrm{SO}%
_{p,q}{}}(M)\times _{\mathrm{ad}^{\prime}}\mathfrak{C}\ell_{p,q}$.

Refs. \cite{daRocha:2011yr,avis0,isham1,petry} introduced 
inequivalent spin structures on 4D Lorentzian spacetimes
with non-trivial topology. Their formal aspects are hereon in this section extended for manifolds of any finite dimension, endowed with a metric of arbitrary signature.  
A spin structure ($\mathcal{P}_{\mathrm{Spin}_{p,q}{}}(M),s$)  is not uniquely defined, when $H^1(M,\mathbb{Z}_2)\neq\{0\}$. Then,  other inequivalent spin structures, labeled by elements of $H^1(M,\mathbb{Z}_2)$, arise from ($\mathcal{P}_{\mathrm{Spin}_{p,q}{}}(M),s$). Hereon quantities having a ring over them are associated to exotic, inequivalent, spin structures. Two spin structures,  ($\mathcal{P}_{\mathrm{Spin}_{p,q}{}}(M),s)$ and $(\mathring{P}_{\mathrm{Spin}_{p,q}{}}(M),\mr{s})$, are said to be equivalent when the mapping $\upzeta: (\mathcal{P}_{\mathrm{Spin}_{p,q}{}}(M),s)\rightarrow (\mathring{P}_{\mathrm{Spin}_{p,q}{}}(M),\mr{s})$ exists, making the following diagram to commute. 
\begin{diagram}
&&\mathcal{P}_{\mathrm{SO}_{p,q}{}}(M)&&\\
&\ruTo_{s} &&\luTo_{\mathring{s}}& \\
(\mathcal{P}_{\mathrm{Spin}_{p,q}{}}(M), s)&&\rTo^\upzeta&&(\mathring{P}_{\mathrm{Spin}_{p,q}{}}(M),\mr{s})
\end{diagram}

To introduce the exotic corrections to the Dirac operator, one needs to define spinor fields as sections the spin bundle.
A $\mathbb{C}$-spinor bundle on $M$ is the bundle
$S_{\mathbb{C}}(M)=\mathcal{P}_{\mathrm{Spin}_{p,q}{}}(M)\times_{\upmu}S_{p,q},%
$ 
where $S_{p,q}$ is a complex left module for $\mathbb{C}\otimes
\cl_{p,q}$, being $\upmu$ a representation of $\mathrm{Spin}_{p,q}{}$ in the endomorphism space of $S_{p,q}$. The particular case where the metric signature has 
$p=1$ and $q=3$ yields $S_{p,q}=\mathbb{C}^{4}$ and $\upmu$ is the $(1/2,0)
\oplus (0,1/2)$ representation of $\mathrm{Spin}_{1,3}{}\simeq
{\textrm SL}(2,\mathbb{C})$ in $\mathrm{End}(\mathbb{C}^{4})$, what complies with the relativistic quantum mechanical standard approach. 
In this case, classical spinor
fields 
are sections of the bundle
$
{P}_{\mathrm{Spin}_{1,3}{}}(M)\times_{\uprho}\mathbb{C}^{4},
$
where $\uprho$ stands for the $(1/2,0)\oplus (0,1/2)$  representation of the 4D Lorentz group in $\mathbb{C}^{4}$. For the most general case for any dimension and any metric $(p,q)$ signature, $S_{p,q}$ is given by the classical spinors given by Table {\color{blue}{1}}, that carry the representations of the respective Clifford algebras in Table {\color{blue}{2}} \cite{Vaz:2016qyw}.

\begin{table}[h!]
\begin{center}
\begin{tabular}{||c|c|c|c|c||}
\toprule[1.5pt]
$\begin{matrix} p-q \\  \end{matrix}$ & 0 (\text{mod} \, 8)& 1 (\text{mod} \, 8)& 2 
(\text{mod} \, 8)& 3 (\text{mod} \, 8)\\ \midrule \midrule
$ S_{p,q}$    & 
$\begin{matrix} \mathbb{R}^{2^{\lfloor(n-1)/2\rfloor}} \\ \oplus \\
\mathbb{R}^{2^{\lfloor(n-1)/2\rfloor}} \end{matrix}$& 
$\mathbb{R}^{2^{\lfloor(n-1)/2\rfloor}}$& 
$\mathbb{C}^{2^{\lfloor(n-1)/2\rfloor}}$& 
$\mathbb{H}^{2^{\lfloor(n-1)/2\rfloor-1}}$ \\ \midrule
$\begin{matrix}p-q \\ \end{matrix}$ & 4 (\text{mod} \, 8)
& 5 (\text{mod} \, 8)& 6 (\text{mod} \, 8)& 7 (\text{mod} \, 8)\\ \midrule\midrule 
  $S_{p,q}$    & 
$\begin{matrix} \mathbb{H}^{2^{\lfloor(n-1)/2\rfloor-1}} \\ \oplus \\ \mathbb{H}^{2^{\lfloor(n-1)/2\rfloor-1}}\end{matrix}$ & 
$\mathbb{H}^{2^{\lfloor(n-1)/2\rfloor-1}}$& 
$\mathbb{C}^{2^{\lfloor(n-1)/2\rfloor}}$ & 
$\mathbb{R}^{2^{\lfloor(n-1)/2\rfloor}}$ \\ \bottomrule[1.5pt]
\end{tabular}
\caption{Classical spinors classification, for $\lfloor n/2\rfloor$ denoting the integer part of $n/2$.}\label{tab00}
\end{center}
\end{table}

{\tiny{\begin{center}
\begin{table}[h]
{
\begin{tabular}{||c|c|c|c|c||} \toprule[1.5pt]\label{tab11}
$\begin{matrix} {\tiny{p-q}} \\  \end{matrix}$ & 0 (\text{mod} \, 8) & 1 (\text{mod} \, 8) & 2  (\text{mod} \, 8)
& 3 (\text{mod} \, 8) \\ \midrule \midrule
$ \cl_{p,q} $ & $M(2^{\lfloor n/2\rfloor},\mathbb{R})$ & 
$\begin{matrix} M(2^{\lfloor n/2\rfloor},\mathbb{R}) \\ \oplus \\
M(2^{\lfloor n/2\rfloor},\mathbb{R}) \end{matrix}$ & 
$M(2^{\lfloor n/2\rfloor},\mathbb{R}) $ & 
$M(2^{\lfloor n/2\rfloor},\mathbb{C}) $ \\ \midrule  
$\begin{matrix}p-q \\ \end{matrix}$ & 4 (\text{mod} \, 8) 
& 5 (\text{mod} \, 8)& 6 (\text{mod} \, 8)& 7 (\text{mod} \, 8)\\ \midrule  \midrule
$\cl_{p,q}$ & $M(2^{\lfloor n/2\rfloor-1},\mathbb{H})$ & 
$\begin{matrix} M(2^{\lfloor n/2\rfloor-1},\mathbb{H}) \\ \oplus \\
M(2^{\lfloor n/2\rfloor-1},\mathbb{H}) \end{matrix} $ & 
$M(2^{\lfloor n/2\rfloor-1},\mathbb{H})$ & 
$M(2^{\lfloor n/2\rfloor},\mathbb{C})$ \\ \bottomrule[1.5pt] 
\end{tabular}
}\\
{
\caption{Real Clifford algebra classification, for $\lfloor n/2\rfloor$ denoting the integer part of $n/2$.}
}
\end{table}
\end{center}}}

Two spin structures ($\mathcal{P}_{\mathrm{Spin}_{p,q}{}}(M),s$) and ($\mr{P}_{\mathrm{Spin}_{p,q}{}}(M),\mr{s}$) are respectively described by the mappings ${h_{jk}}: U_j\cap U_k \to {\textrm Spin}_{p,q}$ and ${\mr{h}_{jk}}: U_j\cap U_k \to {\textrm Spin}_{p,q}$, together with and $u_{jk}: U_j\cap U_k\to {\textrm SO}_{p,q}$, satisfying $
 {h_{ij}}\circ \upsigma=u_{ij},$ ${h_{ij}}\circ {h_{jk}} = {h_{ik}}$ and $ {h_{ii}}=\, {\textrm id}_{U_i\cap U_j }$. 
Now one defines a mapping $c_{jk}$ by the relation $h_{ij}(x) = \mr{h}_{ij}(x)c_{ij}$ such that $c_{ij}:U_i\cap U_j\rightarrow$ ker $\upsigma$ = $\mathbb{Z}_2$, satisfying $c_{ij}\circ c_{jk} = c_{ik}$. Given the irreducible representation $\uprho:\cl_{p,q}\rightarrow M(k,\mathbb{K})$ in $
{P}_{\mathrm{Spin}_{p,q}{}}(M)\times_{\uprho}S_{p,q},
$ one has $\uprho(c_{ij}(x)) = \pm 1$, since $\uprho$ is a faithful representation. 
Deciding 
 the type of the matrix $M(k,\mathbb{K})$ and the representation space, $S_{p,q}$, depends on 
 the dimension of $M$ and on the metric signature as well. In some cases, $M(k,\mathbb{K})$ can also denote the direct sum of two identical matrices, whereas respectively $S_{p,q}$ can also consist of the direct sum of two representation spaces.
 Tables {\color{blue}{1}} and {\color{blue}{2}} make it precise which choice one must take into account, respectively, for $M(k,\mathbb{K})$  and $S_{p,q}$. 
 
When the cohomology group $H^2(M,\mathbb{Z}_2)$ is devoid of  2-torsion, one can define functions $\upxi_i: U_i\rightarrow\CC$,  such that $\upxi_i(x)\in $ U(1)  \cite{daRocha:2011yr,avis0,isham1,petry}, and
  \beq\label{xi1}
  \upxi_i(x) (\upxi_j(x))^{-1} = \uprho(c_{ij}(x)) = \pm 1.\eeq\noi  In addition, ${\upxi}^2_i(x) = {\upxi}^2_j(x)$, for all $x\in U_i\cap U_j$. 
Hence, the (local) scalar fields $\upxi_i$ induce a unique scalar field $\upxi:M\rightarrow\CC$ such that $\upxi(x) = {\upxi}_i^2(x)$, for all $x\in U_i$.
  
  Let one considers an arbitrary spinor field $\uppsi \in$ sec $
{P}_{\mathrm{Spin}_{p,q}{}}(M)\times_{\uprho}S_{p,q}$. There is, then, a one-to-one correspondence between elements of the cohomology group $H^1(M,\mathbb{Z}_2)$ and a Dirac operator,  $D$. A local spinor field component $\uppsi_i: U_i\rightarrow S_{p,q}$ is the mapping that satisfies 
$\uprho(p_i,\uppsi_i(x)) = \uppsi(x)$, for local spin bundle sections $p_i :U_i\rightarrow$ ($\mathcal{P}_{\mathrm{Spin}_{p,q}{}}(M),s$). Therefore the transition rule $\uppsi_i(x) = \uprho(h_{ij}(x))\uppsi_j(x)$, for all $x\in U_i\cap U_j$, does hold. 

Given the local spin bundle sections $p_i :U_i\rightarrow$ ($\mathcal{P}_{\mathrm{Spin}_{p,q}{}}(M),s$), another set constituted of exotic sections $\mr{p}_i: U_i\rightarrow$ $\mr{P}_{\mathrm{Spin}_{p,q}{}}(M)$, corresponding to an inequivalent spin structure, satisfies the following diagram:
\begin{diagram}
\mr{P}_{\mathrm{Spin}_{p,q}{}}(M)&\rTo^{\mr{s}}&{P}_{\mathrm{SO}_{p,q}{}}(M)&\lTo^{s}&{P}_{\mathrm{Spin}_{p,q}{}}(M)\\
&\rdTo^{\mr{p}_i}&\uTo&\ldTo^{p_i}&\\
&&U_i&&
\end{diagram}
Locally, for all $x\in U_i\cap U_j$, it permits the exotic spinor field to have the property
\beq
    \mr\uppsi_i(x) = \uprho(\mr{h}_{ji}) = \uprho[h_{ji}(x)]\uprho[c_{ji}(x)]\mr\uppsi_j(x).\label{ccv}
\eeq
Hence $\uprho(\upxi_j) = \uprho[c_{ji}(x)]\uprho(\upxi_i)$. Comparing it to  Eq.(\ref{ccv}), it is straightforward to realize that the term $\uprho(\upxi_i)\mr\uppsi_i$  behaves as $\uppsi_i\in{P}_{\mathrm{Spin}_{p,q}{}}(M)\times_{\uprho}S_{p,q}$,  inducing a mapping between the exotic spin bundle to the standard spin bundle, given by 
\begin{eqnarray}
\Gamma: \mr{P}_{\mathrm{Spin}_{p,q}{}}(M)\times_{\uprho}S_{p,q}&\rightarrow&{{P}}_{\mathrm{Spin}_{p,q}{}}(M)\times_{\uprho}  S_{p,q} \nonumber\\
\mr\uppsi_i &\mapsto& \Gamma(\mr\uppsi_i)\equiv \uprho(\upxi_i)\mr\uppsi_i = \uppsi_i
\end{eqnarray} such that the exotic Dirac operator can be defined as 
   \beq\mr D_X f(\mr\uppsi) = \Gamma(D_X\mr\uppsi) +\frac{1}{2}(X\lrcorner(\upxi^{-1} {\textrm d}\upxi))\Gamma(\mr\uppsi),\label{xi2}\noi\eeq
 for all sections $\uppsi\in \mathcal{P}_{\mathrm{Spin}_{p,q}{}}(M)\times_{\uprho}S_{p,q}$ and all vector fields $X$ on $M$ \cite{daRocha:2011yr,petry,isham1,avis0}.    
In a chart of coordinates on $M$, Eq. (\ref{xi2}) is equivalent to the exotic Dirac operator to be given in terms of the standard Dirac operator by 
\beq
\mr{D}_X = D_X  -\frac{1}{2}\left[X\lrcorner\left(\upxi^{-1}(x){\textrm d}\upxi(x)\right)\right]\label{covvv}. 
\eeq
As the exotic additional term in Eq.(\ref{covvv}) is an integer in a ${\check{\textrm C}}$ech cohomology class, equivalent to $\frac{1}{2\pi i}\upxi^{-1}(x){\textrm d}\upxi(x)$ \cite{petry,avis0,isham1}, one can rescale $\upxi(x)\mapsto (2\pi)^{-1/2}{\upxi(x)}{}$, yielding 
 \beq
 i\gamma^\mu\mr{D}_\mu = i\gamma^\mu D_\mu + \upxi^{-1}(x){\textrm d}\upxi(x),\label{26}
 \eeq where the $\gamma$ matrices satisfy the Clifford
commutation relations, $
\{\gamma_\mu,\gamma_\nu\}= 2g_{\mu\nu}\,{\textrm id}_{\cl_{p,q}}$ and $g_{\mu\nu}$ denotes the metric tensor components. As  $\upxi(x)\in U(1)$, one can write $\upxi(x)=e^{i\uptheta(x)}$, for a $C^1$ scalar field $\theta:U\subset M\to\mathbb{C}$. The exotic spin structure term, in the exotic Dirac operator (\ref{26}), then reads
\beq\label{tu1}
\upxi^{-1}(x) {\textrm d}\upxi(x) = e^{-i\uptheta(x)}(i\gamma^\mu\partial_\mu\uptheta(x))e^{i\uptheta(x)} = i\gamma^\mu\partial_\mu\uptheta(x).
\eeq

\section{Heat kernel coefficients}
\label{hkc0}
 
Let $
{R^\mu}_{\nu\rho\sigma}$ 
be the Riemann curvature tensor, $R_{\mu\nu}={R^\sigma}_{\mu\nu\sigma}$
the Ricci tensor, and  $R=R^\mu_{\;\,\mu}$ be the scalar
curvature. 
Given a vielbein $\{e_i\}$ for the tangent space $M$, the tetrads are such that $e_j^\mu e_k^\nu g_{\mu\nu}=\delta_{jk}$,
$e_j^\mu e_k^\nu \delta^{jk}=g^{\mu\nu}$. The inverse vielbein
is defined by the relation $e_\mu^j e^\mu_k=\delta_k^j$. It is worth to emphasize that Euclidean spaces make upper and lower indexes to be equivalent. 
The spin connection $\sigma_\mu$ reads $
\nabla_\mu v^j = \partial_\mu v^j +\sigma_\mu^{jk}v_k$, for an arbitrary vector $v_\nu$. 
The condition $\nabla_\mu e_\nu^k=0$ yields \cite{8}
\beq
\sigma_\mu^{kl}=e_l^\nu \Gamma_{\mu\nu}^\rho e_\rho^k -
e^\nu_l \partial_\mu e_\nu^k \,.\label{sc2}
\eeq 
Let
\lq ;\rq\ denote covariant differentiation with respect to 
the Levi-Civita connection of $M$. The quantity $\Omega_{\mu\nu}$ denotes the connection field strength, given by 
$\omega$:
\beq
\Omega_{\mu\nu}=\partial_{[\mu}\omega_{\nu]}+\omega_{[\mu}\omega_{\nu]}\, .
\label{Omega}
\eeq

The action for the spinor fields $\uppsi$ reads  \cite{8}
\beq
\mathcal{L}=\int_M {\textrm d}^n x \sqrt{g} \bar \uppsi\, \Dir\, \uppsi
\label{spinS}
\eeq
where  $\Dir$ is the Dirac operator satisfying $D=\Dir^2$, where $D$ is the Laplace operator.
The matrix $\gamma^5$ represents the volume element, also used to define the chirality operator in QFT. 
The usual Dirac operator reads 
\beq
\Dir =i\gamma^\mu \left(\partial_\mu +\frac 18 [\gamma_\nu ,\gamma_\rho ]
\sigma_\mu^{\nu\rho} +A_\mu +iA_\mu^5 \gamma^5 \right) \,,
\label{dirop}
\eeq
where $A_\mu$ [$A_\mu^5$] denotes the gauge vector [axial gauge vector] field, which are anti-hermitian when gauge indexes are taken into account.
The operator $D=\Dir^2\sim g_{\mu\nu}\nabla^\mu\nabla^\nu + E$ is of Laplace type, where 
\begin{eqnarray}
\!\!\!\!\!\!\!\!\!\!\!\!\omega_\mu &=&\frac 18 [\gamma_\nu ,\gamma_\rho ]
\sigma_\mu^{\nu\rho} +A_\mu +\frac i2 [\gamma_\mu ,\gamma_\nu ]
A^{5\nu}\gamma^5 \,, \\
\!\!\!\!\!\!\!\!\!\!\!\!E&=&-\frac 14 R +\frac 14 [\gamma^\mu ,\gamma^\nu ]F_{\mu\nu}
+i\gamma^5 {\rm d}^\mu A_\mu^5 -(n-2)A_\mu^5A^{5\mu} -\frac 14 (n-3) [\gamma^\mu ,\gamma^\nu ] [A_\mu^5 ,A_\nu^5 ],
\label{oEdir}
\end{eqnarray}
where $F_{\mu\nu}=\partial_\mu A_\nu -
\partial_\nu A_\mu +[A_\mu ,A_\nu ]$, 
$D_\mu A_\nu^5 = \partial_\mu A_\nu^5 -
\Gamma_{\mu\nu}^\rho A_\rho^5 +[A_\mu ,A_\nu^5 ]$.
Besides, 
\begin{eqnarray}
\!\!\!\!\!\!\!\!\!\!\!\!\Omega_{\mu\nu}&=&F_{\mu\nu}-[A_\mu^5 ,A_\nu^5 ]-\frac 14
\gamma^\sigma \gamma^\rho R_{\sigma\rho\mu\nu} 
-i\gamma^5\gamma^\rho (\gamma_\nu D_\mu A_\rho^5
-\gamma_\mu D_\nu A_\rho^5 ) \nonumber \\
&&+i\gamma^5 A_{\mu\nu}^5 
+[A_\mu^5,A_\rho^5] \gamma^\rho \gamma_\nu -
[A_\nu^5,A_\rho^5] \gamma^\rho \gamma_\mu-\gamma^\rho A_\rho^5 \gamma_\mu \gamma^\sigma A_\sigma^5 \gamma_\nu
+ \gamma^\rho A_\rho^5 \gamma_\nu \gamma^\sigma A_\sigma^5 \gamma_\mu~,\label{Omdir}
\end{eqnarray}
where
$
A_{\mu\nu}^5=\partial_{[\mu} A_{\nu]}^5
+[A_{[\mu} ,A_{\nu]}^5 ]$.

When considering compact Riemannian manifolds $M$ with no  boundary, let $V$ be a vector bundle over $M$. Let $f$ denote  a smooth function on $M$. 
The operator
$e^{-tD}$, for $t>0$, represents a trace class on  $L^2(V)$, being the operator function \cite{8}, 
\begin{equation}
K(t,f,D)={\rm Tr}_{L^2}(f\,e^{-tD}),
\label{calor}
\end{equation}
then, well defined, where the operator in Eq. (\ref{calor}) emulates 
the  heat conduction equation solution, 
\begin{equation}
\left(\frac{\partial}{\partial t}+D_x\right) K(t,x,x_1,D)=0,\qquad K(0,x,x_1,D)=\updelta (x,x_1). \label{condeq}
\end{equation}
One can express 
\begin{equation}
K(t,f,D)=\int_M {\rm d}^nx \sqrt{g}\, {\rm Tr}\, \lim_{x\to x_1}K(t,x,x_1,D) f(x).\label{KxxD}
\end{equation}
As $K(t,x,y,D)$ can be thought of as being a tensor with 
gauge group indexes, the operator ${\rm Tr}$ computes the trace over these indexes.

The asymptotic expansion, in the limit $t\to0$ can be usually employed \cite{8}, 
\begin{equation}
{\rm Tr}_{L^2} (f e^{-tD})
\approxeq\sum_{j\in \mathbb{N}\cup 0}t^{\frac{k-n}{2}}a_j(f,D)\,.
\label{asymptotex}
\end{equation}
Heat kernel coefficients with odd index equal zero. The even index heat kernel  coefficients, $a_{2i}(f,D)$ are locally
computable with respect to the geometric invariants of $M$, as  \cite{8}
\beq
\!\!\!\!\!\!a_j(f,D)\!= \!{\textrm{Tr}}_V \iM\, [a_j (x,D) f(x) ]
\!=\!\sum_\ell \,{\textrm{Tr}}_V \iM\,[ f(x) u^\ell(x) \mathcal{B}_j^\ell(x,D)], \label{locinv}
\eeq
where the $\mathcal{B}_j^\ell$ represent all the geometric  invariants of the legnth $j$, composed by the Riemann tensor, by $E$ and $\Omega$ (and their derivatives), and $u^\ell$ are  parameters determined by constraints between heat kernel coefficients. The case $j=2$ yields $E$ and $R$ as the 
only independent invariants.

When $M=M_1\times M_2$ is factorizable, the Laplace type operator $D$ is, therefore, written as 
$D=D_1\otimes {\textrm id}_{M_2} +{\textrm id}_{M_1}\otimes D_2$, where $D_a: M_a\to M_a$, for $a=1,2$, is the Laplace type operator restricted to the manifolds $M_1$ and $M_2$. Therefore, one can write 
$e^{-tD}= e^{-tD_1}\otimes e^{-tD_2}$, yielding  \cite{8},
\beq
a_j(x,D)=\sum_{r+s=j} a_r(x_1,D_1)a_s(x_2,D_2)\,.
\label{product}
\eeq
Eq.\ (\ref{product}) yields the constants  $u^\ell$ to be dependent on the spacetime dimension $n=p+q$. As posed in Ref. \cite{8}, when one considers   $M_1=S^1$, then one can choose 
$D_1=-\partial_{x_1}^2$. 
Hence (\ref{locinv}),
\begin{eqnarray}
a_j(f(x_2),D)
&=&2\pi\int_{M_2}{\textrm d}^{n-1}x\,\sqrt{g}\sum_\ell
\,{\textrm{Tr}}_V \{f(x_2) u^\ell_{n} \mathcal{ A}^\ell_j (D_2) \}\,,
\label{77hk}
\end{eqnarray}
where  the $f$ function was split as  $f(x_1,x_2)=f_1(x_1)f_2(x_2)$, for $x_a\in M_a$. 
The constants $u^\ell_n$ depend on the spacetime dimension
$n$ only by the factor
$(4\pi )^{-n/2}$ . 

The heat kernel coefficients read \cite{8}
\begin{eqnarray}
{a}_{0}(f,D)&=&\frac1{(4\pi)^{n/2}}\iM \,{\textrm{Tr}}_V (f),\label{zeroth}\\
{a}_2(f,D)
         &=&\frac1{6(4\pi)^{n/2}}\iM \,{\textrm{Tr}}_V \left[f\left(6E+R\right)\right],
    \label{second}\\ 
 {a}_4(f,D)&=&\frac1{360(4\pi)^{n/2}}\iM \,{\textrm{Tr}}_V\left[f\left(60E_{;\alpha\alpha}
+12R_{;\alpha\alpha}+5R^2    +60RE    +180E^2\right.\right.\nonumber\\&&\left.\left.\qquad\qquad\qquad\qquad\qquad\qquad-2 R_{\rho\sigma}R_{\rho\sigma}
    +2 R_{\rho\sigma\alpha\beta}R_{\rho\sigma\alpha\beta}+30\Cur_{ \rho\sigma}\Cur_{\rho\sigma}\right)\right],
   \label{fourth} \\
{a}_{6}(f,D)&=&\frac1{(4\pi)^{n/2}}\iM \,{\textrm{Tr}}_V\Bigg\{ \frac f{7!}\left(
    18R_{;\rho\rho \sigma\sigma}+17R_{;\alpha}R_{;\alpha}
    -2R_{\rho\sigma;\alpha}R_{ \rho\sigma;\alpha}-4R_{\sigma\alpha;\zeta}R_{\sigma\zeta;\alpha} \right. \nonumber \\
&&\left.
    +9R_{\rho\sigma \alpha\beta;\zeta}R_{\rho\sigma \alpha\beta;\zeta}+28RR_{;\zeta\zeta}
    -8R_{\sigma\alpha}R_{\sigma\alpha;\zeta\zeta}   +24R_{ \sigma\alpha}R_{\sigma\zeta;\alpha\zeta}
    +12R_{\rho\sigma \alpha\beta}R_{\rho\sigma \alpha\beta ;\zeta\zeta}\right.\nonumber \\
&& \left. 
    +\frac{35}9R^{3}
    -\frac{14}3RR_{\rho\sigma}R_{\rho\sigma} +\frac{14}3R R_{\rho\sigma\alpha\beta}R_{\rho\sigma\alpha\beta}
     -\frac{208}9R_{\sigma\alpha}R_{\sigma\zeta}R_{\alpha\zeta}
     -\frac{64}3R_{ \rho\sigma}R_{\alpha\beta}R_{\rho\alpha \sigma\beta}  \right.\nonumber \\
&&
     \left.-\frac{16}3R_{\sigma\alpha}R_{\sigma\zeta \beta \rho}R_{\alpha\zeta \beta \rho}
     -\frac{44}9R_{\rho\sigma \alpha\zeta}R_{\rho\sigma \beta p}R_{\alpha\zeta \beta \tau}  -\frac{80}9R_{\rho\sigma \alpha\zeta}R_{\rho\beta  \alpha\tau}R_{\sigma\beta  \zeta\tau}\right)\nonumber\\
&&
     +\frac{f}{360}(
      \textcolor{black}{8\Cur_{\rho\sigma;\alpha} \Cur_{\rho\sigma;\alpha}
     +2\Cur_{\rho\sigma;\sigma}\Cur_{ \rho\alpha;\alpha}}  \textcolor{black}{ +12\Cur_{\rho\sigma;\alpha\alpha}\Cur_{\rho\sigma}
     -12\Cur_{\rho\sigma}\Cur_{\sigma\alpha}\Cur_{\alpha\rho}
     -6R_{\rho\sigma \alpha\zeta}\Cur_{\rho\sigma}\Cur_{\alpha\zeta}}\nonumber \\
&&
     -4R_{\sigma\alpha}\Cur_{\sigma\zeta}\Cur_{\alpha\zeta}
       \textcolor{black}{ +5R\Cur_{\alpha\zeta}\Cur_{\alpha\zeta}
     +6E_{;\rho\rho \sigma\sigma}+60EE_{;\rho\rho}
      +30E_{;\rho}E_{;\rho}
     +60E^{3}+30E\Cur_{\rho\sigma}\Cur_{\rho\sigma}}
\nonumber \\
&&
      \textcolor{black}{
     +10R E_{;\alpha\alpha}+4R_{\sigma\alpha}E_{;\sigma\alpha}}
     \textcolor{black}{+12R_{;\alpha}E_{;\alpha}+30E^2R}+12ER_{;\alpha\alpha}+5ER^2
     -2ER_{\rho\sigma}R_{\rho\sigma}
     \nonumber \\
&&
     \textcolor{black}{ +2ER_{\rho\sigma\alpha\beta}R_{\rho\sigma\alpha\beta})}\}.\label{sixth}
    \end{eqnarray}
In the next section we introduce the exotic deviation coefficients, measuring the exotic corrections to the heat kernel coefficients (\ref{zeroth} -- \ref{sixth}).

\section{Exotic heat kernel coefficients}
\label{hkc}
 In Sect. \ref{w2} the exotic additional terms in the spin connection, defining the Dirac operator, was shown to arise from the non-trivial topology of the manifold $M$. These exotic terms, in addition, must be renormalizable. Any spinor field, on which the Dirac operator acts, realizes
the exotic term not only as a shift of the gauge potential that eventually appears in the Dirac equation (or any first order equation of motion that governs the spinor fields \cite{Villalobos:2015xca,daSilvaa:2019kkt}). 
Inequivalent spin structures, arising from manifolds having non-trivial topologies, can also engender deep modifications on  the heat kernel coefficients associated to the manifold. Exotic spinor fields were in 4D spacetimes with non-trivial topology were scrutinized in Refs. \cite{daRocha:2011yr,Bernardini:2012sc}.  

Due to the form of the exotic Dirac operator in Eqs. (\ref{26}, \ref{tu1}), exotic spinor fields, appearing in first order equations of motion, realize it as 
\begin{align}
    \mathring{A}_\mu & \mapsto A_\mu + i\partial_\mu\uptheta\, \label{amu0},\\
    \mathring{A}_\mu^5 & \mapsto {A}_\mu^5 + i\partial_\mu\uptheta^5\,,\label{amu5}
\end{align}
where $\uptheta:U\subset M\to\Upomega^0(M)\simeq\mathbb{R}$ is a scalar field defined on an open subset $U\subset M$, whereas $\uptheta^5:U\subset M\to\Upomega^n(M)$ is a pseudoscalar field. 
Alternatively, Eq. (\ref{amu5}) can be introduced by considering  $\upxi(x)\in $ U(1)$\times$U(1) = $(\mathbb{R}\oplus\mathbb{R})\times {\textrm U}(1)$ in Eq. (\ref{xi2}), what is equivalent to replicate the formal aspects in  Sect. \ref{w2} using the field hyperbolic numbers $\mathbb{R}\oplus\mathbb{R}$, instead of the real numbers. In fact,  employing a second copy of U(1) regards pseudoscalar fields, as shown in Ref. \cite{Vaz:2016qyw,daRocha:2006si}. 
 Therefore, given the usual Dirac operator 
 \begin{align}
  \Dir=  \gamma^\mu {D}_\mu & = i\gamma^\mu{\left(\partial_\mu +\frac{1}{8}\left[\gamma_\mu,\gamma_\nu\right]\sigma^{\nu\rho} + A_\mu + iA_\mu^5\gamma^5\right)},
\end{align} 
we can write the exotic Dirac operator, using Eqs. (\ref{amu0}, \ref{amu5}) as 
\begin{align}
    \gamma^\mu \mathring{D}_\mu & = i\gamma^\mu{\left(\partial_\mu +\frac{1}{8}\left[\gamma_\mu,\gamma_\nu\right]\sigma^{\nu\rho} + A_\mu + iA_\mu^5\gamma^5\right) +i\gamma^\mu \left( i\partial_\mu\uptheta +i^2\partial_\mu\uptheta^5\gamma^5\right)}\nonumber\\
        & = \gamma^\mu \left[D_\mu -\left(\partial_\mu\uptheta + i\partial_\mu\uptheta^5\gamma^5\right)\right]~,\label{do}
\end{align} 
that is, $\mathring{D}_\mu \mapsto D_\mu-\partial_\mu\tilde{\uptheta}$, where 
\beq
\label{tilt}{\tilde{\uptheta}\equiv\uptheta+i\uptheta^5\gamma^5}.
\eeq Thereupon, looking to the exotic field strength, yields 
\begin{align}\label{FmunuExotic}
 \mathring{F}_{\mu\nu} &  = F_{\mu\nu}+\left[A_{[\mu},i\partial_{\nu]}\uptheta\right]~,
\end{align}
where $\left[A_{[\mu},\partial_{\nu]}\uptheta\right]\equiv \left[A_\mu,i\partial_\nu\uptheta\right]-\left[A_\nu,i\partial_\mu\uptheta\right]$~.

On the other hand, the field strength (\ref{Omega}) must be lifted to the inequivalent spin structure. Thus, the exotic connection reads 
\begin{align}
    \mathring{\omega}_\mu &  = \omega_\mu +  \partial_\mu\uptheta-\frac{i}{2}\left[\gamma_\mu,\gamma_\nu\right]\partial_\mu\uptheta^5\gamma^5~.
\end{align}
The complete proof is accomplished in Eq. (\ref{exomeg}) in Appendix \ref{app}. Collecting the computations in  Appendix \ref{app}, implemented by Eq.~\eqref{FmunuExotic} and Eqs.~\eqref{A5comut} --~\eqref{A5A5}, yields the complete form of the exotic full field strength displayed initially in Eq.~\eqref{OmegaExotic}, 
\begin{align}
    \mathring{\Omega}_{\mu\nu} & =\Omega_{\mu\nu} +i\left(A_{[\mu},\partial_{\nu]}\uptheta\right)+i\left[A^5_{[\mu},\partial_{\nu]}\uptheta^5\right]+\gamma^5\gamma^\rho\gamma_{[\nu} D_{\mu]}\partial_\rho\uptheta^5 - i\gamma^5\gamma^\rho\gamma_{[\nu}\partial_{\mu]}\tilde{\uptheta}A^5_\rho
\nonumber\\
    &\quad+ i\gamma^5\left(\left[A_{[\mu},\partial_{\nu]}\uptheta^5\right] + \left[\partial_{[\mu}\uptheta,A_{\nu]}^5\right]\right)+i\left(\left[A^5_{[\mu},\partial_{|\rho|}\uptheta^5\right]-\left[A^5_{|\rho|},\partial_{[\mu}\uptheta^5\right]\right)\gamma^\rho\gamma_{\nu]}
\nonumber\\
    &\quad-i\gamma^\rho A^5_{|\rho|}\gamma_{[\mu}\gamma^\sigma\partial_{|\sigma|}\uptheta^5\gamma_{\nu]}-i\gamma^\rho\partial_{|\rho|}\uptheta^5\gamma_{[\mu}\gamma^\sigma \mr{A}^5_{|\sigma|}\gamma_{\nu]}~.
\end{align}
To recover the usual  field strength (\ref{Omega}), one just sets $\uptheta$ and $\uptheta^5$ to $0$.

Next, let us figure out one of the invariants in the heat kernel coefficients. At first, the exotic $\mathring{E}$ invariant can be written as, 
\begin{align}\label{InvarE}
    \mathring{E} &= -\frac{1}{4}R + \frac{1}{4}\left[\gamma^\mu,\gamma^\nu \right]\mathring{F}_{\mu\nu}
+i\gamma^5 \mathring{D}^\mu \mathring{A}_\mu^5 -(n-2)\mathring{A}_\mu^5\mathring{A}^{5\mu} -\frac{(n-3)}{4}\left[\gamma^\mu ,\gamma^\nu\right]\left[\mathring{A}_\mu^5,\mathring{A}_\nu^5 \right]\nonumber\\
    &= E +\frac{i}{4}\left(A_{[\mu},\partial_{\nu]}\uptheta\right)\left[\gamma^\mu,\gamma^\nu \right] + \gamma^5\left( -D_\mu\partial^\mu\uptheta^5 -i \partial^\mu\tilde{\uptheta}A^5_\mu + \partial^\mu\tilde{\uptheta}\partial_\mu\uptheta^5\right)\nonumber\\
    &\quad-i(n-2)\left(2A^5_\mu\partial^\mu\uptheta^5 -\partial_\mu\uptheta^5\partial^\mu\uptheta^5 \right)  -\frac{i(n-3)}{4}\left[\gamma^\mu ,\gamma^\nu\right] \left[A^5_{[\mu},\partial_{\nu]}\uptheta^5\right]~.
\end{align}
Endowed with the expressions (\ref{FmunuExotic}), (\ref{OmegaExotic}) and \eqref{InvarE}, respectively for the exotic field strengths and the exotic $E$ invariant, one can now compute the exotic heat kernel expansion.

    On the other hand, the exotic counterparts of the heat kernel coefficients take into account all the exotic contributions, that measure the topological non-triviality of $M$.     Clearly, exclusively geometric terms, involving the Riemann tensor and its contractions, remain invariant, as 
    they are not topological terms. The exotic heat kernel coefficients are then given by 
    \begin{eqnarray}
\mr{a}_{0}(f,D)&=&\frac1{(4\pi)^{n/2}}\iM \,{\textrm{Tr}}_V (f),\label{zerothe}\\
\mr{a}_2(f,D)
         &=&\frac1{6(4\pi)^{n/2}}\iM \,{\textrm{Tr}}_V\,\left[f\left(6\mr{E}+R\right)\right],
    \label{seconde}\\ 
    \mr{a}_4(f,D)&=&\frac1{360(4\pi)^{n/2}}\iM \,{\textrm{Tr}}_V \left[f\left(60\mr{E}_{;\alpha\alpha}
+12R_{;\alpha\alpha}+5R^2 +60R\mr{E}    +180\mr{E}^2\right.\right.
\nonumber\\
&&\qquad\qquad
    \left.\left.+12R_{;\alpha\alpha}+5R^2-2 R_{\rho\sigma}R_{\rho\sigma}
    +2 R_{\rho\sigma\alpha\beta}R_{\rho\sigma\alpha\beta}+30\mr\Cur_{ \rho\sigma}\mr\Cur_{\rho\sigma}\right)\right],
   \label{fourthe}
    \end{eqnarray}
        \begin{eqnarray}
 \!\!\!\!\!\!\!\!\!\!\! \mr{a}_{6}(f,D)&=&\frac1{(4\pi)^{n/2}}\iM \,{\textrm{Tr}}_V\Bigg [ \frac f{7!}\left(
    18R_{;\rho\rho \sigma\sigma}+17R_{;\alpha}R_{;\alpha}
    -2R_{\rho\sigma;\alpha}R_{ \rho\sigma;\alpha}\right.
\nonumber \\
&&\qquad
    -4R_{\sigma\alpha;\zeta}R_{\sigma\zeta;\alpha}
    +9R_{\rho\sigma \alpha\beta;\zeta}R_{\rho\sigma \alpha\beta;\zeta}+28RR_{;\zeta\zeta}
    -8R_{\sigma\alpha}R_{\sigma\alpha;\zeta\zeta} \nonumber \\
&&\qquad
    +24R_{ \sigma\alpha}R_{\sigma\zeta;\alpha\zeta}
    +12R_{\rho\sigma \alpha\beta}R_{\rho\sigma \alpha\beta ;\zeta\zeta}
    +\frac{35}9R^{3}
    -\frac{14}3RR_{\rho\sigma}R_{\rho\sigma} \nonumber \\
&&\qquad
    +\frac{14}3R R_{\rho\sigma\alpha\beta}R_{\rho\sigma\alpha\beta}
     -\frac{208}9R_{\sigma\alpha}R_{\sigma\zeta}R_{\alpha\zeta}
     -\frac{64}3R_{ \rho\sigma}R_{\alpha\beta}R_{\rho\alpha \sigma\beta} \nonumber \\
&&\qquad
     -\frac{16}3R_{\sigma\alpha}R_{\sigma\zeta \beta \rho}R_{\alpha\zeta \beta \rho}
     -\frac{44}9R_{\rho\sigma \alpha\zeta}R_{\rho\sigma \beta p}R_{\alpha\zeta \beta \tau} 
     -\frac{80}9R_{\rho\sigma \alpha\zeta}R_{\rho\beta  \alpha\tau}R_{\sigma\beta  \zeta\tau})\nonumber\\&&
     \qquad+\frac{f}{360}(8\mr\Cur_{\rho\sigma;\alpha} \mr\Cur_{\rho\sigma;\alpha}
     +2\mr\Cur_{\rho\sigma;\sigma}\mr\Cur_{ \rho\alpha;\alpha}+12\mr\Cur_{\rho\sigma;\alpha\alpha}\mr\Cur_{\rho\sigma}
     -12\mr\Cur_{\rho\sigma}\mr\Cur_{\sigma\alpha}\mr\Cur_{\alpha\rho}\nonumber \\
&&\qquad
     -6R_{\rho\sigma \alpha\zeta}\mr\Cur_{\rho\sigma}\mr\Cur_{\alpha\zeta}
     -4R_{\sigma\alpha}\mr\Cur_{\sigma\zeta}\mr\Cur_{\alpha\zeta} +5R\mr\Cur_{\alpha\zeta}\mr\Cur_{\alpha\zeta}
     +6\mr{E}_{;\rho\rho \sigma\sigma}+60\mr{E}\mr{E}_{;\rho\rho}
\nonumber \\
&&\qquad
      +30\mr{E}_{;\rho}\mr{E}_{;\rho}
     +60\mr{E}^{3} +30\mr{E}\mr\Cur_{\rho\sigma}\mr\Cur_{\rho\sigma}
     +10R \mr{E}_{;\alpha\alpha}+4R_{\sigma\alpha}\mr{E}_{;\sigma\alpha}
    +12R_{;\alpha}\mr{E}_{;\alpha}
\nonumber \\
&&\qquad
\left.    +30\mr{E}^2R+12\mr{E}R_{;\alpha\alpha}+5\mr{E}R^2
     -2\mr{E}R_{\rho\sigma}R_{\rho\sigma}+2\mr{E}R_{\rho\sigma\alpha\beta}R_{\rho\sigma\alpha\beta}\right)\Bigg].\label{sixthe}
    \end{eqnarray}

To measure the deviations of the exotic heat kernels coefficients (\ref{zerothe} -- \ref{sixthe}) with respect to the standard ones (\ref{zeroth} -- \ref{sixth}), associated to topologically trivial manifolds, we define the exotic deviation coefficient, $\delta{a}_j(f,D)$, by 
\begin{eqnarray}
\delta{a}_j(f,D)=\mr{a}_j(f,D)-{a}_j(f,D).
\end{eqnarray}
The results for the leading heat kernel coefficients are explicit in what follows. First, 
\begin{eqnarray}
\mr{a}_{0}(f,D)&=&\frac1{(4\pi)^{n/2}}\iM \,{\textrm{Tr}}_V\{f\}={a}_{0}(f,D).\label{zeroth1}
 \end{eqnarray}
 Hence $
\delta a_{0}(f,D)=0$ 
and there is no effects due to the exotic topology in the heat kernel coefficient (\ref{zeroth1}). 

Using Eqs. (\ref{second}, \ref{InvarE}), the second exotic deviation coefficient reads 
 \begin{eqnarray}
\mr{a}_2(f,D)    &=&{a}_2(f,D)+\delta{a}_2(f,D),
        \end{eqnarray}
   so that
  \beq\label{dev-a2}
 \delta{a}_2(f,D)= \frac1{(4\pi)^{n/2}}\iM \,{\textrm{Tr}}_V\left[f\left(\delta E\right)\right]~,
     \eeq
where
\begin{align}\label{dE}
    \delta E &=\mr{E}-E\nonumber\\
    &= \frac{i}{4}\left(A_{[\mu},\partial_{\nu]}\uptheta\right)\left[\gamma^\mu,\gamma^\nu \right] + \gamma^5\left( -D_\mu\partial^\mu\uptheta^5 -i \partial^\mu\tilde{\uptheta}A^5_\mu + \partial^\mu\tilde{\uptheta}\partial_\mu\uptheta^5\right)
\nonumber\\
    &-i(n-2)\left(2A^5_\mu\partial^\mu\uptheta^5 -\partial_\mu\uptheta^5\partial^\mu\uptheta^5 \right)-\frac{i(n-3)}{4}\left[\gamma^\mu ,\gamma^\nu\right] \left[A^5_{[\mu},\partial_{\nu]}\uptheta^5\right]~.
\end{align}
    Besides, the exotic correction to the fourth heat kernel coefficient reads 
   \beq\label{dev-a4}
   \mr{a}_4(f,D) ={a}_4(f,D)+\delta{a}_4(f,D),
    \eeq
        where the fourth exotic deviation coefficient is given by 
         \begin{align}\label{al4}
 \!\!\!\!\!\! \delta{a}_4(f,D)=\frac1{6(4\pi)^{n/2}}\iM&\! \,{\textrm{Tr}}_V\Bigg\{f \left[\left(\delta E\right)_{;\alpha\alpha}+R\left(\delta E\right)+3\left(\delta E\right)^2\right.
\nonumber\\
  &\left.+6E\left(\delta E\right)+\frac{1}{2}\left(\delta\Omega_{\rho\sigma}\right)\left(\delta\Omega_{\rho\sigma}\right)+\Omega_{\rho\sigma}\left(\delta\Omega_{\rho\sigma}\right)\right]\Bigg\}~,
        \end{align}
The terms in (\ref{al4}) are expressed as
\begin{align}\label{dOm}
    \delta\Omega_{\rho\sigma}&=\mr{\Omega}_{\rho\sigma} - \Omega_{\rho\sigma}
\nonumber\\
    &=i\left(A_{[\rho},\partial_{\sigma]}\uptheta\right)+i\left[A^5_{[\mu},\partial_{\nu]}\uptheta^5\right]+\gamma^5\gamma^\mu\left(\gamma_{[\sigma}D_{\rho]}\partial_\mu\uptheta^5 - i\gamma_{[\sigma}\partial_{\rho]}\tilde{\uptheta}A^5_\mu\right)
\nonumber\\
    &\quad+ i\gamma^5\left(\left[A_{[\rho},\partial_{\sigma]}\uptheta^5\right] + \left[\partial_{[\rho}\uptheta,A_{\sigma]}^5\right]\right)+i\left(\left[A^5_{[\rho},\partial_{|\mu|}\uptheta^5\right]-\left[A^5_{|\mu|},\partial_{[\rho}\uptheta^5\right]\right)\gamma^\mu\gamma_{\sigma]}
\nonumber\\
    &\quad-i\gamma^\mu\left( A^5_{|\mu|}\gamma_{[\rho}\gamma^\nu\partial_{|\nu|}\uptheta^5\gamma_{\sigma]}+\partial_{|\mu|}\uptheta^5\gamma_{[\rho}\gamma^\nu \mr{A}^5_{|\nu|}\gamma_{\sigma]}\right)~,\\
\label{D2dE}
    \left(\delta E\right)_{;\alpha\alpha}&= \frac{i}{4}\left(A_{[\mu},\partial_{\nu]}\uptheta\right)_{;\alpha\alpha}\left[\gamma^\mu,\gamma^\nu \right] + \gamma^5\left( -D_\mu\partial^\mu\uptheta^5 -i \partial^\mu\tilde{\uptheta}A^5_\mu + \partial^\mu\tilde{\uptheta}\partial_\mu\uptheta^5\right)_{;\alpha\alpha}
\nonumber\\
    &\;\;\;-i(n\!-\!2)\left(2A^5_\mu\partial^\mu\uptheta^5 -\partial_\mu\uptheta^5\partial^\mu\uptheta^5 \right)_{;\alpha\alpha}\!\!-\!\frac{i(n\!-\!3)}{4}\left[\gamma^\mu ,\gamma^\nu\right]\left( \left[A^5_{[\mu},\partial_{\nu]}\uptheta^5\right]\right)_{;\alpha\alpha},\\\label{dE2}
    \left(\delta E\right)^2 &= \Bigg\{\frac{i}{4}\left(A_{[\mu},\partial_{\nu]}\uptheta\right)\left[\gamma^\mu,\gamma^\nu \right] + \gamma^5\left( -D_\mu\partial^\mu\uptheta^5 -i \partial^\mu\tilde{\uptheta}A^5_\mu + \partial^\mu\tilde{\uptheta}\partial_\mu\uptheta^5\right)\nonumber\\
    &\qquad-i(n-2)f\left(2A^5_\mu\partial^\mu\uptheta^5 -\partial_\mu\uptheta^5\partial^\mu\uptheta^5 \right)-\frac{i(n-3)}{4}\left[\gamma^\mu ,\gamma^\nu\right] \left[A^5_{[\mu},\partial_{\nu]}\uptheta^5\right]\Bigg\}
\nonumber\\
    &\times\Bigg\{\frac{i}{4}\left(A_{[\alpha},\partial_{\beta]}\uptheta\right)\left[\gamma^\alpha,\gamma^\beta \right] + \gamma^5\left( -D_\alpha\partial^\alpha\uptheta^5 -i \partial^\alpha\tilde{\uptheta}A^5_\alpha + \partial^\alpha\tilde{\uptheta}\partial_\alpha\uptheta^5\right)
\nonumber\\
    &\;\;\;-i(n\!-\!2)f\left(2A^5_\alpha\partial^\alpha\uptheta^5 -\partial_\alpha\uptheta^5\partial^\alpha\uptheta^5 \right)\!-\!\frac{i(n\!-\!3)}{4}\left[\gamma^\alpha ,\gamma^\beta\right] \left[A^5_{[\alpha},\partial_{\beta]}\uptheta^5\right]\Bigg\}~,\\\label{EdE}
    E\left(\delta E\right)&=\Bigg\{-\frac{1}{4}R + \frac{1}{4}\left[\gamma^\mu,\gamma^\nu \right]F_{\mu\nu}
+i\gamma^5 {\rm d}^\mu A_\mu^5 -(n-2)A_\mu^5 A^{5\mu} -\frac{(n-3)}{4}\left[\gamma^\mu ,\gamma^\nu\right]\left[A_\mu^5,A_\nu^5 \right]\Bigg\}
\nonumber\\
&\times\Bigg\{\frac{i}{4}A_{[\alpha},\partial_{\beta]}\uptheta\left[\gamma^\alpha,\gamma^\beta \right] + \gamma^5\left( -D_\alpha\partial^\alpha\uptheta^5 -i \partial^\alpha\tilde{\uptheta}A^5_\alpha + \partial^\alpha\tilde{\uptheta}\partial_\alpha\uptheta^5\right)
\nonumber\\
    &\;\;\;-i(n-2)\left(2A^5_\alpha\partial^\alpha\uptheta^5 -\partial_\alpha\uptheta^5\partial^\alpha\uptheta^5 \right)-\frac{i(n-3)}{4}\left[\gamma^\alpha ,\gamma^\beta\right]\left[A^5_{[\alpha},\partial_{\beta]}\uptheta^5\right]\Bigg\}~,
\end{align}

\begin{align}\label{dOm2}
\left(\delta\Omega_{\rho\sigma}\right)\left(\delta\Omega_{\rho\sigma}\right) &=\Bigg\{i\left(A_{[\rho},\partial_{\sigma]}\uptheta\right)+i\left[A^5_{[\rho},\partial_{\sigma]}\uptheta^5\right]+\gamma^5\gamma^\mu\left(\gamma_{[\sigma}D_{\rho]}\partial_\mu\uptheta^5 - i\gamma_{[\sigma}\partial_{\rho]}\tilde{\uptheta}A^5_\mu\right)
\nonumber\\
    &\qquad+ i\gamma^5\left(\left[A_{[\rho},\partial_{\sigma]}\uptheta^5\right] + \left[\partial_{[\rho}\uptheta,A_{\sigma]}^5\right]\right)+i\left(\left[A^5_{[\rho},\partial_{|\mu|}\uptheta^5\right]-\left[A^5_{|\mu|},\partial_{[\rho}\uptheta^5\right]\right)\gamma^\mu\gamma_{\sigma]}
\nonumber\\
    &\qquad-i\gamma^\mu\left( A^5_{|\mu|}\gamma_{[\rho}\gamma^\nu\partial_{|\nu|}\uptheta^5\gamma_{\sigma]}+\partial_{|\mu|}\uptheta^5\gamma_{[\rho}\gamma^\nu \mr{A}^5_{|\nu|}\gamma_{\sigma]}\right)\Bigg\}\nonumber\\
&\times \Bigg\{i\left(A_{[\rho},\partial_{\sigma]}\uptheta\right)+i\left[A^5_{[\rho},\partial_{\sigma]}\uptheta^5\right]+\gamma^5\gamma^\alpha\left(\gamma_{[\sigma}D_{\rho]}\partial_\alpha\uptheta^5 - i\gamma_{[\sigma}\partial_{\rho]}\tilde{\uptheta}A^5_\alpha\right)
\nonumber\\
    &\qquad+ i\gamma^5\left(\left[A_{[\rho},\partial_{\sigma]}\uptheta^5\right] + \left[\partial_{[\rho}\uptheta,A_{\sigma]}^5\right]\right)+i\left(\left[A^5_{[\rho},\partial_{|\alpha|}\uptheta^5\right]-\left[A^5_{|\alpha|},\partial_{[\rho}\uptheta^5\right]\right)\gamma^\alpha\gamma_{\sigma]}
\nonumber\\
    &\qquad-i\gamma^\alpha\left( A^5_{|\alpha|}\gamma_{[\rho}\gamma^\beta\partial_{|\beta|}\uptheta^5\gamma_{\sigma]}+\partial_{|\alpha|}\uptheta^5\gamma_{[\rho}\gamma^\beta \mr{A}^5_{|\beta|}\gamma_{\sigma]}\right)\Bigg\}
\end{align}
and 
\begin{align}\label{OmdOm}
\Omega_{\rho\sigma}\left(\delta\Omega_{\rho\sigma}\right)&=\Bigg\{F_{\rho\sigma}-[A_\rho^5 ,A_\sigma^5 ]-\frac 14\gamma^\mu \gamma^\nu R_{\mu\nu\rho\sigma}-i\gamma^5\gamma^\mu\left(\gamma_{[\sigma} D_{\rho]} A_\mu^5\right)
\nonumber\\
    &\qquad+i\gamma^5 A_{\rho\sigma}^5+[A_{[\rho}^5,A_{|\mu|}^5]\gamma^\mu\gamma_{\sigma]} -\gamma^\mu A_\mu^5\gamma_{[\rho} \gamma^\nu A_{|\nu|}^5 \gamma_{\sigma]}\Bigg\}
\nonumber\\
    &\times \Bigg\{i\left(A_{[\rho},\partial_{\sigma]}\uptheta\right)+i\left[A^5_{[\rho},\partial_{\sigma]}\uptheta^5\right]+\gamma^5\gamma^\alpha\left(\gamma_{[\sigma}D_{\rho]}\partial_\alpha\uptheta^5 - i\gamma_{[\sigma}\partial_{\rho]}\tilde{\uptheta}A^5_\alpha\right)
\nonumber\\
    &\qquad+ i\gamma^5\left(\left[A_{[\rho},\partial_{\sigma]}\uptheta^5\right] + \left[\partial_{[\rho}\uptheta,A_{\sigma]}^5\right]\right)+i\left(\left[A^5_{[\rho},\partial_{|\alpha|}\uptheta^5\right]-\left[A^5_{|\alpha|},\partial_{[\rho}\uptheta^5\right]\right)\gamma^\alpha\gamma_{\sigma]}
\nonumber\\
    &\qquad-i\gamma^\alpha\left( A^5_{|\alpha|}\gamma_{[\rho}\gamma^\beta\partial_{|\beta|}\uptheta^5\gamma_{\sigma]}+\partial_{|\alpha|}\uptheta^5\gamma_{[\rho}\gamma^\beta \mr{A}^5_{|\beta|}\gamma_{\sigma]}\right)\Bigg\}
\end{align}

Finally, the sixth exotic deviation coefficient 
can be derived from the analysis of the sixth exotic heat kernel  coefficient, (\ref{sixthe}), as
       \beq\label{dev-a6}
    \mr{a}_6(f,D)={a}_6(f,D)+\delta{a}_6(f,D),
        \eeq
    where 
\begin{align}
 \delta{a}_6(f,D)=&\frac1{360(4\pi)^{n/2}}\iM \,{\textrm{Tr}}_V\Bigg\{f \Big\{8\left[(\delta\Cur_{\rho\sigma})_{;\alpha}(\delta\Cur_{\rho\sigma})_{;\alpha}
+ 2(\delta\Cur_{\rho\sigma})_{;\alpha}\Cur_{\rho\sigma;\alpha}\right]
\nonumber\\
    &+2\left[(\delta\Cur_{\rho\sigma})_{;\sigma}(\delta\Cur_{\rho\alpha})_{;\alpha}+2(\delta\Cur_{\rho\alpha})_{;\alpha}\Cur_{\rho\sigma;\sigma}\right]+12\left[(\delta\Cur_{\rho\sigma})\Cur_{\rho\sigma;\alpha\alpha}+(\delta\Cur_{\rho\sigma})_{;\alpha\alpha}\Cur_{\rho\sigma}\right.\nonumber\\
    &\left.+(\delta\Cur_{\rho\sigma})_{;\alpha\alpha}(\delta\Cur_{\rho\sigma})\right]-12\left[3\Cur_{\rho\sigma}\Cur_{\sigma\alpha}(\delta\Cur_{\alpha\rho}) + 3\Cur_{\rho\sigma}(\delta\Cur_{\sigma\alpha})(\delta\Cur_{\alpha\rho})+12R_{;\alpha}(\delta E)_{;\alpha}\right.\nonumber\\
    & \left.+ (\delta\Cur_{\rho\sigma})(\delta\Cur_{\sigma\alpha})(\delta\Cur_{\alpha\rho})\right]
-6R_{\rho\sigma\alpha\zeta}\left[2\Cur_{\rho\sigma}(\delta\Cur_{\alpha\zeta})+(\delta\Cur_{\rho\sigma})(\delta\Cur_{\alpha\zeta})\right]+4R_{\sigma\alpha}(\delta E)_{;\sigma\alpha}
\nonumber\\
    &-4R_{\sigma\alpha}\left[2\Cur_{\sigma\zeta}(\delta\Cur_{\alpha\zeta})+(\delta\Cur_{\sigma\zeta})(\delta\Cur_{\alpha\zeta})\right]
+5R\left[2\Cur_{\alpha\zeta}(\delta\Cur_{\alpha\zeta})+(\delta\Cur_{\alpha\zeta})(\delta\Cur_{\alpha\zeta})\right]
\nonumber\\
    &+6(\delta E)_{;\rho\rho\sigma\sigma}+60\left[(\delta E)E_{;\rho\rho}+E (\delta E)_{;\rho\rho}+(\delta E)(\delta E)_{;\rho\rho}\right] +10R(\delta E)_{;\alpha\alpha}
\nonumber\\
    &+30\left[2E_{;\rho}(\delta E)_{;\rho}+(\delta E)_{;\rho}(\delta E)_{;\rho}\right]+60\left[3E^2(\delta E)+3E(\delta E)^2+(\delta E)^3\right]
\nonumber\\
    &+30(\delta E)\left[\Cur_{\rho\sigma}\Cur_{\rho\sigma}+2\Cur_{\rho\sigma}(\delta\Cur_{\rho\sigma})+(\delta\Cur_{\rho\sigma})(\delta\Cur_{\rho\sigma})\right]
+30R\left[2E(\delta E)+(\delta E)^2\right]
\nonumber\\
    &+(\delta E)\left(12R_{;\alpha\alpha}+5R^2-2R_{\rho\sigma}R_{\rho\sigma}
+2R_{\rho\sigma\alpha\beta}R_{\rho\sigma\alpha\beta}\right)\Big\}\Bigg\}~,\label{al6}
 \end{align}
where the exotic terms in Eq. (\ref{al6}) are displayed in Appendix \ref{app}, throughout Eqs. (\ref{DdOm2} -- \ref{dEThetaR}).

All the even exotic $\mr{a}_{2i}(f,D)$ heat kernel coefficients can be similarly computed for any manifold with non-trivial topology. 
The further steps consist of computing the coefficient $\mr{a}_8(f,D)$, based on Ref.~\cite{avra}, whereas $\mr{a}_{10}(f,D)$ can be analogously derived, using the general formula of the coefficient ${a}_{10}(f,D)$ in Ref.~\cite{vandeVen:1997pf}. However, 
it is not our aim here, since the computations of the ${a}_{6}(f,D)$ coefficient involved already ten pages of Appendix \ref{app}.

\section{Conclusions}
\label{concl}
\textcolor{black}{This work was motivated by the  heat kernel spectral approach to understand specific geometrical and topological properties of non-trivial manifolds without boundaries, employing exotic spin structures. It was implemented through the exotic additional term described by an integer in a $\check{\textrm C}$ech cohomology.  In fact, the form of the exotic Dirac operator is made explicit in Eq. (\ref{do}). Therefore one can use this form to describe the equations of motion, whatever the spinor field is \cite{daRocha:2011yr,Bernardini:2012sc}.} 
\textcolor{black}{The formalism regarding exotic spinor fields that arise  from inequivalent spin structures, on non-trivial topological manifolds endowed with a metric of arbitrary signature, was extended for any finite dimension. The exotic corrections to the heat kernel coefficients, that relate spectral properties of exotic Dirac operators to the geometric invariants of $M$, were implemented by the exotic heat kernel deviation coefficients. The four first exotic heat kernel even coefficients were derived and their respective deviations from the standard ones were calculated, involving Eq.~\eqref{dev-a2},~\eqref{dev-a4} and~\eqref{dev-a6}.}

\textcolor{black}{The terms carrying the exotic fields are signatures of the influence of the exotic spin structures, which is meticulously demonstrated in Eqs.~\eqref{dE}, \eqref{D2dE} -- \eqref{OmdOm}, and \eqref{DdOm2}--Eq.~\eqref{dEThetaR}. 
These derivations permits, as expected, to recover the standard heat kernel coefficients when such fields go the exotic corrections vanish, when the manifolds are simply connected. One of the most important perspective of this work, consequently, is the future possibility to investigate the respective dynamics of such exotic contributions, which were explicitly computed.}

\acknowledgments

 RdR~is grateful to FAPESP (Grant No.  2017/18897-8) and to the National Council for Scientific and Technological Development  -- CNPq (Grants No. 303390/2019-0, No. 406134/2018-9 and No. 303293/2015-2), for partial financial support. AAT thanks to PNPD -- CAPES -- UFABC (Proc. Ns. 88882.315246/2013-01 and 88882.315253/2019-01) and PNPD -- CAPES -- UFF (Proc. N. 88887.473671/2020-00). The authors are grateful to Prof. D. Vassilevich for fruitful discussions.

 \appendix
 \section{Auxiliary computations to the exotic heat kernel coefficients}
 \label{app}
The exotic field strength can be obtained as
\begin{align}\label{FmunuExotic1}
 \mathring{F}_{\mu\nu} & = \partial_\mu\mathring{A}_\nu - \partial_\nu\mathring{A}_\mu + [\mathring{A}_\mu,\mathring{A}_\nu]\nonumber\\
                       & = \partial_\mu(A_\nu \textcolor{black}{+i\partial_\nu\uptheta})-\partial_\nu(A_\mu \textcolor{black}{+i\partial_\mu\uptheta})+[A_\mu +\textcolor{black}{i\partial_\mu\uptheta},A_\nu \textcolor{black}{+i\partial_\nu\uptheta}]\nonumber\\
                       & = \underbrace{\partial_\mu A_\nu - \partial_\nu A_\mu +[A_\mu,A_\nu]}_{=F_{\mu\nu}} \textcolor{black}{+ \underbrace{i[\partial_\mu,\partial_\nu]\uptheta}_{=0}+[A_\mu,i\partial_\nu\uptheta] + \underbrace{[i\partial_\mu\uptheta,A_\nu]}_{=-[A_\nu,i\partial_\mu\uptheta]} - \underbrace{\left[\partial_\mu,\partial_\nu\right]\uptheta}_{=0}}\nonumber\\
                       & = F_{\mu\nu} \textcolor{black}{+\left[A_{[\mu},i\partial_{\nu]}\uptheta\right]}~,
\end{align}

  Thus, the exotic connection must be defined as
\begin{align}
    \mathring{\omega}_\mu & =\frac{1}{8}\left[\gamma_\nu,\gamma_\rho\right]\sigma_\mu^{\nu\rho} + \mathring{A}_\mu + \frac{i}{2}\left[\gamma_\mu,\gamma_\nu\right]\mathring{A}^{5\nu}\gamma^5~\nonumber\\
        & = \frac{1}{8}\left[\gamma_\nu,\gamma_\rho\right]\sigma_\mu^{\nu\rho} + A_\mu \textcolor{black}{+i\partial_\mu\uptheta} + \frac{i}{2}\left[\gamma_\mu,\gamma_\nu\right]\left({A}^{5\nu} \textcolor{black}{+ i\partial^\nu\uptheta^5}\right)\gamma^5~\nonumber\\
        & = \omega_\mu +  \textcolor{black}{\partial_\mu\uptheta-\frac{i}{2}\left[\gamma_\mu,\gamma_\nu\right]\partial_\mu\uptheta^5\gamma^5}~.\label{exomeg}
\end{align}
 
 Besides, the exotic version of the field strength (\ref{Omega}) reads 
\begin{align}\label{OmegaExotic}
    \mathring{\Omega}_{\mu\nu} & = \mathring{F}_{\mu\nu}-\left[\mathring{A}_\mu^5 ,\mathring{A}_\nu^5\right]-\frac{1}{4}\gamma^\sigma \gamma^\rho R_{\sigma\rho\mu\nu} 
-i\gamma^5\gamma^\rho\gamma_{[\nu}\mathring{D}_{\mu]} \mathring{A}_\rho^5 + i\gamma^5\mathring{A}_{\mu\nu}^5 \nonumber\\
& \qquad\qquad+ \left[\mathring{A}_{[\mu}^5,\mathring{A}_{|\rho|}^5\right]\gamma^\rho \gamma_{\nu]} - \gamma^\rho \mathring{A}_\rho^5 \gamma_{[\mu}\gamma^\sigma \mathring{A}_{|\sigma|}^5 \gamma_{\nu]}~.
\end{align}
As well as the field strength calculated previously, we go now to determine each term of the full field strength separately, as follows:
\begin{align}\label{A5comut}
    \left[\mathring{A}_\mu^5 ,\mathring{A}_\nu^5\right] &= \left[A^5_\mu + i\partial_\mu\uptheta^5,A^5_\nu + i\partial_\nu\uptheta^5\right]\nonumber\\
    &= \left[A^5_\mu,A^5_\nu\right] \textcolor{black}{+ i\left(\left[A^5_\mu,\partial_\nu\uptheta^5\right] + \left[\partial_\mu\uptheta^5,A^5_\nu\right]\right) - \underbrace{\left[\partial_\mu,\partial_\nu\right]\uptheta^5}_{=0}}\nonumber\\
    &= \left[A^5_\mu,A^5_\nu\right] \textcolor{black}{+i\left[A^5_{[\mu},\partial_{\nu]}\uptheta^5\right]}~;
\end{align}

First, the commutator,
\begin{align}\label{A5munu}
    \mathring{A}_{\mu\nu}^5 &= \partial_{[\mu}\mathring{A}_{\nu]}^5 +\left[\mathring{A}_{[\mu},\mathring{A}_{\nu]}^5\right]\nonumber\\
     &= \partial_{[\mu}\mathring{A}_{\nu]}^5 \textcolor{black}{+\underbrace{i\partial_{[\mu}\partial_{\nu]}\uptheta^5}_{=0}} + \left[A_{[\mu},A^5_{\nu]}\right]  \textcolor{black}{+i\left[A_{[\mu},\partial_{\nu]}\uptheta^5\right] + i\left[\partial_{[\mu}\uptheta,A_{\nu]}^5\right] - \underbrace{\left[\partial_{[\mu}\uptheta,\partial_{\nu]}\uptheta^5\right]}_{=0}}\nonumber\\
      &= A_{\mu\nu}^5 \textcolor{black}{+ i\left(\left[A_{[\mu},\partial_{\nu]}\uptheta^5\right] + \left[\partial_{[\mu}\uptheta,A_{\nu]}^5\right]\right)}~.
\end{align}
Second, considering
\begin{align}
    \mathring{D}_\mu\mathring{A}^5_\rho &= \left(D_\mu - \partial_\mu\tilde{\uptheta}\right) \left(A^5_\rho + i\partial_\rho\uptheta^5\right)\nonumber\\
     &= D_\mu A^5_\rho+ iD_\mu\partial_\rho\uptheta^5 - \partial_\mu\tilde{\uptheta}A^5_\rho - i\partial_\mu\tilde{\uptheta}\partial_\rho\uptheta^5~,
\end{align}
yields 
\begin{align}\label{DA5}
    -i\gamma^5\gamma^\rho\gamma_{[\nu}\mathring{D}_{\mu]}\mathring{A}^5_\rho &= -i\gamma^5\gamma^\rho\gamma_{[\nu}\left(D_{\mu]}A^5_\rho + iD_{\mu]}\partial_\rho\uptheta^5 -  \partial_{\mu]}\tilde{\uptheta}A^5_\rho\right)\nonumber\\
     &= -i\gamma^5\gamma^\rho\gamma_{[\nu}D_{\mu]}A^5_\rho +\gamma^5\gamma^\rho\gamma_{[\nu}D_{\mu]}\partial_\rho\uptheta^5 - i\gamma^5\gamma^\rho\gamma_{[\nu}\partial_{\mu]}\tilde{\uptheta}A^5_\rho~.
\end{align}
Using the commutator obtained in \eqref{A5comut}, we get
\beq\label{A5comut2}
    \left[\mathring{A}^5_{[\mu},\mathring{A}^5_{|\rho|}\right]\gamma^\rho\gamma_{\nu]} = \left[A^5_{[\mu},A^5_{|\rho|}\right]\gamma^\rho\gamma_{\nu]} \textcolor{black}{+i\left(\left[A^5_{[\mu},\partial_{|\rho|}\uptheta^5\right]-\left[A^5_{|\rho|},\partial_{[\mu}\uptheta^5\right]\right)\gamma^\rho\gamma_{\nu]}}~.
\eeq
The last computation becomes
\begin{align}\label{A5A5}
    -\gamma^\rho\mathring{A}^5_{|\rho|}\gamma_{[\mu}\gamma^\sigma\mathring{A}^5_{|\sigma|}\gamma_{\nu]} &= -\gamma^\rho\left(A^5_{|\rho|}+i\partial_{|\rho|}\uptheta^5\right)\gamma_{[\mu}\gamma^\sigma\left(A^5_{|\sigma|}+i\partial_{|\sigma|}\uptheta^5\right)\gamma_{\nu]}\nonumber\\
     &= -\gamma^\rho A^5_{|\rho|}\gamma_{[\mu}\gamma^\sigma A^5_{|\sigma|}\gamma_{\nu]} \textcolor{black}{-i\gamma^\rho A^5_{|\rho|}\gamma_{[\mu}\gamma^\sigma\partial_{|\sigma|}\uptheta^5\gamma_{\nu]}-i\gamma^\rho\partial_{|\rho|}\uptheta^5\gamma_{[\mu}\gamma^\sigma\mathring{A}^5_{|\sigma|}\gamma_{\nu]}}~.
\end{align}

Now, the terms involving sixth exotic deviation coefficient (\ref{sixthe}), read 
\begin{align}\label{DdOm2}
    (\delta\Cur_{\rho\sigma})_{;\alpha}(\delta\Cur_{\rho\sigma})_{;\alpha}&=\Bigg\{i\left(A_{[\rho},\partial_{\sigma]}\uptheta\right)_{;\alpha}\!+\!i\left[A^5_{[\rho},\partial_{\sigma]}\uptheta^5\right]_{;\alpha}\!+\!\gamma^5\gamma^\beta\left(\gamma_{[\sigma}D_{\rho]}\partial_\beta\uptheta^5\right)_{;\alpha} \!-\! i\gamma^5\gamma^\beta\left(\gamma_{[\sigma}\partial_{\rho]}\tilde{\uptheta}A^5_\beta\right)_{;\alpha}
\nonumber\\
    &\;\;+ i\gamma^5\left(\left[A_{[\rho},\partial_{\sigma]}\uptheta^5\right]_{;\alpha} \!+\! \left[\partial_{[\rho}\uptheta,A_{\sigma]}^5\right]_{;\alpha}\right)\!+\!i\left(\left[A^5_{[\rho},\partial_{|\beta|}\uptheta^5\right]_{;\alpha}\!-\!\left[A^5_{|\beta|},\partial_{[\rho}\uptheta^5\right]_{;\alpha}\right)\gamma^\beta\gamma_{\sigma]}
\nonumber\\
    &\quad-i\gamma^\beta\left( A^5_{|\beta|}\gamma_{[\rho}\gamma^\zeta\partial_{|\zeta|}\uptheta^5\gamma_{\sigma]}+\partial_{|\beta|}\uptheta^5\gamma_{[\rho}\gamma^\zeta\mr{A}^5_{|\zeta|}\gamma_{\sigma]}\right)_{;\alpha}\Bigg\}
\nonumber\\
    &\times \Bigg\{i\left(A_{[\rho},\partial_{\sigma]}\uptheta\right)_{;\alpha}+i\left[A^5_{[\rho},\partial_{\sigma]}\uptheta^5\right]_{;\alpha}+\gamma^5\gamma^\lambda\left(\gamma_{[\sigma}D_{\rho]}\partial_\lambda\uptheta^5\right)_{;\alpha} - i\gamma^5\gamma^\lambda\left(\gamma_{[\sigma}\partial_{\rho]}\tilde{\uptheta}A^5_\lambda\right)_{;\alpha}
\nonumber\\
    &\;\;+ i\gamma^5\left(\left[A_{[\rho},\partial_{\sigma]}\uptheta^5\right]_{;\alpha} \!+\! \left[\partial_{[\rho}\uptheta,A_{\sigma]}^5\right]_{;\alpha}\right)\!+\!i\left(\left[A^5_{[\rho},\partial_{|\lambda|}\uptheta^5\right]_{;\alpha}\!-\!\left[A^5_{|\lambda|},\partial_{[\rho}\uptheta^5\right]_{;\alpha}\right)\gamma^\lambda\gamma_{\sigma]}
\nonumber\\
    &\;\;-i\gamma^\lambda\left(A^5_{|\lambda|}\gamma_{[\rho}\gamma^\chi\partial_{|\chi|}\uptheta^5\gamma_{\sigma]}\!+\!\partial_{|\lambda|}\uptheta^5\gamma_{[\rho}\gamma^\chi\mr{A}^5_{|\chi|}\gamma_{\sigma]}\right)_{;\alpha}\Bigg\}~,
\end{align}

\begin{align}\label{DdOmDOm}
    (\delta\Cur_{\rho\sigma})_{;\alpha}\Cur_{\rho\sigma;\alpha}&=\Bigg\{i\left(A_{[\rho},\partial_{\sigma]}\uptheta\right)_{;\alpha}+i\left[A^5_{[\rho},\partial_{\sigma]}\uptheta^5\right]_{;\alpha}+\gamma^5\gamma^\beta\left(\gamma_{[\sigma}D_{\rho]}\partial_\beta\uptheta^5\right)_{;\alpha} - i\gamma^5\gamma^\beta\left(\gamma_{[\sigma}\partial_{\rho]}\tilde{\uptheta}A^5_\beta\right)_{;\alpha}
\nonumber\\
    &\quad+ i\gamma^5\left(\left[A_{[\rho},\partial_{\sigma]}\uptheta^5\right]_{;\alpha} + \left[\partial_{[\rho}\uptheta,A_{\sigma]}^5\right]_{;\alpha}\right)+i\left(\left[A^5_{[\rho},\partial_{|\beta|}\uptheta^5\right]_{;\alpha}-\left[A^5_{|\beta|},\partial_{[\rho}\uptheta^5\right]_{;\alpha}\right)\gamma^\beta\gamma_{\sigma]}
\nonumber\\
    &\quad-i\gamma^\beta\left( A^5_{|\beta|}\gamma_{[\rho}\gamma^\zeta\partial_{|\zeta|}\uptheta^5\gamma_{\sigma]}+\partial_{|\beta|}\uptheta^5\gamma_{[\rho}\gamma^\zeta\mr{A}^5_{|\zeta|}\gamma_{\sigma]}\right)_{;\alpha}\Bigg\}
\nonumber\\
    &\times \Bigg\{F_{\rho\sigma;\alpha}-[A_\rho^5 ,A_\sigma^5 ]_{;\alpha}-\frac 14\gamma^\mu \gamma^\nu R_{\mu\nu\rho\sigma;\alpha}-i\gamma^5\gamma^\mu\left(\gamma_{[\sigma} D_{\rho]} A_\mu^5\right)_{;\alpha}
\nonumber\\
    &+i\gamma^5 A_{\rho\sigma;\alpha}^5+\left([A_{[\rho}^5,A_{|\mu|}^5]\gamma^\mu\gamma_{\sigma]}\right)_{;\alpha} -\gamma^\mu\left( A_\mu^5\gamma_{[\rho} \gamma^\nu A_{|\nu|}^5 \gamma_{\sigma]}\right)_{;\alpha}\Bigg\}~,
\end{align}

\begin{align}\label{DdOmDdOm}
    (\delta\Cur_{\rho\sigma})_{;\sigma}(\delta\Cur_{\rho\alpha})_{;\alpha}
    &=\Bigg\{i\left(A_{[\rho},\partial_{\sigma]}\uptheta\right)_{;\sigma}+i\left[A^5_{[\rho},\partial_{\sigma]}\uptheta^5\right]_{;\sigma}+\gamma^5\gamma^\beta\left(\gamma_{[\sigma}D_{\rho]}\partial_\beta\uptheta^5 - i\gamma_{[\sigma}\partial_{\rho]}\tilde{\uptheta}A^5_\beta\right)_{;\sigma}
\nonumber\\
    &\quad+ i\gamma^5\left(\left[A_{[\rho},\partial_{\sigma]}\uptheta^5\right]_{;\sigma} + \left[\partial_{[\rho}\uptheta,A_{\sigma]}^5\right]_{;\sigma}\right)+i\left(\left[A^5_{[\rho},\partial_{|\beta|}\uptheta^5\right]_{;\sigma}-\left[A^5_{|\beta|},\partial_{[\rho}\uptheta^5\right]_{;\sigma}\right)\gamma^\beta\gamma_{\sigma]}
\nonumber\\
    &\quad-i\gamma^\beta\left( A^5_{|\beta|}\gamma_{[\rho}\gamma^\zeta\partial_{|\zeta|}\uptheta^5\gamma_{\sigma]}+\partial_{|\beta|}\uptheta^5\gamma_{[\rho}\gamma^\zeta\mr{A}^5_{|\zeta|}\gamma_{\sigma]}\right)_{;\sigma}\Bigg\}
\nonumber\\
    &\times \Bigg\{i\left(A_{[\rho},\partial_{\alpha]}\uptheta\right)_{;\alpha}+i\left[A^5_{[\rho},\partial_{\alpha]}\uptheta^5\right]_{;\alpha}+\gamma^5\gamma^\lambda\left(\gamma_{[\alpha}D_{\rho]}\partial_\lambda\uptheta^5\right)_{;\alpha} - i\gamma^5\gamma^\lambda\left(\gamma_{[\alpha}\partial_{\rho]}\tilde{\uptheta}A^5_\lambda\right)_{;\alpha}
\nonumber\\
    &\quad+ i\gamma^5\left(\left[A_{[\rho},\partial_{\alpha]}\uptheta^5\right]_{;\alpha} + \left[\partial_{[\rho}\uptheta,A_{\alpha]}^5\right]_{;\alpha}\right)+i\left(\left[A^5_{[\rho},\partial_{|\lambda|}\uptheta^5\right]_{;\alpha}-\left[A^5_{|\lambda|},\partial_{[\rho}\uptheta^5\right]_{;\alpha}\right)\gamma^\lambda\gamma_{\alpha]}
\nonumber\\
    &\quad-i\gamma^\lambda\left( A^5_{|\lambda|}\gamma_{[\rho}\gamma^\chi\partial_{|\chi|}\uptheta^5\gamma_{\alpha]}+\partial_{|\lambda|}\uptheta^5\gamma_{[\rho}\gamma^\chi\mr{A}^5_{|\chi|}\gamma_{\alpha]}\right)_{;\alpha}\Bigg\}~,
\end{align}

\begin{align}\label{DdOm-Om}
    (\delta\Cur_{\rho\alpha})_{;\alpha}\Cur_{\rho\sigma;\sigma}&=\Bigg\{i\left(A_{[\rho},\partial_{\alpha]}\uptheta\right)_{;\alpha}+i\left[A^5_{[\rho},\partial_{\alpha]}\uptheta^5\right]_{;\alpha}+\gamma^5\gamma^\beta\left(\gamma_{[\alpha}D_{\rho]}\partial_\beta\uptheta^5 - i\gamma_{[\alpha}\partial_{\rho]}\tilde{\uptheta}A^5_\beta\right)_{;\alpha}
\nonumber\\
    &\quad+ i\gamma^5\left(\left[A_{[\rho},\partial_{\alpha]}\uptheta^5\right]_{;\alpha} + \left[\partial_{[\rho}\uptheta,A_{\alpha]}^5\right]_{;\alpha}\right)+i\left(\left[A^5_{[\rho},\partial_{|\beta|}\uptheta^5\right]_{;\alpha}-\left[A^5_{|\beta|},\partial_{[\rho}\uptheta^5\right]_{;\alpha}\right)\gamma^\beta\gamma_{\alpha]}
\nonumber\\
    &\quad-i\gamma^\beta\left( A^5_{|\beta|}\gamma_{[\rho}\gamma^\zeta\partial_{|\zeta|}\uptheta^5\gamma_{\alpha]}+\partial_{|\beta|}\uptheta^5\gamma_{[\rho}\gamma^\zeta\mr{A}^5_{|\zeta|}\gamma_{\alpha]}\right)_{;\alpha}\Bigg\}
\nonumber\\
    &\times \Bigg\{F_{\rho\sigma;\sigma}-[A_\rho^5 ,A_\sigma^5 ]_{;\sigma}-\frac 14\gamma^\mu \gamma^\nu R_{\mu\nu\rho\sigma;\sigma}-i\gamma^5\gamma^\mu\left(\gamma_{[\sigma} D_{\rho]} A_\mu^5\right)_{;\sigma}
\nonumber\\
    &+i\gamma^5 A_{\rho\sigma;\sigma}^5+\left([A_{[\rho}^5,A_{|\mu|}^5]\gamma^\mu\gamma_{\sigma]}\right)_{;\alpha} -\gamma^\mu\left(A_\mu^5 \gamma_{[\rho} \gamma^\nu A_{|\nu|}^5 \gamma_{\sigma]}\right)_{;\sigma}\Bigg\}~,
\end{align}

\begin{align}
    (\delta\Cur_{\rho\sigma})\Cur_{\rho\sigma;\alpha\alpha}&=\Bigg\{i\left(A_{[\rho},\partial_{\sigma]}\uptheta\right)+i\left[A^5_{[\rho},\partial_{\sigma]}\uptheta^5\right]+\gamma^5\gamma^\beta\left(\gamma_{[\sigma}D_{\rho]}\partial_\beta\uptheta^5 - i\gamma_{[\sigma}\partial_{\rho]}\tilde{\uptheta}A^5_\beta\right)
\nonumber\\
    &\quad+ i\gamma^5\left(\left[A_{[\rho},\partial_{\sigma]}\uptheta^5\right] + \left[\partial_{[\rho}\uptheta,A_{\sigma]}^5\right]\right)+i\left(\left[A^5_{[\rho},\partial_{|\beta|}\uptheta^5\right]-\left[A^5_{|\beta|},\partial_{[\rho}\uptheta^5\right]\right)\gamma^\beta\gamma_{\sigma]}
\nonumber\\
    &\quad-i\gamma^\beta\left( A^5_{|\beta|}\gamma_{[\rho}\gamma^\lambda\partial_{|\lambda|}\uptheta^5\gamma_{\sigma]}+\partial_{|\beta|}\uptheta^5\gamma_{[\rho}\gamma^\lambda \mr{A}^5_{|\lambda|}\gamma_{\sigma]}\right)\Bigg\}
\nonumber\\
    &\times \Bigg\{F_{\rho\sigma;\alpha\alpha}-[A_\rho^5 ,A_\sigma^5 ]_{;\alpha\alpha}-\frac 14\gamma^\mu \gamma^\nu R_{\mu\nu\rho\sigma;\alpha\alpha}-i\gamma^5\gamma^\mu\left(\gamma_{[\sigma} D_{\rho]} A_\mu^5\right)_{;\alpha\alpha}
\nonumber\\
    &+i\gamma^5 A_{\rho\sigma;\alpha\alpha}^5+\left([A_{[\rho}^5,A_{|\mu|}^5]\gamma^\mu\gamma_{\sigma]}\right)_{;\alpha\alpha} -\gamma^\mu\left( A_\mu^5\gamma_{[\rho} \gamma^\nu A_{|\nu|}^5 \gamma_{\sigma]}\right)_{;\alpha\alpha}\Bigg\}~, 
\end{align}

\begin{align}\label{D2dOmOm}
    (\delta\Cur_{\rho\sigma})_{;\alpha\alpha}\Cur_{\rho\sigma}&=\Bigg\{i\left(A_{[\rho},\partial_{\sigma]}\uptheta\right)_{;\alpha\alpha}+i\left[A^5_{[\rho},\partial_{\sigma]}\uptheta^5\right]_{;\alpha\alpha}+\gamma^5\gamma^\beta\left(\gamma_{[\sigma}D_{\rho]}\partial_\beta\uptheta^5 - i\gamma_{[\sigma}\partial_{\rho]}\tilde{\uptheta}A^5_\beta\right)_{;\alpha\alpha}
\nonumber\\
    &\quad+ i\gamma^5\left(\left[A_{[\rho},\partial_{\sigma]}\uptheta^5\right]_{;\alpha\alpha} + \left[\partial_{[\rho}\uptheta,A_{\sigma]}^5\right]_{;\alpha\alpha}\right)+i\left(\left[A^5_{[\rho},\partial_{|\beta|}\uptheta^5\right]_{;\alpha\alpha}-\left[A^5_{|\beta|},\partial_{[\rho}\uptheta^5\right]_{;\alpha\alpha}\right)\gamma^\beta\gamma_{\sigma]}
\nonumber\\
    &\quad-i\gamma^\beta\left( A^5_{|\beta|}\gamma_{[\rho}\gamma^\zeta\partial_{|\zeta|}\uptheta^5\gamma_{\sigma]}+\partial_{|\beta|}\uptheta^5\gamma_{[\rho}\gamma^\zeta\mr{A}^5_{|\zeta|}\gamma_{\sigma]}\right)_{;\alpha\alpha}\Bigg\}
\nonumber\\
    &\times \Bigg\{F_{\rho\sigma}-[A_\rho^5 ,A_\sigma^5 ]-\frac 14\gamma^\mu \gamma^\nu R_{\mu\nu\rho\sigma}-i\gamma^5\gamma^\mu\gamma_{[\sigma} D_{\rho]} A_\mu^5
\nonumber\\
    &\qquad+i\gamma^5 A_{\rho\sigma}^5+[A_{[\rho}^5,A_{|\mu|}^5]\gamma^\mu\gamma_{\sigma]} -\gamma^\mu A_\mu^5\gamma_{[\rho} \gamma^\nu A_{|\nu|}^5 \gamma_{\sigma]}\Bigg\}~, 
\end{align}

\begin{align}\label{D2dOmdOm}
    (\delta\Cur_{\rho\sigma})_{;\alpha\alpha}(\delta\Cur_{\rho\sigma})&=\Bigg\{i\left(A_{[\rho},\partial_{\sigma]}\uptheta\right)_{;\alpha\alpha}+i\left[A^5_{[\rho},\partial_{\sigma]}\uptheta^5\right]_{;\alpha\alpha}+\gamma^5\gamma^\beta\left(\gamma_{[\sigma}D_{\rho]}\partial_\beta\uptheta^5 - i\gamma_{[\sigma}\partial_{\rho]}\tilde{\uptheta}A^5_\beta\right)_{;\alpha\alpha}
\nonumber\\
    &\quad+ i\gamma^5\left(\left[A_{[\rho},\partial_{\sigma]}\uptheta^5\right]_{;\alpha\alpha} + \left[\partial_{[\rho}\uptheta,A_{\sigma]}^5\right]_{;\alpha\alpha}\right)+i\left(\left[A^5_{[\rho},\partial_{|\beta|}\uptheta^5\right]_{;\alpha\alpha}-\left[A^5_{|\beta|},\partial_{[\rho}\uptheta^5\right]_{;\alpha\alpha}\right)\gamma^\beta\gamma_{\sigma]}
\nonumber\\
    &\quad-i\gamma^\beta\left( A^5_{|\beta|}\gamma_{[\rho}\gamma^\mu\partial_{|\mu|}\uptheta^5\gamma_{\sigma]}+\partial_{|\beta|}\uptheta^5\gamma_{[\rho}\gamma^\mu\mr{A}^5_{|\mu|}\gamma_{\sigma]}\right)_{;\alpha\alpha}\Bigg\}
\nonumber\\
    &\times \Bigg\{i\left(A_{[\rho},\partial_{\sigma]}\uptheta\right)+i\left[A^5_{[\rho},\partial_{\sigma]}\uptheta^5\right]+\gamma^5\gamma^\nu\left(\gamma_{[\sigma}D_{\rho]}\partial_\nu\uptheta^5 - i\gamma_{[\sigma}\partial_{\rho]}\tilde{\uptheta}A^5_\nu\right)
\nonumber\\
    &\quad+ i\gamma^5\left(\left[A_{[\rho},\partial_{\sigma]}\uptheta^5\right] + \left[\partial_{[\rho}\uptheta,A_{\sigma]}^5\right]\right)+i\left(\left[A^5_{[\rho},\partial_{|\nu|}\uptheta^5\right]-\left[A^5_{|\nu|},\partial_{[\rho}\uptheta^5\right]\right)\gamma^\nu\gamma_{\sigma]}
\nonumber\\
    &\quad-i\gamma^\nu\left( A^5_{|\nu|}\gamma_{[\rho}\gamma^\lambda\partial_{|\lambda|}\uptheta^5\gamma_{\sigma]}+\partial_{|\nu|}\uptheta^5\gamma_{[\rho}\gamma^\lambda \mr{A}^5_{|\lambda|}\gamma_{\sigma]}\right)\Bigg\}~, 
\end{align}

\begin{align}\label{Om-Om-dOm}
    \Cur_{\rho\sigma}\Cur_{\sigma\alpha}(\delta\Cur_{\alpha\rho})&=\Bigg\{F_{\rho\sigma}-[A_\rho^5 ,A_\sigma^5 ]-\frac 14\gamma^\mu\gamma^\nu R_{\mu\nu\rho\sigma}-i\gamma^5\gamma^\mu\left(\gamma_{[\sigma} D_{\rho]} A_\mu^5\right)
\nonumber\\
    &\qquad+i\gamma^5 A_{\rho\sigma}^5+[A_{[\rho}^5,A_{|\mu|}^5]\gamma^\mu\gamma_{\sigma]} -\gamma^\mu A_\mu^5\gamma_{[\rho} \gamma^\nu A_{|\nu|}^5 \gamma_{\sigma]}\Bigg\}
\nonumber\\
    &\times\Bigg\{F_{\sigma\alpha}-[A_\sigma^5 ,A_\alpha^5 ]-\frac 14\gamma^\xi\gamma^\lambda R_{\xi\lambda\sigma\alpha}-i\gamma^5\gamma^\xi\left(\gamma_{[\alpha}D_{\sigma]}A_\xi^5\right)
\nonumber\\
    &\qquad+i\gamma^5 A_{\sigma\alpha}^5+[A_{[\sigma}^5,A_{|\xi|}^5]\gamma^\xi\gamma_{\alpha]} -\gamma^\xi A_\xi^5\gamma_{[\sigma}\gamma^\lambda A_{|\lambda|}^5 \gamma_{\alpha]}\Bigg\}
\nonumber\\
    &\times\Bigg\{i\left(A_{[\alpha},\partial_{\rho]}\uptheta\right)+i\left[A^5_{[\alpha},\partial_{\rho]}\uptheta^5\right]+\gamma^5\gamma^\beta\left(\gamma_{[\rho}D_{\alpha]}\partial_\beta\uptheta^5 - i\gamma_{[\rho}\partial_{\alpha]}\tilde{\uptheta}A^5_\beta\right)
\nonumber\\
    &\qquad+i\gamma^5\left(\left[A_{[\alpha},\partial_{\rho]}\uptheta^5\right] + \left[\partial_{[\alpha}\uptheta,A_{\rho]}^5\right]\right)+i\left(\left[A^5_{[\alpha},\partial_{|\beta|}\uptheta^5\right]-\left[A^5_{|\beta|},\partial_{[\rho}\uptheta^5\right]\right)\gamma^\beta\gamma_{\rho]}
\nonumber\\
    &\qquad-i\gamma^\beta\left( A^5_{|\beta|}\gamma_{[\alpha}\gamma^\chi\partial_{|\chi|}\uptheta^5\gamma_{\rho]}+\partial_{|\beta|}\uptheta^5\gamma_{[\alpha}\gamma^\chi \mr{A}^5_{|\chi|}\gamma_{\rho]}\right)\Bigg\}~,
\end{align}

\begin{align}\label{Om-dOm-dOm}
     \Cur_{\rho\sigma}(\delta\Cur_{\sigma\alpha})(\delta\Cur_{\alpha\rho})&=\Bigg\{F_{\rho\sigma}-[A_\rho^5 ,A_\sigma^5 ]-\frac 14\gamma^\mu \gamma^\nu R_{\mu\nu\rho\sigma}-i\gamma^5\gamma^\mu\left(\gamma_{[\sigma} D_{\rho]} A_\mu^5\right)
\nonumber\\
    &+i\gamma^5 A_{\rho\sigma}^5+[A_{[\rho}^5,A_{|\mu|}^5]\gamma^\mu\gamma_{\sigma]} -\gamma^\mu A_\mu^5\gamma_{[\rho} \gamma^\nu A_{|\nu|}^5 \gamma_{\sigma]}\Bigg\}
\nonumber\\
    &\times\Bigg\{i\left(A_{[\sigma},\partial_{\alpha]}\uptheta\right)+i\left[A^5_{[\sigma},\partial_{\alpha]}\uptheta^5\right]+\gamma^5\gamma^\xi\left(\gamma_{[\alpha}D_{\sigma]}\partial_\xi\uptheta^5 - i\gamma_{[\alpha}\partial_{\sigma]}\tilde{\uptheta}A^5_\xi\right)
\nonumber\\
    &\quad+ i\gamma^5\left(\left[A_{[\sigma},\partial_{\alpha]}\uptheta^5\right] + \left[\partial_{[\sigma}\uptheta,A_{\alpha]}^5\right]\right)+i\left(\left[A^5_{[\sigma},\partial_{|\xi|}\uptheta^5\right]-\left[A^5_{|\xi|},\partial_{[\sigma}\uptheta^5\right]\right)\gamma^\xi\gamma_{\alpha]}
\nonumber\\
    &\quad-i\gamma^\xi\left( A^5_{|\xi|}\gamma_{[\sigma}\gamma^\lambda\partial_{|\lambda|}\uptheta^5\gamma_{\alpha]}+\partial_{|\xi|}\uptheta^5\gamma_{[\sigma}\gamma^\lambda \mr{A}^5_{|\lambda|}\gamma_{\alpha]}\right)\Bigg\}
\nonumber\\
    &\times\Bigg\{i\left(A_{[\alpha},\partial_{\rho]}\uptheta\right)+i\left[A^5_{[\alpha},\partial_{\rho]}\uptheta^5\right]+\gamma^5\gamma^\beta\left(\gamma_{[\rho}D_{\alpha]}\partial_\beta\uptheta^5 - i\gamma_{[\rho}\partial_{\alpha]}\tilde{\uptheta}A^5_\beta\right)
\nonumber\\
    &\qquad+i\gamma^5\left(\left[A_{[\alpha},\partial_{\rho]}\uptheta^5\right] + \left[\partial_{[\alpha}\uptheta,A_{\rho]}^5\right]\right)+i\left(\left[A^5_{[\alpha},\partial_{|\beta|}\uptheta^5\right]-\left[A^5_{|\beta|},\partial_{[\alpha}\uptheta^5\right]\right)\gamma^\beta\gamma_{\rho]}
\nonumber\\
    &\qquad-i\gamma^\beta\left( A^5_{|\beta|}\gamma_{[\alpha}\gamma^\chi\partial_{|\chi|}\uptheta^5\gamma_{\rho]}+\partial_{|\beta|}\uptheta^5\gamma_{[\alpha}\gamma^\chi \mr{A}^5_{|\chi|}\gamma_{\rho]}\right)\Bigg\}~,
\end{align}

\begin{align}\label{dOm-dOm-dOm}
     (\delta\Cur_{\rho\sigma})(\delta\Cur_{\sigma\alpha})(\delta\Cur_{\alpha\rho})&=\Bigg\{i\left(A_{[\rho},\partial_{\sigma]}\uptheta\right)+i\left[A^5_{[\rho},\partial_{\sigma]}\uptheta^5\right]+\gamma^5\gamma^\mu\left(\gamma_{[\sigma}D_{\rho]}\partial_\mu\uptheta^5 - i\gamma_{[\sigma}\partial_{\rho]}\tilde{\uptheta}A^5_\mu\right)
\nonumber\\
    &\qquad+ i\gamma^5\left(\left[A_{[\rho},\partial_{\sigma]}\uptheta^5\right] + \left[\partial_{[\rho}\uptheta,A_{\sigma]}^5\right]\right)+i\left(\left[A^5_{[\rho},\partial_{|\mu|}\uptheta^5\right]-\left[A^5_{|\mu|},\partial_{[\rho}\uptheta^5\right]\right)\gamma^\mu\gamma_{\sigma]}
\nonumber\\
    &\qquad-i\gamma^\mu\left( A^5_{|\mu|}\gamma_{[\rho}\gamma^\nu\partial_{|\nu|}\uptheta^5\gamma_{\sigma]}+\partial_{|\mu|}\uptheta^5\gamma_{[\rho}\gamma^\nu \mr{A}^5_{|\nu|}\gamma_{\sigma]}\right)\Bigg\}
\nonumber\\
     &\times\Bigg\{i\left(A_{[\sigma},\partial_{\alpha]}\uptheta\right)+i\left[A^5_{[\sigma},\partial_{\alpha]}\uptheta^5\right]+\gamma^5\gamma^\xi\left(\gamma_{[\alpha}D_{\sigma]}\partial_\xi\uptheta^5 - i\gamma_{[\alpha}\partial_{\sigma]}\tilde{\uptheta}A^5_\xi\right)
\nonumber\\
    &\quad+ i\gamma^5\left(\left[A_{[\sigma},\partial_{\alpha]}\uptheta^5\right] + \left[\partial_{[\sigma}\uptheta,A_{\alpha]}^5\right]\right)+i\left(\left[A^5_{[\sigma},\partial_{|\xi|}\uptheta^5\right]-\left[A^5_{|\xi|},\partial_{[\sigma}\uptheta^5\right]\right)\gamma^\xi\gamma_{\alpha]}
\nonumber\\
    &\quad-i\gamma^\xi\left( A^5_{|\xi|}\gamma_{[\sigma}\gamma^\lambda\partial_{|\lambda|}\uptheta^5\gamma_{\alpha]}+\partial_{|\xi|}\uptheta^5\gamma_{[\sigma}\gamma^\lambda \mr{A}^5_{|\lambda|}\gamma_{\alpha]}\right)\Bigg\}
\nonumber\\
    &\times\Bigg\{i\left(A_{[\alpha},\partial_{\rho]}\uptheta\right)+i\left[A^5_{[\alpha},\partial_{\rho]}\uptheta^5\right]+\gamma^5\gamma^\beta\left(\gamma_{[\rho}D_{\alpha]}\partial_\beta\uptheta^5 - i\gamma_{[\rho}\partial_{\alpha]}\tilde{\uptheta}A^5_\beta\right)
\nonumber\\
    &\qquad+i\gamma^5\left(\left[A_{[\alpha},\partial_{\rho]}\uptheta^5\right] + \left[\partial_{[\alpha}\uptheta,A_{\rho]}^5\right]\right)+i\left(\left[A^5_{[\alpha},\partial_{|\beta|}\uptheta^5\right]-\left[A^5_{|\beta|},\partial_{[\alpha}\uptheta^5\right]\right)\gamma^\beta\gamma_{\rho]}
\nonumber\\
    &\qquad-i\gamma^\beta\left( A^5_{|\beta|}\gamma_{[\alpha}\gamma^\chi\partial_{|\chi|}\uptheta^5\gamma_{\rho]}+\partial_{|\beta|}\uptheta^5\gamma_{[\alpha}\gamma^\chi \mr{A}^5_{|\chi|}\gamma_{\rho]}\right)\Bigg\}~,
\end{align}

\begin{align}\label{Riem-Om-dOm}
    R_{\rho\sigma\alpha\zeta}\left[2\Cur_{\rho\sigma}(\delta\Cur_{\alpha\zeta})\right]&=2R_{\rho\sigma\alpha\zeta}\Bigg\{F_{\rho\sigma}-[A_\rho^5 ,A_\sigma^5 ]-\frac 14\gamma^\mu \gamma^\nu R_{\mu\nu\rho\sigma}-i\gamma^5\gamma^\mu\left(\gamma_{[\sigma} D_{\rho]} A_\mu^5\right)
\nonumber\\
    &\qquad+i\gamma^5 A_{\rho\sigma}^5+[A_{[\rho}^5,A_{|\mu|}^5]\gamma^\mu\gamma_{\sigma]} -\gamma^\mu A_\mu^5\gamma_{[\rho} \gamma^\nu A_{|\nu|}^5 \gamma_{\sigma]}\Bigg\}
\nonumber\\
    &\times\Bigg\{i\left(A_{[\alpha},\partial_{\zeta]}\uptheta\right)+i\left[A^5_{[\alpha},\partial_{\zeta]}\uptheta^5\right]+\gamma^5\gamma^\xi\left(\gamma_{[\zeta}D_{\alpha]}\partial_\xi\uptheta^5 - i\gamma_{[\zeta}\partial_{\alpha]}\tilde{\uptheta}A^5_\xi\right)
\nonumber\\
    &\qquad+ i\gamma^5\left(\left[A_{[\alpha},\partial_{\zeta]}\uptheta^5\right] + \left[\partial_{[\alpha}\uptheta,A_{\zeta]}^5\right]\right)+i\left(\left[A^5_{[\alpha},\partial_{|\xi|}\uptheta^5\right]-\left[A^5_{|\xi|},\partial_{[\alpha}\uptheta^5\right]\right)\gamma^\xi\gamma_{\zeta]}
\nonumber\\
    &\qquad-i\gamma^\xi\left( A^5_{|\xi|}\gamma_{[\alpha}\gamma^\beta\partial_{|\beta|}\uptheta^5\gamma_{\zeta]}+\partial_{|\xi|}\uptheta^5\gamma_{[\alpha}\gamma^\beta \mr{A}^5_{|\beta|}\gamma_{\zeta]}\right)\Bigg\}~,
\end{align}

\begin{align}\label{Riem-dOm-dOm}
R_{\rho\sigma\alpha\zeta}(\delta\Cur_{\rho\sigma})(\delta\Cur_{\alpha\zeta})&=R_{\rho\sigma\alpha\zeta}\Bigg\{i\left(A_{[\rho},\partial_{\sigma]}\uptheta\right)+i\left[A^5_{[\rho},\partial_{\sigma]}\uptheta^5\right]+\gamma^5\gamma^\mu\left(\gamma_{[\sigma}D_{\rho]}\partial_\mu\uptheta^5 - i\gamma_{[\sigma}\partial_{\rho]}\tilde{\uptheta}A^5_\mu\right)
\nonumber\\
    &\qquad+ i\gamma^5\left(\left[A_{[\rho},\partial_{\sigma]}\uptheta^5\right] + \left[\partial_{[\rho}\uptheta,A_{\sigma]}^5\right]\right)+i\left(\left[A^5_{[\rho},\partial_{|\mu|}\uptheta^5\right]-\left[A^5_{|\mu|},\partial_{[\rho}\uptheta^5\right]\right)\gamma^\mu\gamma_{\sigma]}
\nonumber\\
    &\qquad-i\gamma^\mu\left( A^5_{|\mu|}\gamma_{[\rho}\gamma^\nu\partial_{|\nu|}\uptheta^5\gamma_{\sigma]}+\partial_{|\mu|}\uptheta^5\gamma_{[\rho}\gamma^\nu \mr{A}^5_{|\nu|}\gamma_{\sigma]}\right)\Bigg\}\nonumber\\
&\times \Bigg\{i\left(A_{[\alpha},\partial_{\zeta]}\uptheta\right)+i\left[A^5_{[\alpha},\partial_{\zeta]}\uptheta^5\right]+\gamma^5\gamma^\xi\left(\gamma_{[\zeta}D_{\alpha]}\partial_\xi\uptheta^5 - i\gamma_{[\zeta}\partial_{\alpha]}\tilde{\uptheta}A^5_\xi\right)
\nonumber\\
    &\qquad+ i\gamma^5\left(\left[A_{[\alpha},\partial_{\zeta]}\uptheta^5\right] + \left[\partial_{[\alpha}\uptheta,A_{\zeta]}^5\right]\right)+i\left(\left[A^5_{[\alpha},\partial_{|\xi|}\uptheta^5\right]-\left[A^5_{|\xi|},\partial_{[\alpha}\uptheta^5\right]\right)\gamma^\xi\gamma_{\zeta]}
\nonumber\\
    &\qquad-i\gamma^\xi\left( A^5_{|\xi|}\gamma_{[\alpha}\gamma^\beta\partial_{|\beta|}\uptheta^5\gamma_{\zeta]}+\partial_{|\xi|}\uptheta^5\gamma_{[\alpha}\gamma^\beta \mr{A}^5_{|\beta|}\gamma_{\zeta]}\right)\Bigg\}~,
\end{align}

\begin{align}\label{R-Om-dOm2}
    R_{\sigma\alpha}\left[2\Cur_{\sigma\zeta}(\delta\Cur_{\alpha\zeta})\right]&=2R_{\sigma\alpha}\Bigg\{F_{\sigma\zeta}-[A_\sigma^5 ,A_\zeta^5 ]-\frac 14\gamma^\mu \gamma^\nu R_{\mu\nu\sigma\zeta}-i\gamma^5\gamma^\mu\left(\gamma_{[\zeta} D_{\sigma]} A_\mu^5\right)
\nonumber\\
    &\qquad+i\gamma^5 A_{\sigma\zeta}^5+[A_{[\sigma}^5,A_{|\mu|}^5]\gamma^\mu\gamma_{\zeta]} -\gamma^\mu A_\mu^5\gamma_{[\sigma} \gamma^\nu A_{|\nu|}^5 \gamma_{\zeta]}\Bigg\}
\nonumber\\
    &\times\Bigg\{i\left(A_{[\alpha},\partial_{\zeta]}\uptheta\right)+i\left[A^5_{[\alpha},\partial_{\zeta]}\uptheta^5\right]+\gamma^5\gamma^\xi\left(\gamma_{[\zeta}D_{\alpha]}\partial_\xi\uptheta^5 - i\gamma_{[\zeta}\partial_{\alpha]}\tilde{\uptheta}A^5_\xi\right)
\nonumber\\
    &\qquad+ i\gamma^5\left(\left[A_{[\alpha},\partial_{\zeta]}\uptheta^5\right] + \left[\partial_{[\alpha}\uptheta,A_{\zeta]}^5\right]\right)+i\left(\left[A^5_{[\alpha},\partial_{|\xi|}\uptheta^5\right]-\left[A^5_{|\xi|},\partial_{[\alpha}\uptheta^5\right]\right)\gamma^\xi\gamma_{\zeta]}
\nonumber\\
    &\qquad-i\gamma^\xi\left( A^5_{|\xi|}\gamma_{[\alpha}\gamma^\beta\partial_{|\beta|}\uptheta^5\gamma_{\zeta]}+\partial_{|\xi|}\uptheta^5\gamma_{[\alpha}\gamma^\beta \mr{A}^5_{|\beta|}\gamma_{\zeta]}\right)\Bigg\}~,
\end{align}

\begin{align}\label{R-dOm-dOm}
    R_{\sigma\alpha}(\delta\Cur_{\sigma\zeta})(\delta\Cur_{\alpha\zeta})&=R_{\sigma\alpha}\Bigg\{i\left(A_{[\sigma},\partial_{\zeta]}\uptheta\right)+i\left[A^5_{[\sigma},\partial_{\zeta]}\uptheta^5\right]+\gamma^5\gamma^\mu\left(\gamma_{[\zeta}D_{\sigma]}\partial_\mu\uptheta^5 - i\gamma_{[\zeta}\partial_{\sigma]}\tilde{\uptheta}A^5_\mu\right)
\nonumber\\
    &\qquad+ i\gamma^5\left(\left[A_{[\sigma},\partial_{\zeta]}\uptheta^5\right] + \left[\partial_{[\sigma}\uptheta,A_{\zeta]}^5\right]\right)+i\left(\left[A^5_{[\sigma},\partial_{|\mu|}\uptheta^5\right]-\left[A^5_{|\mu|},\partial_{[\sigma}\uptheta^5\right]\right)\gamma^\mu\gamma_{\zeta]}
\nonumber\\
    &\qquad-i\gamma^\mu\left( A^5_{|\mu|}\gamma_{[\sigma}\gamma^\nu\partial_{|\nu|}\uptheta^5\gamma_{\zeta]}+\partial_{|\mu|}\uptheta^5\gamma_{[\sigma}\gamma^\nu \mr{A}^5_{|\nu|}\gamma_{\zeta]}\right)\Bigg\}\nonumber\\
&\times \Bigg\{i\left(A_{[\alpha},\partial_{\zeta]}\uptheta\right)+i\left[A^5_{[\alpha},\partial_{\zeta]}\uptheta^5\right]+\gamma^5\gamma^\xi\left(\gamma_{[\zeta}D_{\alpha]}\partial_\xi\uptheta^5 - i\gamma_{[\zeta}\partial_{\alpha]}\tilde{\uptheta}A^5_\xi\right)
\nonumber\\
    &\qquad+ i\gamma^5\left(\left[A_{[\alpha},\partial_{\zeta]}\uptheta^5\right] + \left[\partial_{[\alpha}\uptheta,A_{\zeta]}^5\right]\right)+i\left(\left[A^5_{[\alpha},\partial_{|\xi|}\uptheta^5\right]-\left[A^5_{|\xi|},\partial_{[\alpha}\uptheta^5\right]\right)\gamma^\xi\gamma_{\zeta]}
\nonumber\\
    &\qquad-i\gamma^\xi\left( A^5_{|\xi|}\gamma_{[\alpha}\gamma^\beta\partial_{|\beta|}\uptheta^5\gamma_{\zeta]}+\partial_{|\xi|}\uptheta^5\gamma_{[\alpha}\gamma^\beta \mr{A}^5_{|\beta|}\gamma_{\zeta]}\right)\Bigg\}~,
\end{align}

\begin{align}\label{R-Om-dOm}
    R\left[2\Cur_{\alpha\zeta}(\delta\Cur_{\alpha\zeta})\right]&=2R\Bigg\{F_{\alpha\zeta}-[A_\alpha^5 ,A_\zeta^5 ]-\frac 14\gamma^\mu \gamma^\nu R_{\mu\nu\alpha\zeta}-i\gamma^5\gamma^\mu\left(\gamma_{[\zeta} D_{\alpha]} A_\mu^5\right)
\nonumber\\
    &\qquad+i\gamma^5 A_{\alpha\zeta}^5+[A_{[\alpha}^5,A_{|\mu|}^5]\gamma^\mu\gamma_{\zeta]} -\gamma^\mu A_\mu^5\gamma_{[\alpha} \gamma^\nu A_{|\nu|}^5 \gamma_{\zeta]}\Bigg\}
\nonumber\\
    &\times \Bigg\{i\left(A_{[\alpha},\partial_{\zeta]}\uptheta\right)+i\left[A^5_{[\alpha},\partial_{\zeta]}\uptheta^5\right]+\gamma^5\gamma^\xi\left(\gamma_{[\zeta}D_{\alpha]}\partial_\xi\uptheta^5 - i\gamma_{[\zeta}\partial_{\alpha]}\tilde{\uptheta}A^5_\xi\right)
\nonumber\\
    &\qquad+ i\gamma^5\left(\left[A_{[\alpha},\partial_{\zeta]}\uptheta^5\right] + \left[\partial_{[\alpha}\uptheta,A_{\zeta]}^5\right]\right)+i\left(\left[A^5_{[\alpha},\partial_{|\xi|}\uptheta^5\right]-\left[A^5_{|\xi|},\partial_{[\alpha}\uptheta^5\right]\right)\gamma^\xi\gamma_{\zeta]}
\nonumber\\
    &\qquad-i\gamma^\xi\left( A^5_{|\xi|}\gamma_{[\alpha}\gamma^\beta\partial_{|\beta|}\uptheta^5\gamma_{\zeta]}+\partial_{|\xi|}\uptheta^5\gamma_{[\alpha}\gamma^\beta \mr{A}^5_{|\beta|}\gamma_{\zeta]}\right)\Bigg\}
\end{align}

\begin{align}\label{RdOmdOm}
    R(\delta\Cur_{\alpha\zeta})(\delta\Cur_{\alpha\zeta})&=R\Bigg\{i\left(A_{[\alpha},\partial_{\zeta]}\uptheta\right)+i\left[A^5_{[\alpha},\partial_{\zeta]}\uptheta^5\right]+\gamma^5\gamma^\xi\left(\gamma_{[\zeta}D_{\alpha]}\partial_\xi\uptheta^5 - i\gamma_{[\zeta}\partial_{\alpha]}\tilde{\uptheta}A^5_\xi\right)
\nonumber\\
    &\qquad+ i\gamma^5\left(\left[A_{[\alpha},\partial_{\zeta]}\uptheta^5\right] + \left[\partial_{[\alpha}\uptheta,A_{\zeta]}^5\right]\right)+i\left(\left[A^5_{[\alpha},\partial_{|\xi|}\uptheta^5\right]-\left[A^5_{|\xi|},\partial_{[\alpha}\uptheta^5\right]\right)\gamma^\xi\gamma_{\zeta]}
\nonumber\\
    &\qquad-i\gamma^\xi\left( A^5_{|\xi|}\gamma_{[\alpha}\gamma^\beta\partial_{|\beta|}\uptheta^5\gamma_{\zeta]}+\partial_{|\xi|}\uptheta^5\gamma_{[\alpha}\gamma^\beta \mr{A}^5_{|\beta|}\gamma_{\zeta]}\right)\Bigg\}
\nonumber\\    
    &\times \Bigg\{i\left(A_{[\alpha},\partial_{\zeta]}\uptheta\right)+i\left[A^5_{[\alpha},\partial_{\zeta]}\uptheta^5\right]+\gamma^5\gamma^\nu\left(\gamma_{[\zeta}D_{\alpha]}\partial_\nu\uptheta^5 - i\gamma_{[\zeta}\partial_{\alpha]}\tilde{\uptheta}A^5_\nu\right)
\nonumber\\
    &\qquad+ i\gamma^5\left(\left[A_{[\alpha},\partial_{\zeta]}\uptheta^5\right] + \left[\partial_{[\alpha}\uptheta,A_{\zeta]}^5\right]\right)+i\left(\left[A^5_{[\alpha},\partial_{|\nu|}\uptheta^5\right]-\left[A^5_{|\nu|},\partial_{[\alpha}\uptheta^5\right]\right)\gamma^\nu\gamma_{\zeta]}
\nonumber\\
    &\qquad-i\gamma^\nu\left( A^5_{|\nu|}\gamma_{[\alpha}\gamma^\chi\partial_{|\chi|}\uptheta^5\gamma_{\zeta]}+\partial_{|\nu|}\uptheta^5\gamma_{[\alpha}\gamma^\chi \mr{A}^5_{|\chi|}\gamma_{\zeta]}\right)\Bigg\}~,
\end{align}

\begin{align}\label{D4dE}
    (\delta E)_{;\rho\rho\sigma\sigma}&=\frac{i}{4}\left(A_{[\mu},\partial_{\nu]}\uptheta\right)_{;\rho\rho\sigma\sigma}\left[\gamma^\mu,\gamma^\nu \right] + \gamma^5\left(-D_\mu\partial^\mu\uptheta^5 -i \partial^\mu\tilde{\uptheta}A^5_\mu + \partial^\mu\tilde{\uptheta}\partial_\mu\uptheta^5\right)_{;\rho\rho\sigma\sigma}
\nonumber\\
    &-i(n-2)\left(2A^5_\mu\partial^\mu\uptheta^5 -\partial_\mu\uptheta^5\partial^\mu\uptheta^5 \right)_{;\rho\rho\sigma\sigma}-\frac{i(n-3)}{4}\left[\gamma^\mu ,\gamma^\nu\right] \left[A^5_{[\mu},\partial_{\nu]}\uptheta^5\right]_{;\rho\rho\sigma\sigma}~,
\end{align}

\begin{align}\label{dED2E}
    (\delta E)E_{;\rho\rho}&=\Bigg\{\frac{i}{4}A_{[\alpha},\partial_{\beta]}\uptheta\left[\gamma^\alpha,\gamma^\beta \right] + \gamma^5\left( -D_\alpha\partial^\alpha\uptheta^5 -i \partial^\alpha\tilde{\uptheta}A^5_\alpha + \partial^\alpha\tilde{\uptheta}\partial_\alpha\uptheta^5\right)
\nonumber\\
    &\qquad-i(n-2)\left(2A^5_\alpha\partial^\alpha\uptheta^5 -\partial_\alpha\uptheta^5\partial^\alpha\uptheta^5 \right)-\frac{i(n-3)}{4}\left[\gamma^\alpha ,\gamma^\beta\right]\left[A^5_{[\alpha},\partial_{\beta]}\uptheta^5\right]\Bigg\}
\nonumber\\    
    &\times\Bigg\{-\frac{1}{4}R_{;\rho\rho} + \frac{1}{4}\left[\gamma^\mu,\gamma^\nu \right]F_{\mu\nu;\rho\rho}
+i\gamma^5 \left({\rm d}^\mu A_\mu^5\right)_{;\rho\rho} -(n-2)\left(A_\mu^5 A^{5\mu}\right)_{;\rho\rho} 
\nonumber\\
    &\qquad-\frac{(n-3)}{4}\left[\gamma^\mu ,\gamma^\nu\right]\left[A_\mu^5,A_\nu^5 \right]_{;\rho\rho}\Bigg\}~,
\end{align}

\begin{align}\label{ED2dE}
    E(\delta E)_{;\rho\rho}&=\Bigg\{-\frac{1}{4}R + \frac{1}{4}\left[\gamma^\mu,\gamma^\nu \right]F_{\mu\nu}
+i\gamma^5 {\rm d}^\mu A_\mu^5 -(n-2)A_\mu^5 A^{5\mu} -\frac{(n-3)}{4}\left[\gamma^\mu ,\gamma^\nu\right]\left[A_\mu^5,A_\nu^5 \right]\Bigg\}
\nonumber\\
&\times\Bigg\{\frac{i}{4}\left(A_{[\alpha},\partial_{\beta]}\uptheta\right)_{;\rho\rho}\left[\gamma^\alpha,\gamma^\beta \right] + \gamma^5\left( -D_\alpha\partial^\alpha\uptheta^5 -i \partial^\alpha\tilde{\uptheta}A^5_\alpha + \partial^\alpha\tilde{\uptheta}\partial_\alpha\uptheta^5\right)_{;\rho\rho}
\nonumber\\
    &\qquad-i(n-2)\left(2A^5_\alpha\partial^\alpha\uptheta^5 -\partial_\alpha\uptheta^5\partial^\alpha\uptheta^5 \right)_{;\rho\rho}-\frac{i(n-3)}{4}\left[\gamma^\alpha ,\gamma^\beta\right]\left[A^5_{[\alpha},\partial_{\beta]}\uptheta^5\right]_{;\rho\rho}\Bigg\}~,
\end{align}

\begin{align}\label{dED2dE}
    (\delta E)(\delta E)_{;\rho\rho}&=\Bigg\{\frac{i}{4}\left(A_{[\mu},\partial_{\nu]}\uptheta\right)\left[\gamma^\mu,\gamma^\nu \right] + \gamma^5\left( -D_\mu\partial^\mu\uptheta^5 -i \partial^\mu\tilde{\uptheta}A^5_\mu + \partial^\mu\tilde{\uptheta}\partial_\mu\uptheta^5\right)\nonumber\\
    &\qquad-i(n-2)f\left(2A^5_\mu\partial^\mu\uptheta^5 -\partial_\mu\uptheta^5\partial^\mu\uptheta^5 \right)-\frac{i(n-3)}{4}\left[\gamma^\mu ,\gamma^\nu\right] \left[A^5_{[\mu},\partial_{\nu]}\uptheta^5\right]\Bigg\}
\nonumber\\
&\times\Bigg\{\frac{i}{4}\left(A_{[\alpha},\partial_{\beta]}\uptheta\right)_{;\rho\rho}\left[\gamma^\alpha,\gamma^\beta \right] + \gamma^5\left( -D_\alpha\partial^\alpha\uptheta^5 -i \partial^\alpha\tilde{\uptheta}A^5_\alpha + \partial^\alpha\tilde{\uptheta}\partial_\alpha\uptheta^5\right)_{;\rho\rho}
\nonumber\\
    &\qquad-i(n-2)\left(2A^5_\alpha\partial^\alpha\uptheta^5 -\partial_\alpha\uptheta^5\partial^\alpha\uptheta^5 \right)_{;\rho\rho}-\frac{i(n-3)}{4}\left[\gamma^\alpha ,\gamma^\beta\right]\left[A^5_{[\alpha},\partial_{\beta]}\uptheta^5\right]_{;\rho\rho}\Bigg\}~,
\end{align}

\begin{align}\label{DEDdE}
    2E_{;\rho}(\delta E)_{;\rho}&=2\Bigg\{-\frac{1}{4}R_{;\rho} + \frac{1}{4}\left[\gamma^\mu,\gamma^\nu \right]F_{\mu\nu;\rho}
+i\gamma^5\left({\rm d}^\mu A_\mu^5\right)_{;\rho} -(n-2)\left(A_\mu^5 A^{5\mu}\right)_{;\rho}
\nonumber\\
    &\qquad-\frac{(n-3)}{4}\left[\gamma^\mu ,\gamma^\nu\right]\left[A_\mu^5,A_\nu^5 \right]_{;\rho}\Bigg\}
\nonumber\\
&\quad\times\Bigg\{\frac{i}{4}\left(A_{[\alpha},\partial_{\beta]}\uptheta\right)_{;\rho}\left[\gamma^\alpha,\gamma^\beta \right] + \gamma^5\left( -D_\alpha\partial^\alpha\uptheta^5 -i \partial^\alpha\tilde{\uptheta}A^5_\alpha + \partial^\alpha\tilde{\uptheta}\partial_\alpha\uptheta^5\right)_{;\rho}
\nonumber\\
    &\qquad-i(n-2)\left(2A^5_\alpha\partial^\alpha\uptheta^5 -\partial_\alpha\uptheta^5\partial^\alpha\uptheta^5 \right)_{;\rho}-\frac{i(n-3)}{4}\left[\gamma^\alpha ,\gamma^\beta\right]\left[A^5_{[\alpha},\partial_{\beta]}\uptheta^5\right]_{;\rho}\Bigg\}~,
\end{align}

\begin{align}\label{DdEDdE}
    (\delta E)_{;\rho}(\delta E)_{;\rho}&=\Bigg\{\frac{i}{4}\left(A_{[\mu},\partial_{\nu]}\uptheta\right)_{;\rho}\left[\gamma^\mu,\gamma^\nu \right] + \gamma^5\left( -D_\mu\partial^\mu\uptheta^5 -i \partial^\mu\tilde{\uptheta}A^5_\mu + \partial^\mu\tilde{\uptheta}\partial_\mu\uptheta^5\right)_{;\rho}\nonumber\\
    &\qquad-i(n-2)f\left(2A^5_\mu\partial^\mu\uptheta^5 -\partial_\mu\uptheta^5\partial^\mu\uptheta^5 \right)_{;\rho}-\frac{i(n-3)}{4}\left[\gamma^\mu ,\gamma^\nu\right] \left[A^5_{[\mu},\partial_{\nu]}\uptheta^5\right]_{;\rho}\Bigg\}
\nonumber\\    
    &\times\Bigg\{\frac{i}{4}\left(A_{[\alpha},\partial_{\beta]}\uptheta\right)_{;\rho}\left[\gamma^\alpha,\gamma^\beta \right] + \gamma^5\left( -D_\alpha\partial^\alpha\uptheta^5 -i \partial^\alpha\tilde{\uptheta}A^5_\alpha + \partial^\alpha\tilde{\uptheta}\partial_\alpha\uptheta^5\right)_{;\rho}
\nonumber\\
    &\qquad-i(n-2)\left(2A^5_\alpha\partial^\alpha\uptheta^5 -\partial_\alpha\uptheta^5\partial^\alpha\uptheta^5 \right)_{;\rho}-\frac{i(n-3)}{4}\left[\gamma^\alpha ,\gamma^\beta\right]\left[A^5_{[\alpha},\partial_{\beta]}\uptheta^5\right]_{;\rho}\Bigg\}~,
\end{align}

\begin{align}\label{E2dE}
    3E^2(\delta E)&=3\Bigg\{-\frac{1}{4}R + \frac{1}{4}\left[\gamma^\mu,\gamma^\nu \right]F_{\mu\nu}
+i\gamma^5 {\rm d}^\mu A_\mu^5 -(n-2)A_\mu^5 A^{5\mu} -\frac{(n-3)}{4}\left[\gamma^\mu ,\gamma^\nu\right]\left[A_\mu^5,A_\nu^5 \right]\Bigg\}
\nonumber\\
    &\times\Bigg\{-\frac{1}{4}R + \frac{1}{4}\left[\gamma^\xi,\gamma^\chi \right]F_{\xi\chi}
+i\gamma^5 {\rm d}^\xi A_\xi^5 -(n-2)A_\xi^5 A^{5\xi} -\frac{(n-3)}{4}\left[\gamma^\xi ,\gamma^\chi\right]\left[A_\xi^5,A_\chi^5 \right]\Bigg\}
\nonumber\\
    &\times\Bigg\{\frac{i}{4}A_{[\alpha},\partial_{\beta]}\uptheta\left[\gamma^\alpha,\gamma^\beta \right] + \gamma^5\left( -D_\alpha\partial^\alpha\uptheta^5 -i \partial^\alpha\tilde{\uptheta}A^5_\alpha + \partial^\alpha\tilde{\uptheta}\partial_\alpha\uptheta^5\right)
\nonumber\\
    &\qquad-i(n-2)\left(2A^5_\alpha\partial^\alpha\uptheta^5 -\partial_\alpha\uptheta^5\partial^\alpha\uptheta^5 \right)-\frac{i(n-3)}{4}\left[\gamma^\alpha ,\gamma^\beta\right]\left[A^5_{[\alpha},\partial_{\beta]}\uptheta^5\right]\Bigg\}~,
\end{align}

\begin{align}\label{EdE2}
    3E(\delta E)^2&=3\Bigg\{-\frac{1}{4}R + \frac{1}{4}\left[\gamma^\xi,\gamma^\chi \right]F_{\xi\chi}
+i\gamma^5 {\rm d}^\xi A_\xi^5 -(n-2)A_\xi^5 A^{5\xi} -\frac{(n-3)}{4}\left[\gamma^\xi ,\gamma^\chi\right]\left[A_\xi^5,A_\chi^5 \right]\Bigg\}
\nonumber\\
    &\times\Bigg\{\frac{i}{4}\left(A_{[\mu},\partial_{\nu]}\uptheta\right)\left[\gamma^\mu,\gamma^\nu \right] + \gamma^5\left( -D_\mu\partial^\mu\uptheta^5 -i \partial^\mu\tilde{\uptheta}A^5_\mu + \partial^\mu\tilde{\uptheta}\partial_\mu\uptheta^5\right)\nonumber\\
    &\qquad-i(n-2)f\left(2A^5_\mu\partial^\mu\uptheta^5 -\partial_\mu\uptheta^5\partial^\mu\uptheta^5 \right)-\frac{i(n-3)}{4}\left[\gamma^\mu ,\gamma^\nu\right] \left[A^5_{[\mu},\partial_{\nu]}\uptheta^5\right]\Bigg\}
\nonumber\\
    &\times\Bigg\{\frac{i}{4}\left(A_{[\alpha},\partial_{\beta]}\uptheta\right)\left[\gamma^\alpha,\gamma^\beta \right] + \gamma^5\left( -D_\alpha\partial^\alpha\uptheta^5 -i \partial^\alpha\tilde{\uptheta}A^5_\alpha + \partial^\alpha\tilde{\uptheta}\partial_\alpha\uptheta^5\right)
\nonumber\\
    &\qquad-i(n-2)f\left(2A^5_\alpha\partial^\alpha\uptheta^5 -\partial_\alpha\uptheta^5\partial^\alpha\uptheta^5 \right)-\frac{i(n-3)}{4}\left[\gamma^\alpha ,\gamma^\beta\right] \left[A^5_{[\alpha},\partial_{\beta]}\uptheta^5\right]\Bigg\}~,
\end{align}

\begin{align}\label{dE3}
    (\delta E)^3&=\Bigg\{\frac{i}{4}\left(A_{[\xi},\partial_{\chi]}\uptheta\right)\left[\gamma^\xi,\gamma^\chi \right] + \gamma^5\left( -D_\xi\partial^\xi\uptheta^5 -i \partial^\xi\tilde{\uptheta}A^5_\xi + \partial^\xi\tilde{\uptheta}\partial_\xi\uptheta^5\right)
\nonumber\\
    &\qquad-i(n-2)f\left(2A^5_\xi\partial^\xi\uptheta^5 -\partial_\xi\uptheta^5\partial^\xi\uptheta^5 \right)-\frac{i(n-3)}{4}\left[\gamma^\xi ,\gamma^\chi\right] \left[A^5_{[\xi},\partial_{\chi]}\uptheta^5\right]\Bigg\}
\nonumber\\
    &\times\Bigg\{\frac{i}{4}\left(A_{[\mu},\partial_{\nu]}\uptheta\right)\left[\gamma^\mu,\gamma^\nu \right] + \gamma^5\left( -D_\mu\partial^\mu\uptheta^5 -i \partial^\mu\tilde{\uptheta}A^5_\mu + \partial^\mu\tilde{\uptheta}\partial_\mu\uptheta^5\right)
\nonumber\\
    &\qquad-i(n-2)f\left(2A^5_\mu\partial^\mu\uptheta^5 -\partial_\mu\uptheta^5\partial^\mu\uptheta^5 \right)-\frac{i(n-3)}{4}\left[\gamma^\mu ,\gamma^\nu\right] \left[A^5_{[\mu},\partial_{\nu]}\uptheta^5\right]\Bigg\}
\nonumber\\
    &\times\Bigg\{\frac{i}{4}\left(A_{[\alpha},\partial_{\beta]}\uptheta\right)\left[\gamma^\alpha,\gamma^\beta \right] + \gamma^5\left( -D_\alpha\partial^\alpha\uptheta^5 -i \partial^\alpha\tilde{\uptheta}A^5_\alpha + \partial^\alpha\tilde{\uptheta}\partial_\alpha\uptheta^5\right)
\nonumber\\
    &\qquad-i(n-2)f\left(2A^5_\alpha\partial^\alpha\uptheta^5 -\partial_\alpha\uptheta^5\partial^\alpha\uptheta^5 \right)-\frac{i(n-3)}{4}\left[\gamma^\alpha ,\gamma^\beta\right] \left[A^5_{[\alpha},\partial_{\beta]}\uptheta^5\right]\Bigg\}~,
\end{align}

\begin{align}\label{dEOmOm}
(\delta E)\Cur_{\rho\sigma}\Cur_{\rho\sigma}&=\Bigg\{\frac{i}{4}\left(A_{[\alpha},\partial_{\beta]}\uptheta\right)\left[\gamma^\alpha,\gamma^\beta \right] + \gamma^5\left( -D_\alpha\partial^\alpha\uptheta^5 -i \partial^\alpha\tilde{\uptheta}A^5_\alpha + \partial^\alpha\tilde{\uptheta}\partial_\alpha\uptheta^5\right)
\nonumber\\
    &\qquad-i(n-2)f\left(2A^5_\alpha\partial^\alpha\uptheta^5 -\partial_\alpha\uptheta^5\partial^\alpha\uptheta^5 \right)-\frac{i(n-3)}{4}\left[\gamma^\alpha ,\gamma^\beta\right] \left[A^5_{[\alpha},\partial_{\beta]}\uptheta^5\right]\Bigg\}
\nonumber\\    
    &\times\Bigg\{F_{\rho\sigma}-[A_\rho^5 ,A_\sigma^5 ]-\frac 14\gamma^\mu\gamma^\nu R_{\mu\nu\rho\sigma}-i\gamma^5\gamma^\mu\left(\gamma_{[\sigma} D_{\rho]} A_\mu^5\right)
\nonumber\\
    &\qquad+i\gamma^5 A_{\rho\sigma}^5+[A_{[\rho}^5,A_{|\mu|}^5]\gamma^\mu\gamma_{\sigma]} -\gamma^\mu A_\mu^5\gamma_{[\rho} \gamma^\nu A_{|\nu|}^5 \gamma_{\sigma]}\Bigg\}
\nonumber\\
    &\times\Bigg\{F_{\rho\sigma}-[A_\rho^5 ,A_\sigma^5 ]-\frac 14\gamma^\xi\gamma^\lambda R_{\xi\lambda\rho\sigma}-i\gamma^5\gamma^\xi\left(\gamma_{[\sigma}D_{\rho]}A_\xi^5\right)
\nonumber\\
    &\qquad+i\gamma^5 A_{\rho\sigma}^5+[A_{[\rho}^5,A_{|\xi|}^5]\gamma^\xi\gamma_{\sigma]} -\gamma^\xi A_\xi^5\gamma_{[\rho}\gamma^\lambda A_{|\lambda|}^5 \gamma_{\sigma]}\Bigg\}~,
\end{align}

\begin{align}\label{dEOmdOm}
(\delta E)\left[2\Cur_{\rho\sigma}(\delta\Cur_{\rho\sigma})\right]&=2\Bigg\{\frac{i}{4}\left(A_{[\alpha},\partial_{\beta]}\uptheta\right)\left[\gamma^\alpha,\gamma^\beta \right] + \gamma^5\left( -D_\alpha\partial^\alpha\uptheta^5 -i \partial^\alpha\tilde{\uptheta}A^5_\alpha + \partial^\alpha\tilde{\uptheta}\partial_\alpha\uptheta^5\right)
\nonumber\\
    &\qquad-i(n-2)f\left(2A^5_\alpha\partial^\alpha\uptheta^5 -\partial_\alpha\uptheta^5\partial^\alpha\uptheta^5 \right)-\frac{i(n-3)}{4}\left[\gamma^\alpha ,\gamma^\beta\right] \left[A^5_{[\alpha},\partial_{\beta]}\uptheta^5\right]\Bigg\}
\nonumber\\    
    &\times\Bigg\{F_{\rho\sigma}-[A_\rho^5 ,A_\sigma^5 ]-\frac 14\gamma^\mu\gamma^\nu R_{\mu\nu\rho\sigma}-i\gamma^5\gamma^\mu\left(\gamma_{[\sigma} D_{\rho]} A_\mu^5\right)
\nonumber\\
    &\qquad+i\gamma^5 A_{\rho\sigma}^5+[A_{[\rho}^5,A_{|\mu|}^5]\gamma^\mu\gamma_{\sigma]} -\gamma^\mu A_\mu^5\gamma_{[\rho} \gamma^\nu A_{|\nu|}^5 \gamma_{\sigma]}\Bigg\}
\nonumber\\
    &\times\Bigg\{i\left(A_{[\rho},\partial_{\sigma]}\uptheta\right)+i\left[A^5_{[\rho},\partial_{\sigma]}\uptheta^5\right]+\gamma^5\gamma^\xi\gamma_{[\sigma}D_{\rho]}\partial_\xi\uptheta^5 - i\gamma^5\gamma^\xi\gamma_{[\sigma}\partial_{\rho]}\tilde{\uptheta}A^5_\xi
\nonumber\\
    &\quad+ i\gamma^5\left(\left[A_{[\rho},\partial_{\sigma]}\uptheta^5\right] + \left[\partial_{[\rho}\uptheta,A_{\sigma]}^5\right]\right)+i\left(\left[A^5_{[\rho},\partial_{|\xi|}\uptheta^5\right]-\left[A^5_{|\xi|},\partial_{[\rho}\uptheta^5\right]\right)\gamma^\xi\gamma_{\sigma]}
\nonumber\\
    &\quad-i\gamma^\xi\left( A^5_{|\xi|}\gamma_{[\rho}\gamma^\lambda\partial_{|\lambda|}\uptheta^5\gamma_{\sigma]}+\partial_{|\xi|}\uptheta^5\gamma_{[\rho}\gamma^\lambda \mr{A}^5_{|\lambda|}\gamma_{\sigma]}\right)\Bigg\}
\end{align}

\begin{align}\label{dEdOmdOm}
    (\delta E)(\delta\Cur_{\rho\sigma})(\delta\Cur_{\rho\sigma})&=\Bigg\{\frac{i}{4}\left(A_{[\xi},\partial_{\chi]}\uptheta\right)\left[\gamma^\xi,\gamma^\chi \right] + \gamma^5\left( -D_\xi\partial^\xi\uptheta^5 -i \partial^\xi\tilde{\uptheta}A^5_\xi + \partial^\xi\tilde{\uptheta}\partial_\xi\uptheta^5\right)
\nonumber\\
    &\qquad-i(n-2)f\left(2A^5_\xi\partial^\xi\uptheta^5 -\partial_\xi\uptheta^5\partial^\xi\uptheta^5 \right)-\frac{i(n-3)}{4}\left[\gamma^\xi ,\gamma^\chi\right] \left[A^5_{[\xi},\partial_{\chi]}\uptheta^5\right]\Bigg\}
\nonumber\\    
    &\times\Bigg\{i\left(A_{[\rho},\partial_{\sigma]}\uptheta\right)+i\left[A^5_{[\rho},\partial_{\sigma]}\uptheta^5\right]+\gamma^5\gamma^\mu\left(\gamma_{[\sigma}D_{\rho]}\partial_\mu\uptheta^5 - i\gamma_{[\sigma}\partial_{\rho]}\tilde{\uptheta}A^5_\mu\right)
\nonumber\\
    &\qquad+ i\gamma^5\left(\left[A_{[\rho},\partial_{\sigma]}\uptheta^5\right] + \left[\partial_{[\rho}\uptheta,A_{\sigma]}^5\right]\right)+i\left(\left[A^5_{[\rho},\partial_{|\mu|}\uptheta^5\right]-\left[A^5_{|\mu|},\partial_{[\rho}\uptheta^5\right]\right)\gamma^\mu\gamma_{\sigma]}
\nonumber\\
    &\qquad-i\gamma^\mu\left( A^5_{|\mu|}\gamma_{[\rho}\gamma^\nu\partial_{|\nu|}\uptheta^5\gamma_{\sigma]}+\partial_{|\mu|}\uptheta^5\gamma_{[\rho}\gamma^\nu \mr{A}^5_{|\nu|}\gamma_{\sigma]}\right)\Bigg\}\nonumber\\
&\times \Bigg\{i\left(A_{[\rho},\partial_{\sigma]}\uptheta\right)+i\left[A^5_{[\rho},\partial_{\sigma]}\uptheta^5\right]+\gamma^5\gamma^\alpha\left(\gamma_{[\sigma}D_{\rho]}\partial_\alpha\uptheta^5 - i\gamma_{[\sigma}\partial_{\rho]}\tilde{\uptheta}A^5_\alpha\right)
\nonumber\\
    &\qquad+ i\gamma^5\left(\left[A_{[\rho},\partial_{\sigma]}\uptheta^5\right] + \left[\partial_{[\rho}\uptheta,A_{\sigma]}^5\right]\right)+i\left(\left[A^5_{[\rho},\partial_{|\alpha|}\uptheta^5\right]-\left[A^5_{|\alpha|},\partial_{[\rho}\uptheta^5\right]\right)\gamma^\alpha\gamma_{\sigma]}
\nonumber\\
    &\qquad-i\gamma^\alpha\left( A^5_{|\alpha|}\gamma_{[\rho}\gamma^\beta\partial_{|\beta|}\uptheta^5\gamma_{\sigma]}+\partial_{|\alpha|}\uptheta^5\gamma_{[\rho}\gamma^\beta \mr{A}^5_{|\beta|}\gamma_{\sigma]}\right)\Bigg\}
\end{align}

\begin{align}\label{RD2dE}
    R(\delta E)_{;\alpha\alpha}&=R\Bigg\{\frac{i}{4}\left(A_{[\mu},\partial_{\nu]}\uptheta\right)_{;\alpha\alpha}\left[\gamma^\mu,\gamma^\nu \right] + \gamma^5\left( -D_\mu\partial^\mu\uptheta^5 -i \partial^\mu\tilde{\uptheta}A^5_\mu + \partial^\mu\tilde{\uptheta}\partial_\mu\uptheta^5\right)_{;\alpha\alpha}
\nonumber\\
    &\quad-i(n-2)\left(2A^5_\mu\partial^\mu\uptheta^5 -\partial_\mu\uptheta^5\partial^\mu\uptheta^5 \right)_{;\alpha\alpha}-\frac{i(n-3)}{4}\left[\gamma^\mu ,\gamma^\nu\right]\left( \left[A^5_{[\mu},\partial_{\nu]}\uptheta^5\right]\right)_{;\alpha\alpha}\Bigg\}
\end{align}

\begin{align}\label{RicDDdE}
    R_{\sigma\alpha}(\delta E)_{;\sigma\alpha}&=R_{\sigma\alpha}\Bigg\{\frac{i}{4}\left(A_{[\mu},\partial_{\nu]}\uptheta\right)_{;\sigma\alpha}\left[\gamma^\mu,\gamma^\nu \right] + \gamma^5\left( -D_\mu\partial^\mu\uptheta^5 -i \partial^\mu\tilde{\uptheta}A^5_\mu + \partial^\mu\tilde{\uptheta}\partial_\mu\uptheta^5\right)_{;\sigma\alpha}
\nonumber\\
    &\quad-i(n-2)\left(2A^5_\mu\partial^\mu\uptheta^5 -\partial_\mu\uptheta^5\partial^\mu\uptheta^5 \right)_{;\sigma\alpha}-\frac{i(n-3)}{4}\left[\gamma^\mu ,\gamma^\nu\right]\left( \left[A^5_{[\mu},\partial_{\nu]}\uptheta^5\right]\right)_{;\sigma\alpha}\Bigg\}~,
\end{align}

\begin{align}\label{DRDdE}
    R_{;\alpha}(\delta E)_{;\alpha}&=R_{;\alpha}\Bigg\{\frac{i}{4}\left(A_{[\mu},\partial_{\nu]}\uptheta\right)_{;\alpha}\left[\gamma^\mu,\gamma^\nu \right] + \gamma^5\left( -D_\mu\partial^\mu\uptheta^5 -i \partial^\mu\tilde{\uptheta}A^5_\mu + \partial^\mu\tilde{\uptheta}\partial_\mu\uptheta^5\right)_{;\alpha}
\nonumber\\
    &\quad-i(n-2)\left(2A^5_\mu\partial^\mu\uptheta^5 -\partial_\mu\uptheta^5\partial^\mu\uptheta^5 \right)_{;\alpha}-\frac{i(n-3)}{4}\left[\gamma^\mu ,\gamma^\nu\right]\left( \left[A^5_{[\mu},\partial_{\nu]}\uptheta^5\right]\right)_{;\alpha}\Bigg\}~,
\end{align}

\begin{align}\label{REdE}
    R\left[2E(\delta E)\right]&=2R\Bigg\{-\frac{1}{4}R + \frac{1}{4}\left[\gamma^\xi,\gamma^\chi \right]F_{\xi\chi}
+i\gamma^5 {\rm d}^\xi A_\xi^5 -(n-2)A_\xi^5 A^{5\xi}
\nonumber\\
    &\qquad-\frac{(n-3)}{4}\left[\gamma^\xi ,\gamma^\chi\right]\left[A_\xi^5,A_\chi^5 \right]\Bigg\}
\nonumber\\
    &\times\Bigg\{\frac{i}{4}\left(A_{[\mu},\partial_{\nu]}\uptheta\right)\left[\gamma^\mu,\gamma^\nu \right] + \gamma^5\left( -D_\mu\partial^\mu\uptheta^5 -i \partial^\mu\tilde{\uptheta}A^5_\mu + \partial^\mu\tilde{\uptheta}\partial_\mu\uptheta^5\right)\nonumber\\
    &\qquad-i(n-2)f\left(2A^5_\mu\partial^\mu\uptheta^5 -\partial_\mu\uptheta^5\partial^\mu\uptheta^5 \right)-\frac{i(n-3)}{4}\left[\gamma^\mu ,\gamma^\nu\right] \left[A^5_{[\mu},\partial_{\nu]}\uptheta^5\right]\Bigg\}~,
\end{align}

\begin{align}\label{RdE2}
    R(\delta E)^2&=R\Bigg\{\frac{i}{4}\left(A_{[\xi},\partial_{\chi]}\uptheta\right)\left[\gamma^\xi,\gamma^\chi \right] + \gamma^5\left( -D_\xi\partial^\xi\uptheta^5 -i \partial^\xi\tilde{\uptheta}A^5_\xi + \partial^\xi\tilde{\uptheta}\partial_\xi\uptheta^5\right)
\nonumber\\
    &\qquad-i(n-2)f\left(2A^5_\xi\partial^\xi\uptheta^5 -\partial_\xi\uptheta^5\partial^\xi\uptheta^5 \right)-\frac{i(n-3)}{4}\left[\gamma^\xi ,\gamma^\chi\right] \left[A^5_{[\xi},\partial_{\chi]}\uptheta^5\right]\Bigg\}
\nonumber\\
    &\times\Bigg\{\frac{i}{4}\left(A_{[\mu},\partial_{\nu]}\uptheta\right)\left[\gamma^\mu,\gamma^\nu \right] + \gamma^5\left( -D_\mu\partial^\mu\uptheta^5 -i \partial^\mu\tilde{\uptheta}A^5_\mu + \partial^\mu\tilde{\uptheta}\partial_\mu\uptheta^5\right)
\nonumber\\
    &\qquad-i(n-2)f\left(2A^5_\mu\partial^\mu\uptheta^5 -\partial_\mu\uptheta^5\partial^\mu\uptheta^5 \right)-\frac{i(n-3)}{4}\left[\gamma^\mu ,\gamma^\nu\right] \left[A^5_{[\mu},\partial_{\nu]}\uptheta^5\right]\Bigg\}~,
\end{align}

\begin{align}\label{dEThetaR}
    (\delta E)\Theta\left(\mathcal{R}\right)&=\Bigg\{\frac{i}{4}\left(A_{[\xi},\partial_{\chi]}\uptheta\right)\left[\gamma^\xi,\gamma^\chi \right] + \gamma^5\left( -D_\xi\partial^\xi\uptheta^5 -i \partial^\xi\tilde{\uptheta}A^5_\xi + \partial^\xi\tilde{\uptheta}\partial_\xi\uptheta^5\right)
\nonumber\\
    &\quad-i(n-2)f\left(2A^5_\xi\partial^\xi\uptheta^5 -\partial_\xi\uptheta^5\partial^\xi\uptheta^5 \right)-\frac{i(n-3)}{4}\left[\gamma^\xi ,\gamma^\chi\right] \left[A^5_{[\xi},\partial_{\chi]}\uptheta^5\right]\Bigg\}\times\Theta\left(\mathcal{R}\right)~,
\end{align}
since $\Theta\left(\mathcal{R}\right)=12R_{;\alpha\alpha}+5R^2-2R_{\rho\sigma}R_{\rho\sigma}
+2R_{\rho\sigma\alpha\beta}R_{\rho\sigma\alpha\beta}$~.

\end{document}